\documentclass[twocolumn]{aastex631}

\newcommand{\HII}{H\,{\scriptsize II}}

\newcommand{\HI}{H\,{\scriptsize I}}
\newcommand{\CII}{[C\,{\scriptsize II}]}
\newcommand{\dCII}{[$^{\rm 13}$C\,{\scriptsize II}]}

\newcommand{\OI}{[O\,{\scriptsize I}]}

\newcommand{\two}{2$\to$1}
\shorttitle{DR21}
\shortauthors{Bonne et al.}
\graphicspath{{./}{figures/}}
\begin{document}

%\title{Unveiling the formation of the massive DR21 ridge with low- and high-density tracers}
\title{Unveiling the formation of the massive DR21 ridge}

\correspondingauthor{Lars Bonne}
\email{lbonne@usra.edu}

\author{L. Bonne}
\affiliation{SOFIA Science Center, NASA Ames Research Center, Moffett Field, CA 94 045, USA}
\author{S. Bontemps}
\affiliation{Laboratoire d'Astrophysique de Bordeaux, Universit\'e de Bordeaux, CNRS, B18N, all\'ee Geoffroy Saint-Hilaire, 33615 Pessac, France}
\author{N. Schneider}
\affiliation{I. Physikalisches Institut, Universit\"at zu K\"oln, Z\"ulpicher Str. 77, 50937 K\"oln}
\author{R. Simon}
\affiliation{I. Physikalisches Institut, Universit\"at zu K\"oln, Z\"ulpicher Str. 77, 50937 K\"oln}
\author{S. D. Clarke}
\affiliation{Academia Sinica, Institute of Astronomy and Astrophysics, Taipei, Taiwan}
\author{T. Csengeri}
\affiliation{Laboratoire d'Astrophysique de Bordeaux, Universit\'e de Bordeaux, CNRS, B18N, all\'ee Geoffroy Saint-Hilaire, 33615 Pessac, France}
\author{E. Chambers}
\affiliation{SOFIA Science Center, NASA Ames Research Center, Moffett Field, CA 94 045, USA}
\author{U. Graf}
\affiliation{I. Physikalisches Institut, Universit\"at zu K\"oln, Z\"ulpicher Str. 77, 50937 K\"oln}
\author{J. M. Jackson}
\affiliation{SOFIA Science Center, NASA Ames Research Center, Moffett Field, CA 94 045, USA}
\affiliation{Green Bank Observatory, PO Box 2, Green Bank, WV 24 944, USA}
\author{R. Klein}
\affiliation{SOFIA Science Center, NASA Ames Research Center, Moffett Field, CA 94 045, USA}
%\author{F. Motte}
%\affiliation{Univ. Grenoble Alpes, CNRS, IPAG, 38000 Grenoble, France}
\author{Y. Okada}
\affiliation{I. Physikalisches Institut, Universit\"at zu K\"oln, Z\"ulpicher Str. 77, 50937 K\"oln}
\author{A. G. G. M. Tielens}
\affiliation{University of Maryland, Department of Astronomy, College Park, MD 20742-2421, USA}
\affiliation{Leiden Observatory, PO Box 9513, 2300 RA Leiden, The Netherlands}
\author{M. Tiwari}
\affiliation{University of Maryland, Department of Astronomy, College Park, MD 20742-2421, USA}



\begin{abstract}
%To constrain the physical processes that initiate high-mass star formation, the formation and evolution of dense filamentary gas in molecular clouds has to be better understood. 
We present new $^{13}$CO(1-0), C$^{18}$O(1-0), HCO$^{+}$(1-0) and H$^{13}$CO$^{+}$(1-0) maps from the IRAM 30m telescope, and a spectrally-resolved \CII\ 158 $\mu$m map observed with the SOFIA telescope towards the massive DR21 cloud. This traces the kinematics from low- to high-density gas in the cloud which allows to constrain the formation scenario of the high-mass star forming DR21 ridge. The molecular line data reveals that the sub-filaments are systematically redshifted relative to the dense ridge. We demonstrate that \CII\ unveils the surrounding CO-poor gas of the dense filaments in the DR21 cloud. We also show that this surrounding gas is organized in a flattened cloud with curved redshifted dynamics perpendicular to the ridge. The sub-filaments thus form in this curved and flattened mass reservoir. A virial analysis of the different lines indicates that self-gravity should drive the evolution of the ridge and surrounding cloud.
Combining all results we propose that bending of the magnetic field, due to the interaction with a mostly atomic colliding cloud, explains the velocity field and resulting mass accretion on the ridge. This is remarkably similar to what was found for at least two nearby low-mass filaments. We tentatively propose that this scenario might be a widespread mechanism to initiate star formation in the Milky Way. However, in contrast to low-mass clouds, gravitational collapse plays a role on the pc scale of the DR21 ridge because of the higher density. This allows more effective mass collection at the centers of collapse and should facilitate massive cluster formation. 
\end{abstract}



\section{Introduction} \label{intro} 
\begin{figure*}
\begin{center}
\-\hspace{-1.6 cm} \includegraphics[width=0.8\hsize]{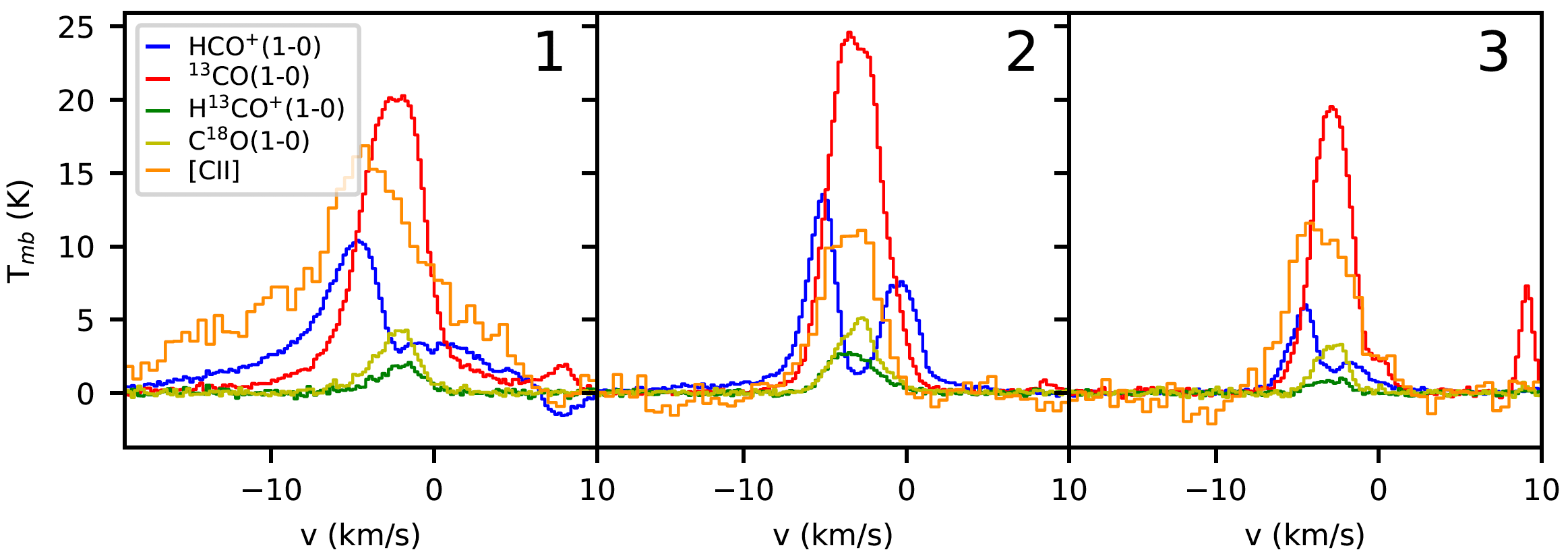}
\includegraphics[width=0.45\hsize]{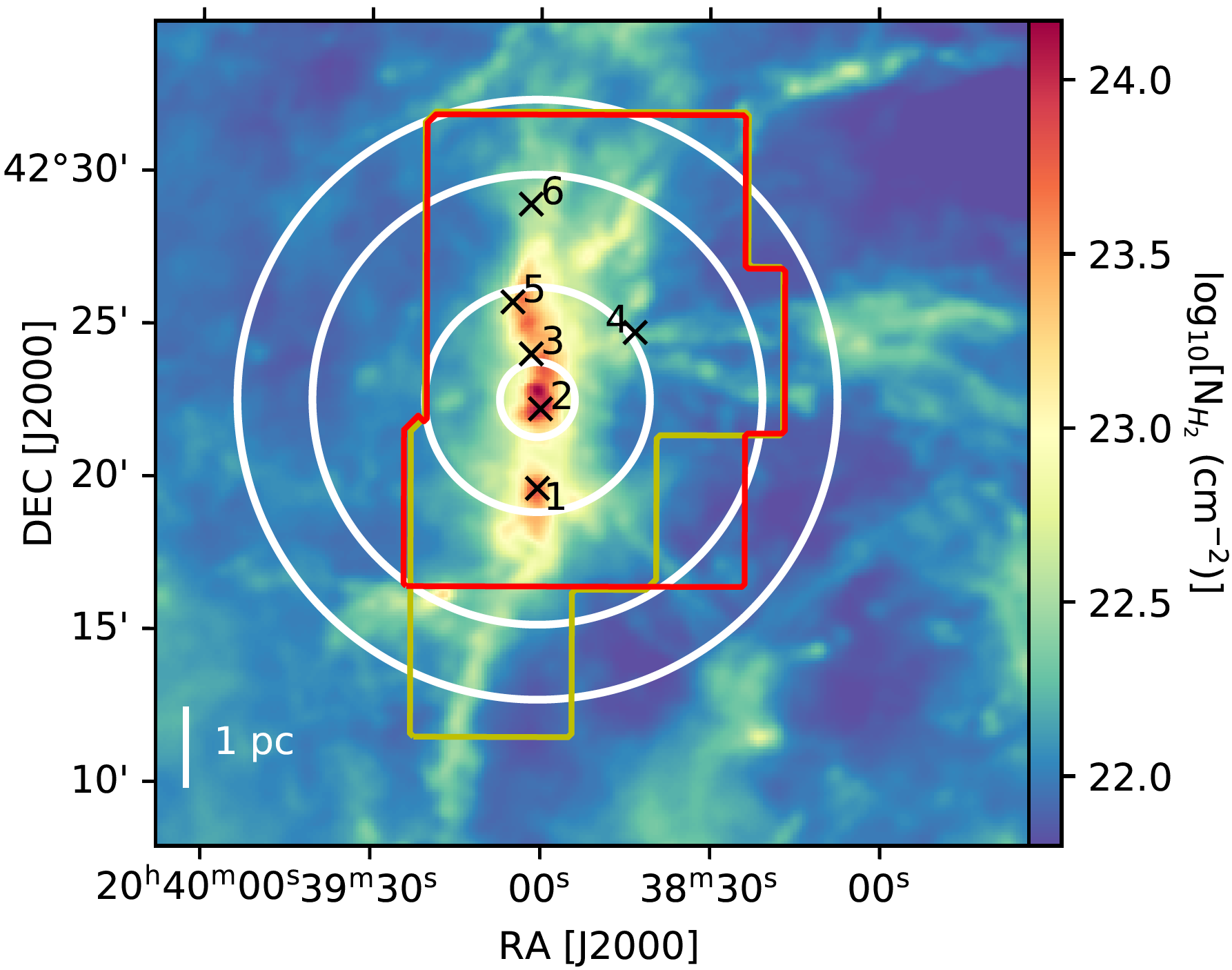}
\includegraphics[width=0.45\hsize]{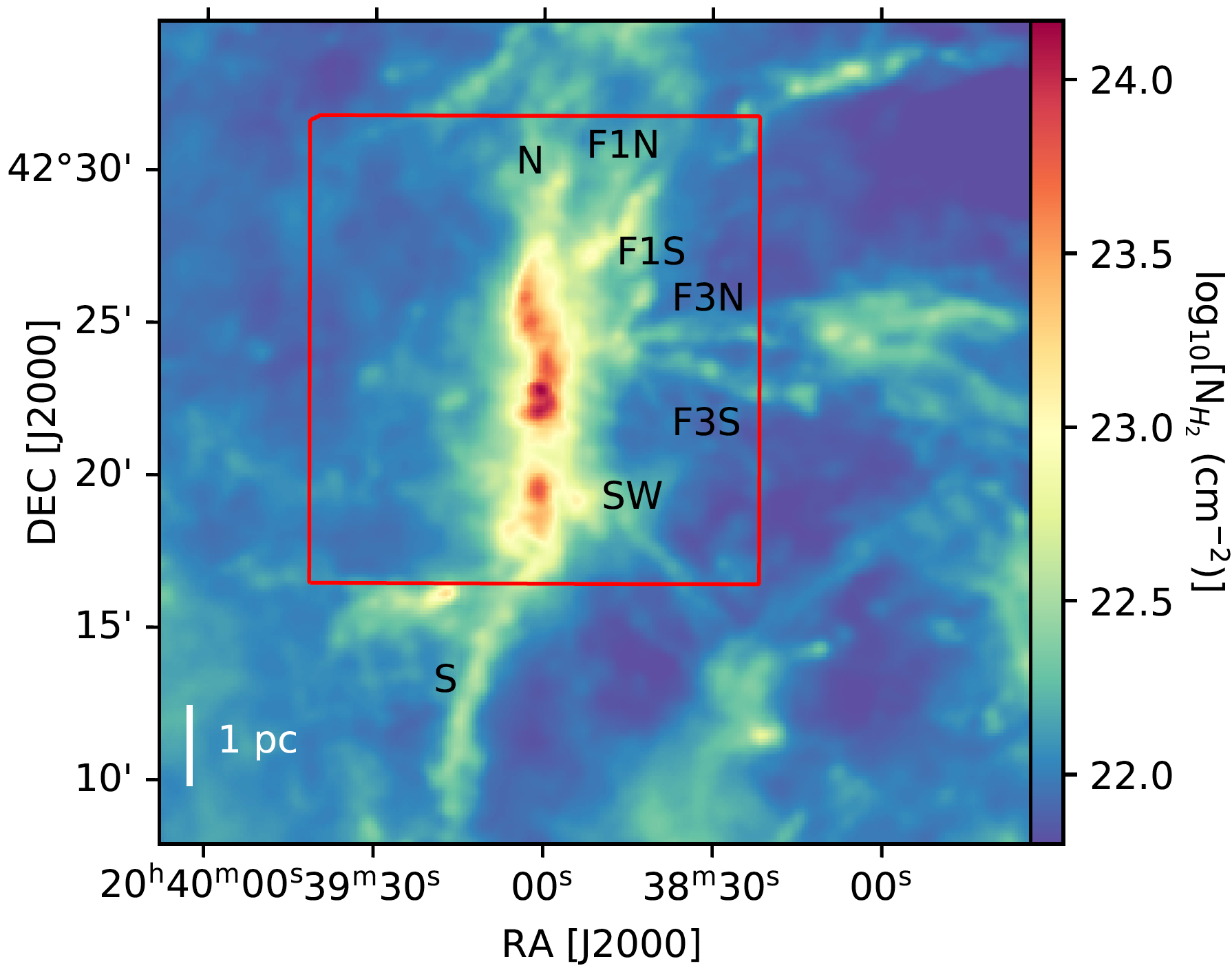}\\
\-\hspace{-1.4 cm}\includegraphics[width=0.8\hsize]{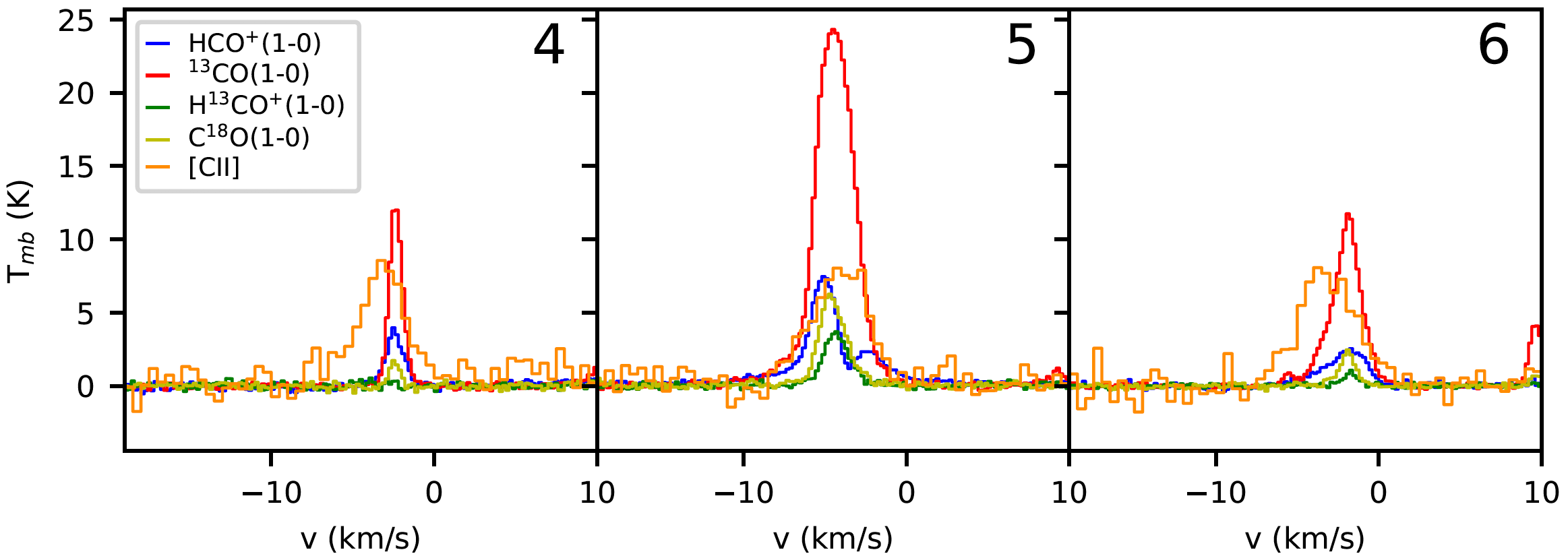}
\caption{{\bf Middle left}: \textit{Herschel} column density map \citep{Hennemann2012} of the DR21 cloud (ridge and surrounding sub-filaments). The yellow and red polygons indicate the area covered by the two different backend setups of the IRAM 30m observations, i.e. setup1 (yellow) and setup2 (red). The southern sub-filament is only covered by the first setup which contains HCO$^{+}$(1-0) and H$^{13}$CO$^{+}$(1-0). The crosses indicate the location of the spectra displayed in the bottom and top panels. The white circles  are centered on the DR21(OH) clump and indicate radii of 0.5, 1.5, 3 and 4 pc at a distance of 1.4 kpc, which will provide reference for the virial analysis presented in Sect.~\ref{sec:analysis}. {\bf Middle right}: The red box outlines the area that we use from the available SOFIA \CII\ observations that cover a significantly larger region in Cygnus-X North \citep{Schneider2023}. The nomenclature indicated for sub-filaments is adopted from \cite{Schneider2010,Hennemann2012}: N (north), F1N, F1S, F3N, F3S, SW (south-west), S (south). {\bf Bottom and top}: $^{13}$CO(1-0), C$^{18}$O(1-0), HCO$^{+}$(1-0), H$^{13}$CO$^{+}$(1-0) and \CII\ spectra from the locations indicated in the left middle panel. %{\bf NICOLA: In the text, you mention a 10$^{23}$cm$^{-2}$ contour that is not shown here.}
}
\label{iramCoverage}
\end{center}
\end{figure*}
The formation of massive stars is still a matter of intense debate \citep[e.g.][]{Zinnecker2007,Tan2014,Motte2018}. The very short, or potentially non-existing, massive prestellar core phase \citep[e.g.][]{Motte2007,Motte2010,Russeil2010,Csengeri2014,Tige2017,Sanhueza2019} suggests that high-mass stars might not form from the quasi-static evolution toward collapse of individual prestellar cores. This opens the question of how massive stars can form. From observations of molecular clouds that form massive stars it was proposed that some large-scale dynamics, partly driven by self-gravity, can provide the fast concentration of mass necessary to form massive stars  \citep[e.g.][]{Peretto2006,Peretto2013,Peretto2014,Hartmann2007,Schneider2010,Schneider2015,Csengeri2011a,Csengeri2011b,GalvanMadrid2010,Wyrowski2012,Wyrowski2016,Beuther2015,Williams2018,Jackson2019,Bonne2022a}.\\
This probable coupling to the molecular cloud dynamics is also put forward by several theoretical models. In the competitive accretion model \citep{Bonnell2001,Bonnell2004,Bonnell2006,Wang2010}, the gravitational pull of low-mass protostellar cores drives the mass accretion from the surrounding cloud to determine their final mass. %The drawback of this view is that mostly single massive stars without associated low-mass stars are formed, while observations show that massive stars form typically in clusters. 
%This limitation is solved in t
The fragmentation-induced starvation scenario proposes that massive stars form through gravitational fragmentation at the center of dense filamentary accretion flows \citep{Peters2010,Peters2011,Girichidis2012}, where the fragmentation in the inflowing filaments sets a limit on the accretion of the most massive stars at the center of the accretion flow. 
Another variant proposes that gravitational collapse on multiple scales after the thermal instability \citep{VazquezSemadeni2009,VazquezSemadeni2017,VazquezSemadeni2019} drives the required dynamics and mass concentration responsible for the formation of massive stars which is then halted by stellar feedback. From simulations it was also proposed that compression by the collision of atomic flows or fully developed molecular clouds at high velocities can form dense filamentary structures that host massive star formation \citep[e.g.][]{Dobbs2012,Dobbs2020,Inoue2013,Wu2015,Balfour2017,Bisbas2017}. The velocity and density of the \HI\ flows or molecular clouds would then be decisive to determine whether massive stars can form. Typically, collision velocities $>$10 km s$^{-1}$ would be needed to form massive stars \citep{Haworth2015,Dobbs2020}. Based on observed bridging with CO lines between separated velocity components in several clouds, it has been proposed that cloud-cloud collisions (CCCs) might play a role in high-mass star formation \citep[e.g.][]{Bisbas2018,Fukui2021,Lim2021}. From simulations, it was also proposed that an oblique shock associated with collision velocities above 7 km s$^{-1}$ could bend the magnetic field around the filaments and drive subsequent mass inflow that enables high-mass star formation \citep{Inoue2018,Abe2020}.\\\\
%Nicola: I propose some small changes:
Analyzing observations of the low-mass Musca filament and the surrounding Chamaeleon-Musca regions, \citet{Bonne2020b,Bonne2020a} concluded that continuous mass accretion on the Musca filament was driven by such bending of the magnetic field due to a 50 pc scale collision at $\sim$ 7 km s$^{-1}$ between \HI\ clouds in the Chamaeleon-Musca complex. This mechanism was proposed for the formation of high-mass star forming filaments \citep{Inoue2018} but might also be applicable for low-mass star forming filaments. Specifically, the Musca filament would be the result of a turbulent overdensity that is compressed and guided by the bended magnetic field during interaction with more diffuse gas in the colliding cloud. %which we call a $^{\prime}$soft$^{\prime}$ cloud collision.
%Analyzing observations of the low-mass Musca filament and the surrounding Chamaeleon-Musca regions, it was concluded that continuous mass accretion on the Musca filament was driven by bending of the magnetic field due to a 50 pc scale \HI\ cloud collision, at $\sim$ 7-10 km s$^{-1}$, in the Chamaeleon-Musca complex \citep{Bonne2020b,Bonne2020a}. It thus appears that a mechanism, proposed for high-mass star formation \citep{Inoue2018}, might also be at the origin of low-mass star forming filaments. Specifically, the Musca filament would be the result of a turbulent overdensity that is compressed during interaction with more diffuse gas in the colliding cloud. %which we call a $^{\prime}$soft$^{\prime}$ cloud collision.
Interestingly, Faraday rotation measurements in the radio domain, which trace the magnetic field properties in the interstellar medium (ISM), unveiled curved magnetic fields around several nearby low- to intermediate mass star forming regions \citep{Tahani2019,Tahani2022}. It also unveiled a correlation of the cold (CNM) \& lukewarm (LNM) neutral medium with the magnetic field structure in the diffuse ISM \citep{Bracco2020}. These results would have a straightforward explanation in the proposed scenario of colliding \HI\ clouds that bend the magnetic field. Furthermore, recent observations of the massive star forming cloud NGC 6334 found kinematic structure resembling the results in Musca \citep{Arzoumanian2022}. %Tracing the kinematics to probe the role of the magnetic field is important as \citet{Reissl2021} indicates that magnetic field measurements alone might be difficult to conclude on the actual magnetic field morphology. 
%Based on the kinematics in the low-mass Musca cloud it was proposed that this bending of the magnetic field is responsible for the inflow towards the dense filament that will form stars. 
%Several physical processes can thus play a role in the formation of high-mass stars. Their role at different densities in the cloud therefore has to be understood in high-mass star forming regions.\\
%These different results thus raise the question whether bending of the magnetic field, triggered by the $^{\prime}$soft$^{\prime}$ collision of \HI\ clouds, might be rather universal and initiate both low- and high-mass star formation.

\begin{table}[htbp] 
 \centering 
  \caption{Summary of the observational datasets.}
 \begin{tabular}{lccccc} 
\hline 
\hline 
 {\small Instrument}  &  {\small Species}            & $\lambda$ &  $\nu$  & $\Delta$ {\rm v} & $\Theta$ \\ 
            &                                        &  [$\mu$m] &  [GHz]  &   [km/s]     &  [$''$]   \\  
\hline 
\multicolumn{2}{l}{{\sl \bf {IRAM }}} &           &         &   \\
 {\small EMIR}        & {\small $^{13}$CO(1-0)}      & 2720.4    & 110.20  &  0.2   &  23  \\  
 {\small EMIR}        & {\small C$^{18}$O(1-0)}      & 2730.8    & 109.78  &  0.2   &  23  \\        
 {\small EMIR}        & {\small HCO$^+$(1-0)}        & 3361.3    & 89.19   &  0.2   &  28 \\        
 {\small EMIR}        & {\small H$^{13}$CO$^+$(1-0)} & 3455.7   & 86.75   &  0.2   &  29 \\
\hline 
\multicolumn{2}{l}{{\sl \bf {SOFIA}}}                &           &         &          \\   
 {\small upGREAT}     & {\small \CII}                & 157.74    & 1900.54 &  0.5  &  14  \\  
 %{\small 4GREAT}      & {\small \CI}                 & 609.1     &  492.16 &  0.20  &  52  \\        
%\hline 
%\multicolumn{2}{l}{{\sl \bf {FCRAO}}}                &           &         &    \\   
 %{\small SEQUOIA}     & {\small $^{13}$CO 1$\to$0}   & 2720.4     & 110.20  &  0.1  &  45       \\        
\hline 
\multicolumn{2}{l}{{\sl \bf {JCMT}}}                 &           &         &            \\   
 {\small HARP}        & {\small $^{12}$CO  3$\to$2}  & 869.0     &  345.80  &   0.42 &   15       \\
            &                              &           &             &         &          \\  
\end{tabular} 
%\tablefoot{For the atomic and molecular lines, the line transition wavelength ($\lambda$) and frequency ($\nu$) are given in columns 3 and 4, respectively. The effective velocity resolution is indicated in column 5 and the angular resolution of the original data is displayed in column 6.}
For the atomic and molecular lines, the line transition wavelength ($\lambda$) and frequency ($\nu$) are given in columns 3 and 4, respectively. The effective velocity resolution is indicated in column 5 and the angular resolution of the original data is displayed in column 6.
\label{table-obs} 
\end{table} 
%In this paper, we present a study the formation of dense filamentary gas in Cygnus-X, which is one of the most %nearby active high-mass star forming regions, in particular focusing on the massive DR21 ridge. \\ 
In this paper, we present a multi-wavelength study of the DR21 cloud in molecular and atomic lines and dust continuum to follow up on the link between molecular cloud evolution and the formation of dense filamentary structures. Specifically, after focusing on Musca, we now focus on a region where massive stars are presently forming. The DR21 cloud is located in the north of the Cygnus-X molecular cloud complex \citep{Reipurth2008} at an estimated distance of $\sim$ 1.4 kpc \citep{Rygl2012}. This cloud hosts the DR21 ridge which is the filamentary structure with a length of $\sim$ 4 pc and a mass of $\sim$ 10$^{4}$ M$_{\odot}$ that is typically defined by high column densities, i.e., $N_{\rm H_{2}} >$ 10$^{23}$ cm$^{-2}$ \citep{Schneider2010,Hennemann2012}. The DR21 ridge is the densest and most massive filament in the entire Cygnus-X complex and one of the most active high-mass star forming regions within 2 kpc from the sun  \citep{Motte2007,Kumar2007,Bontemps2010,Beerer2010,Csengeri2011a,Csengeri2011b,DuarteCabral2013,DuarteCabral2014}. 
In this paper, we will use the term $^{\prime}$ridge$^{\prime}$ for the inner, high-column density part ($N_{\rm H_{2}} >$ 10$^{23}$cm$^{-2}$) of the molecular cloud \citep{Hill2011,Hill2012}, $^{\prime}$sub-filaments$^{\prime}$ for the lower density filaments connected to the ridge and $^{\prime}$cloud$^{\prime}$ for the parsec scale surrounding cloud that includes the ridge. %, and 'complex' for the ensemble of giant molecular clouds (GMCs). 
Detailed kinematic studies of the DR21 region by \citet{Schneider2010} demonstrated that the ridge is experiencing large scale gravitational collapse and mass accretion by sub-filaments that run parallel to the magnetic field \citep{Vallee2006,Ching2022}. Figure~\ref{iramCoverage} shows the DR21 cloud and ridge ($N_{\rm H_{2}}>$10$^{23}$cm$^{-2}$) and the sub-filaments defined in \citet{Schneider2010} and \citet{Hennemann2012}. It is also proposed that the DR21 cloud is part of the Cygnus-X complex which forms a single region with multiple velocity components \citep{Schneider2006,Schneider2007}. This was later confirmed by \citet{Rygl2012} and provided arguments that DR21 is the result of a collision between two molecular clouds with velocity components at $\sim$ -3 km s$^{-1}$ and $\sim$ 9 km s$^{-1}$ \citep{Dickel1978,Dobashi2019}. This collision is revisited and discussed in \citet{Schneider2023} using new insight from the SOFIA \CII\ data.

In this paper, we aim to establish the main processes, i.e. gravity, magnetic field and turbulence that drive the evolution of this massive star forming molecular cloud from low-density ($n_{\rm H_{2}} \lesssim$ 10$^{3}$ cm$^{-3}$) gas in the surrounding cloud to high-density ($n_{\rm H_{2}} >$ 10$^{5}$ cm$^{-3}$) gas in the ridge. In Sect.~\ref{sec:obs}, we give observational details. Section~\ref{sec:dr21} presents the observational results of the DR21 cloud, followed by the inflow and virial analysis of these observational results in Sect. \ref{sec:analysis}. Lastly, in Sect.~\ref{sec:discuss} we 
%discuss 
put these results into context to propose a scenario for the DR21 cloud evolution and discuss how it compares to low-mass star forming clouds.

%-----------------------------------------------------------------------------------------------------------------------
\section{Observations} \label{sec:obs} 
The observations tracing different density regimes in the DR21 cloud are described in the following subsections and are summarized in Table~\ref{table-obs}. 

\subsection{IRAM 30m observations} \label{obs-iram} 
\begin{figure*}
\begin{center}
\includegraphics[width=0.37\hsize]{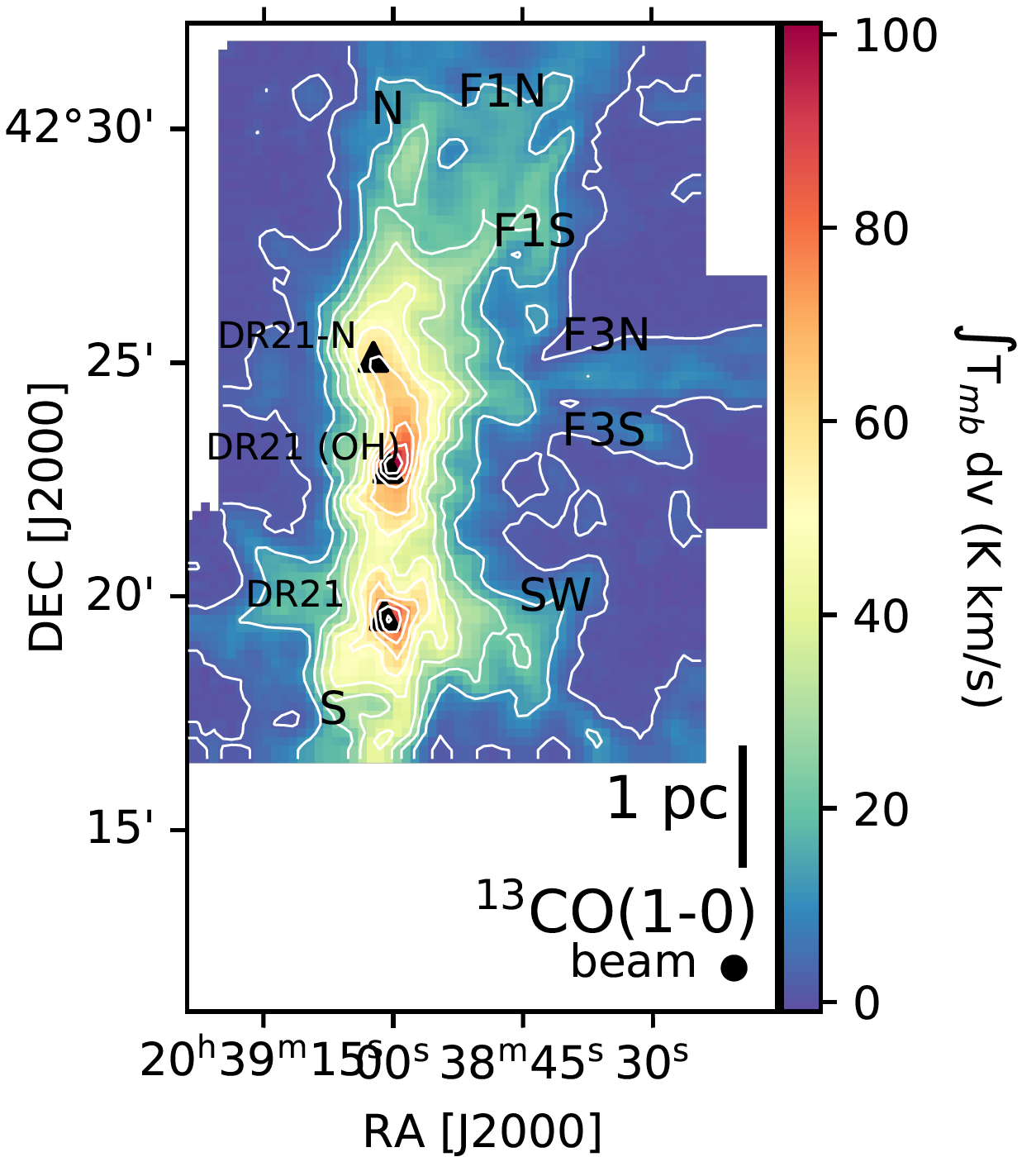}
\includegraphics[width=0.37\hsize]{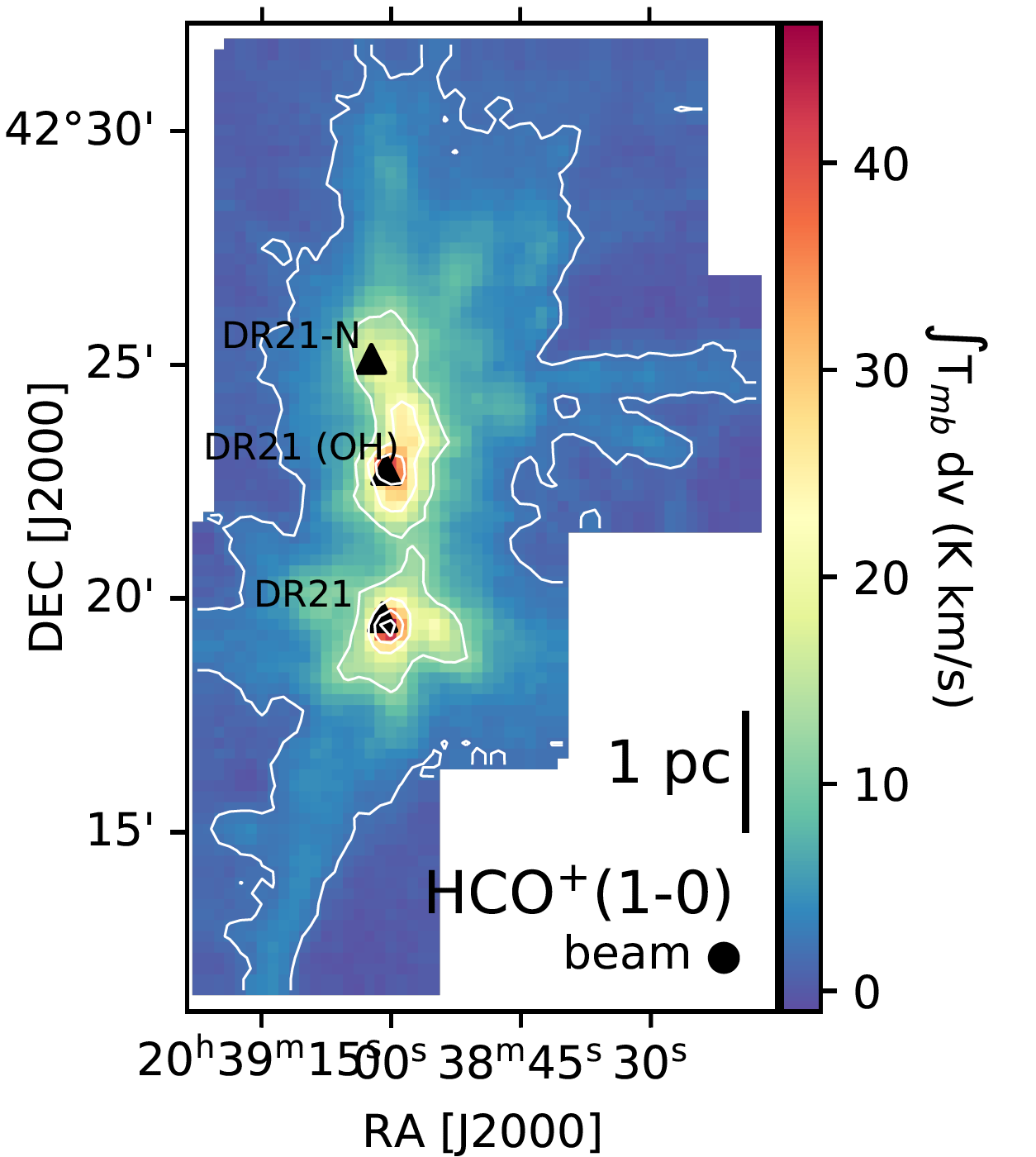}
\includegraphics[width=0.37\hsize]{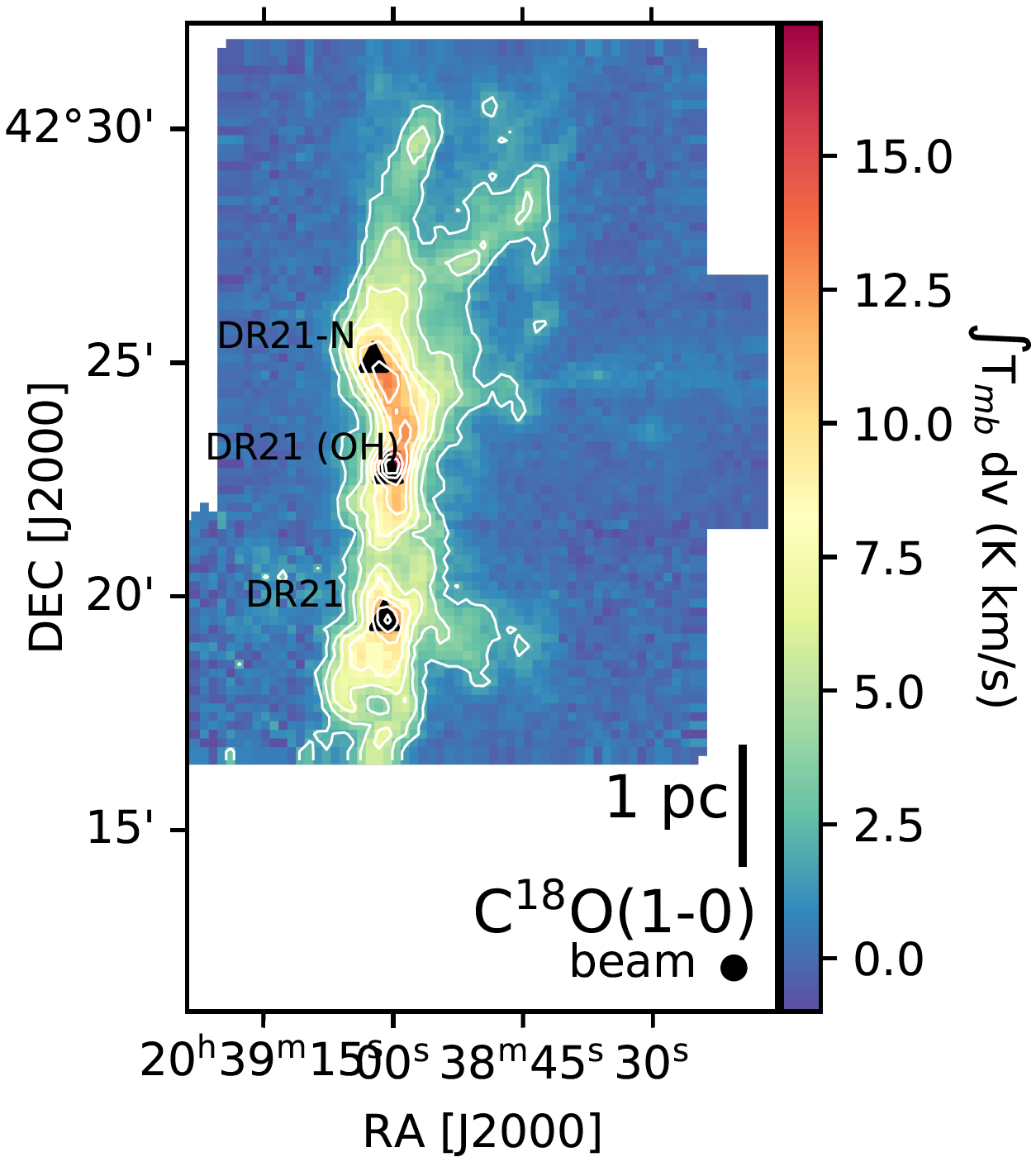}
\includegraphics[width=0.37\hsize]{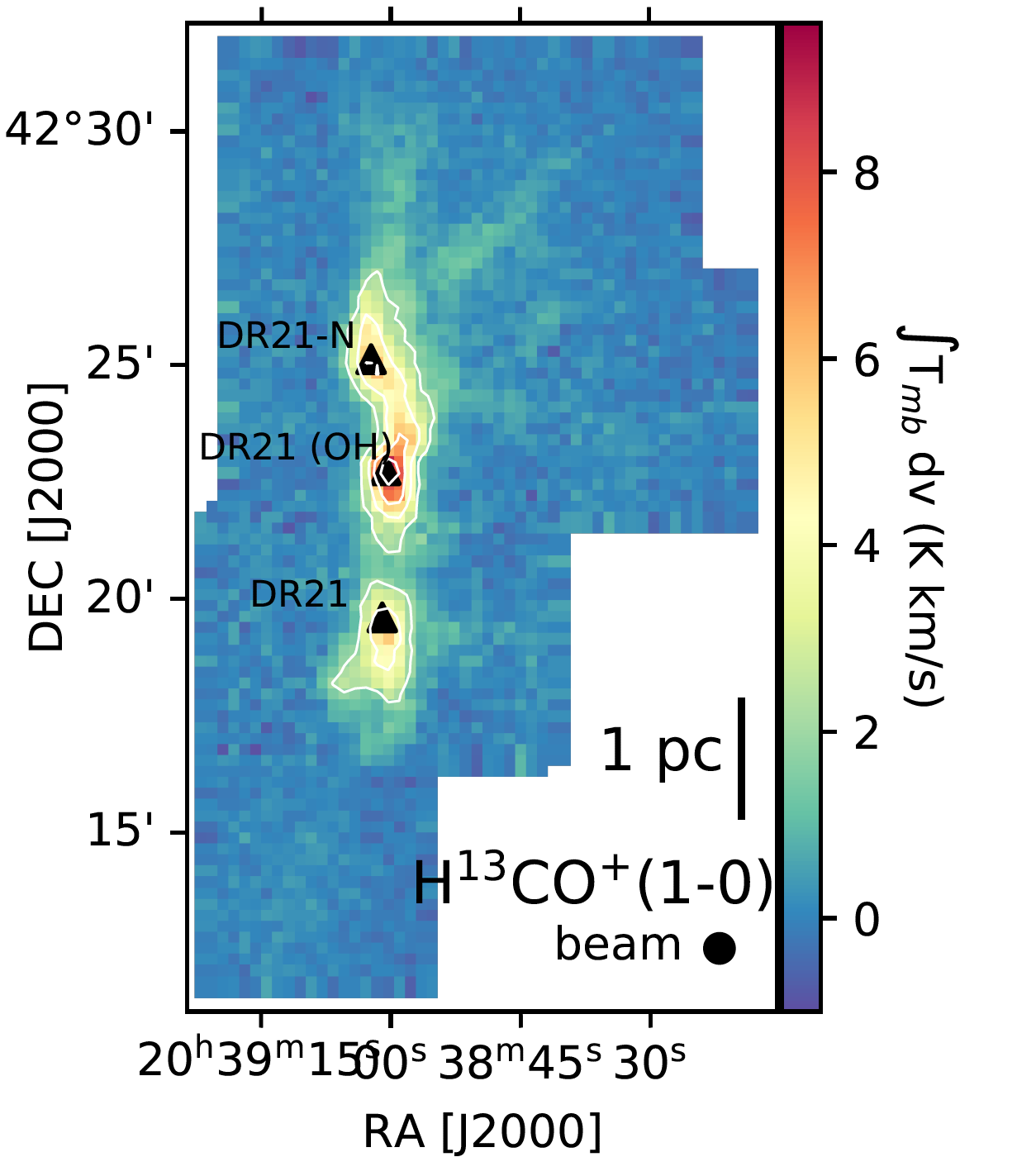}
\end{center}
\caption{\textbf{Top left:} Integrated brightness map of $^{13}$CO(1-0) from -8 to 1 km s$^{-1}$ towards the DR21 cloud. Overlaid on the map are the integrated intensity contours starting at 2 K km s$^{-1}$ with increments of 10 K km s$^{-1}$. The locations of the DR21 radio continuum source \citep{Downes1966}, and the DR21(OH) and DR21-N massive clumps \citep[e.g.][]{Motte2007} are indicated with black triangles. The subfilaments in cloud are indicated with their names. \textbf{Top right:} The same for HCO$^{+}$(1-0) with increments of 5 K km s$^{-1}$. The triangles indicating DR21, DR21(OH) and DR21-N help to compare with the $^{13}$CO and C$^{18}$O maps, from setup2, which have a different map size. \textbf{Bottom left:} The same for C$^{18}$O(1-0), with contour increments of 2 K km s$^{-1}$. \textbf{Bottom right:} The same for H$^{13}$CO$^{+}$(1-0), with contour increments of 2 K km s$^{-1}$.}
\label{intIntensityMaps}
\end{figure*}
\begin{figure}
\centering
\includegraphics[width=\hsize]{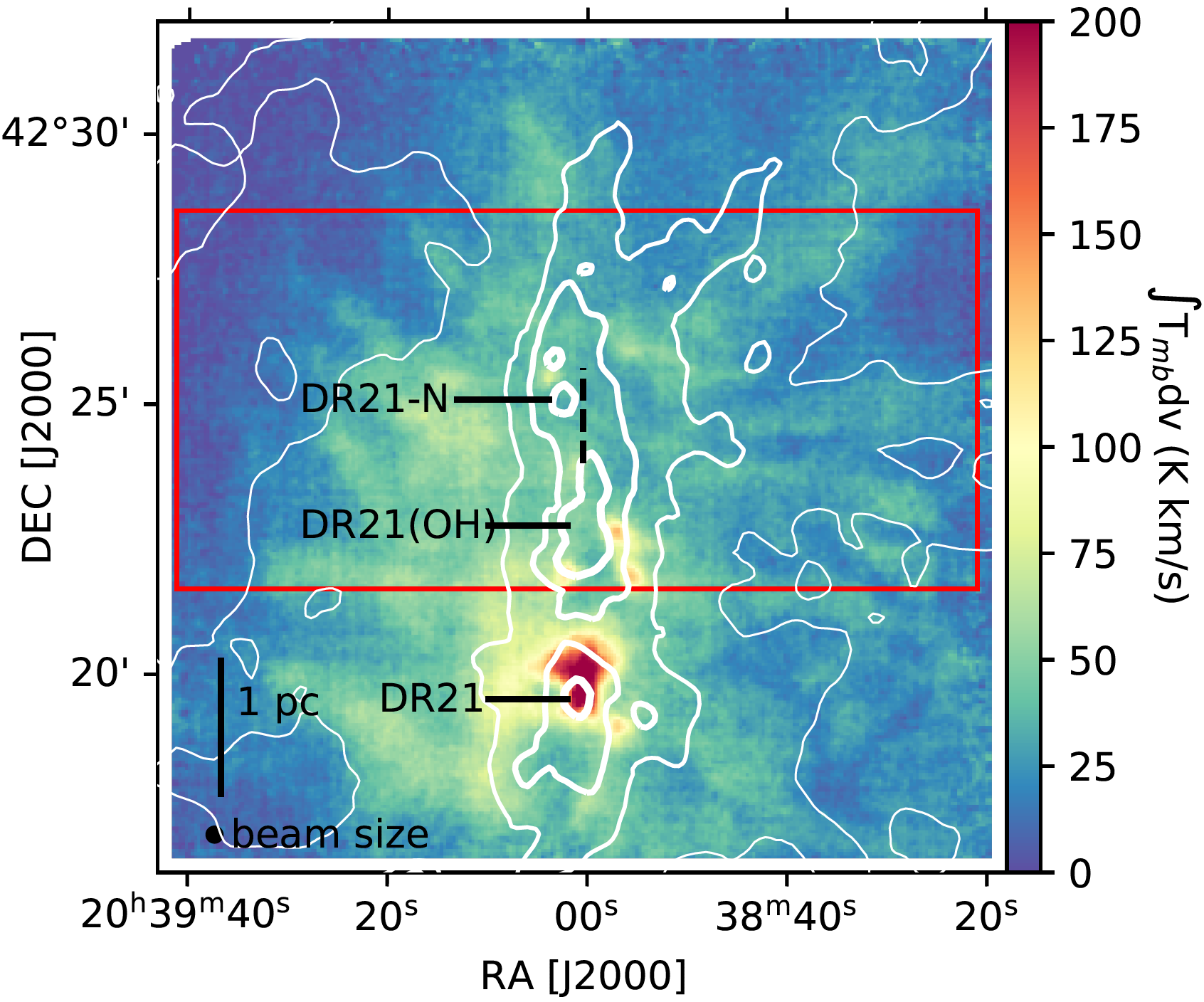}
\caption{Line integrated emission from -8 to 1 km s$^{-1}$ for \CII.
The \CII\ emission shows that there is extended emission around the ridge and sub-filaments. The white contours indicate the \textit{Herschel} column density map at $N_{\rm H_{2}}$ = 10$^{22}$ cm$^{-2}$, 3$\times$10$^{22}$ cm$^{-2}$, 10$^{23}$ cm$^{-2}$ and 4$\times$10$^{23}$ cm$^{-2}$. The red box outlines the area used to make the position-velocity (PV) diagram perpendicular to the DR21 ridge in Fig. \ref{pvDiagrams}. The vertical dashed black line defines the center (r = 0 pc) used for the PV diagram. The locations of the DR21 radio continuum source \citep{Downes1966}, and the DR21(OH) and DR21-N massive clumps \citep[e.g.][]{Motte2007} are indicated.} \label{integratedCIICI}
\end{figure}

\begin{figure}
\centering
\includegraphics[width=\hsize]{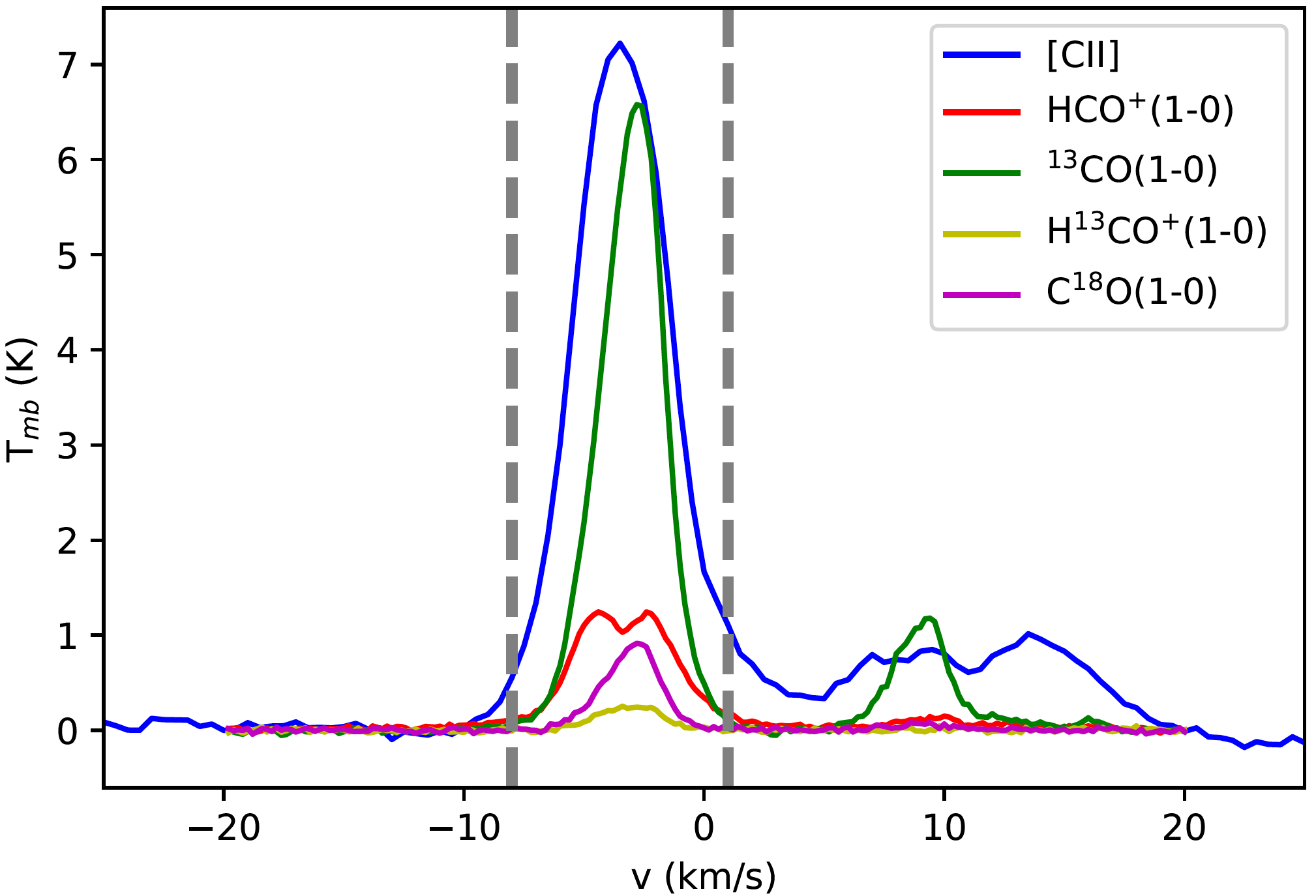}
\caption{Average spectrum of the DR21 ridge (excluding the DR21 continuum source) displaying multiple velocity components between -10 and 20 km s$^{-1}$.
%in multiple lines which will be the focus of Schneider et al. (in prep.). NICOLA: Not necessary to repeat here.
The focus of this paper is on the emission between -8 and 1 km s$^{-1}$ alone, indicated by the dashed vertical lines, which is the velocity range associated with the DR21 ridge \citep{Schneider2006,Schneider2010}.}
\label{fig:specFocus}
\end{figure}

\begin{figure*}
\begin{center}
\includegraphics[width=0.85\hsize]{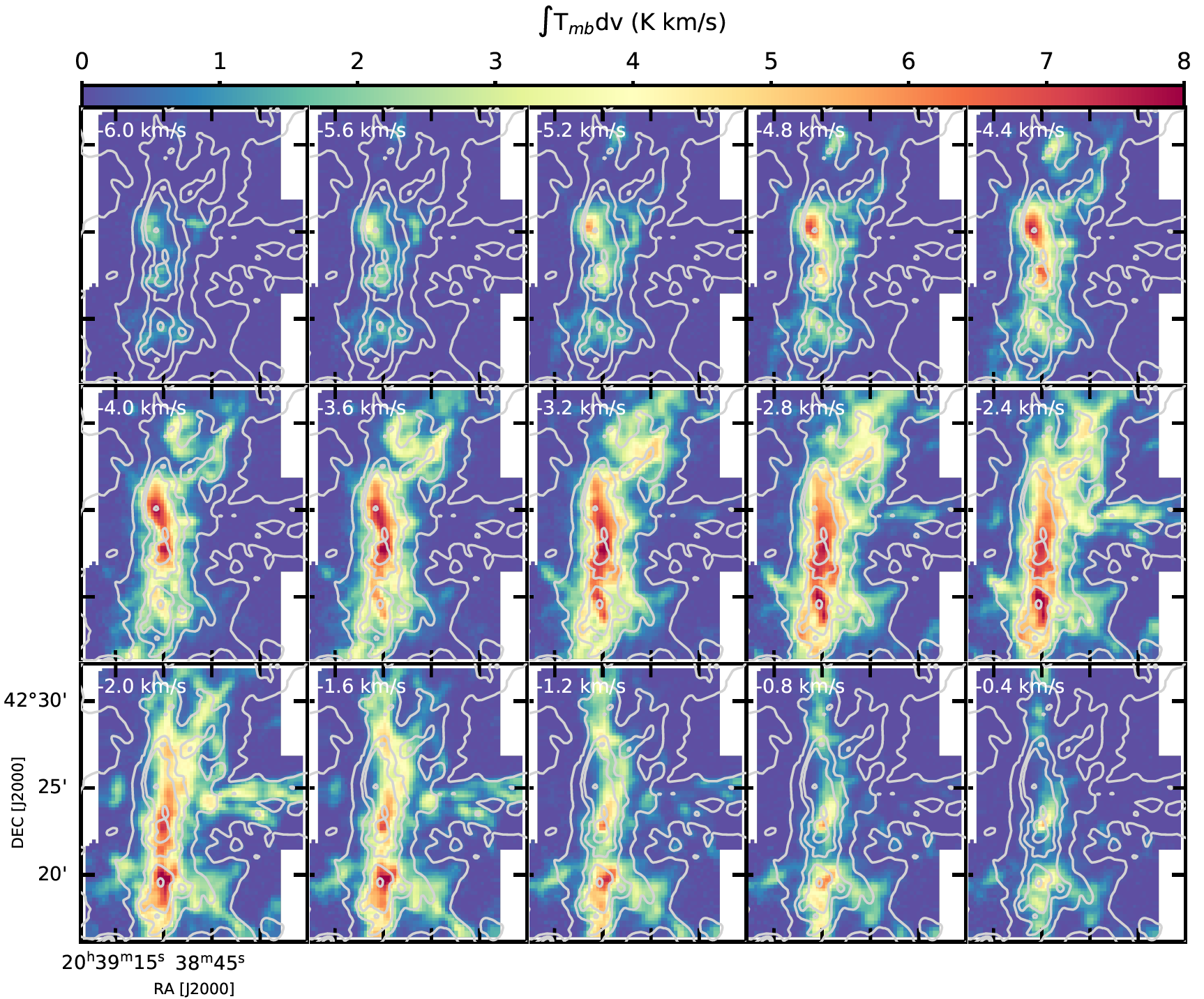}
\caption{$^{13}$CO(1-0) channel maps (IRAM) between -6.0 and -0.4 km s$^{-1}$ of the DR21 cloud. The grey contours trace the \textit{Herschel} column density at $N_{\rm H_{2}}$ = 10$^{22}$, 2$\times$10$^{22}$, 5$\times$10$^{22}$, 10$^{23}$ and 5$\times$10$^{23}$ cm$^{-2}$, to highlight the ridge and sub-filaments. Most sub-filaments become particularly visible at v$_{lsr} >$ -3 km s$^{-1}$.}
\label{13coChanMap}
\end{center}
\end{figure*}
Observations with the IRAM 30m telescope of the DR21 region (surrounding cloud, ridge and sub-filaments) were carried out in February 2015. These observations were performed with two different spectral setups using the FTS50 configuration of the EMIR receiver. Setup 1, which contains HCO$^{+}$(1-0) and H$^{13}$CO$^{+}$(1-0), covers the spectral ranges $\sim$ 85.2 GHz - 87.2 GHz, 88.5 GHz - 90.5 GHz, 101.0 GHz - 102.8 GHz and 104.2 GHz - 106.0 GHz. Setup 2, which contains $^{13}$CO(1-0) and C$^{18}$O(1-0), covers the spectral ranges $\sim$ 89.6 GHz - 91.6 GHz, 93.0 GHz - 95.0 GHz, 105.4 GHz - 107.2 GHz and 108.8 GHz - 110.6 GHz. These two setups allow to observe a variety of molecular lines (e.g. C$^{18}$O(1-0), N$_{2}$H$^{+}$(1-0), HCN(1-0), HCO$^{+}$(1-0), NH$_{2}$D[1(1,1)0-1(0,1)0], etc...), which allows to follow the kinematic and chemical evolution of the dense gas in the DR21 ridge and the connected sub-filaments. 
%, similar to what was done for the Orion-B cloud \citep{Pety2017}. NICOLA: Why mentioning Orion B here? Does not fit, in particular because they perform a rather different study. 
The obtained antenna temperature (T$_{A}^{*}$) noise rms within a spectral resolution of 0.2 km s$^{-1}$ is $\sim$0.14 K around 90 GHz and $\sim$0.20 K around 110 GHz, respectively.  
%NICOLA: Such a small variation? Then I would talk of averages... varies between 0.13-0.15 K (around 90 GHz) and 0.19-0.2 K (around 110 GHz). 
A main beam efficiency of 0.81 for frequencies around 90 GHz and 0.78 for frequencies around 110 GHz was used \citep{Kramer2013}. The region on the sky that is covered by the two different setups is displayed in Fig. \ref{iramCoverage}. The mapping was performed in the On-the-Fly (OTF) observing mode, and the resulting data cubes have a spatial resolution between $\sim$23$''$ (110 GHz) and $\sim$29$''$ (90 GHz), see Tab. \ref{table-obs}. To produce the data cubes, a first order baseline was fitted to the data with CLASS in GILDAS\footnote{https://www.iram.fr/IRAMFR/GILDAS/} as the baselines are generally well behaved at the observed frequencies. A full analysis of all the lines that are covered will be the topic of a future paper. Here, the first results from $^{13}$CO(1-0), C$^{18}$O(1-0), HCO$^{+}$(1-0), and H$^{13}$CO$^{+}$(1-0) will be presented.

\subsection{SOFIA \CII\ observations with upGREAT}
To obtain a more complete view on the kinematics of the DR21 cloud, the IRAM observations are combined with observations of the \CII\ 158 $\mu$m line by the Stratospheric Observatory for Infrared Astronomy \citep[SOFIA; ][]{Young2012}.
%NICOLA: this is not adequate for the observations sections (and more complicated as stated here.... which is expected to trace the kinematics of the outer layers of this molecular cloud. 
The \CII\ observations at 158 $\mu$m were taken as part of the SOFIA Legacy program FEEDBACK\footnote{https://feedback.astro.umd.edu}, which maps 11 Galactic high-mass star-forming regions \citep{Schneider2020}, including a part of the Cygnus-X north region, in the \CII\ 158 $\mu$m line and the \OI\ 63 $\mu$m line. The mapping of Cygnus-X north with FEEDBACK is now completed\footnote{https://astro.uni-koeln.de/index.php?id=18130} and covers the DR21 cloud, the Diamond Ring \citep{Marston2004} and the W75N molecular cloud. The data can be found on the IRSA archive with project number 07\_0077\footnote{https://irsa.ipac.caltech.edu/applications/sofia}. We use data that covers the DR21 cloud, see Fig.~\ref{iramCoverage}. The observations were carried out with the dual-frequency heterodyne array upGREAT receiver \citep{Risacher2018} in OTF mapping mode and calibrated with the GREAT pipeline \citep{Guan2012}. We used an emission-free reference position at RA(2000)=20$^h$39$^m$48.34$^s$, Dec(2000)=42$^\circ$57$'$39.11$''$. The 2$\times$7 pixel LFA array was tuned to the \CII\ 158 $\mu$m line  and the 7 pixel high-frequency array (HFA) was tuned to the \OI\ 63 $\mu$m line (data not used here). The beam size at 158 (63) $\mu$m is 14.1$^{\prime\prime}$ (6$''$). Here we employ \CII\ data smoothed to an angular resolution of 20$''$ and a velocity  resolution of 0.5 km s$^{-1}$, which results in a typical noise rms of $\sim$ 1 K per beam. In order to improve the baseline removal to reduce striping in the data cube we employ the Principal Component Analysis (PCA) method described in \citet{Tiwari2021,Kabanovic2021,Schneider2023}. More details on the SOFIA FEEDBACK observational scheme and data reduction are found in \cite{Schneider2020}.

%\subsubsection{\CI\ observations with 4GREAT}
%The \CI\ observations at 492 $\mu$m were carried out with the 4GREAT receiver \citep{Duran2020} during one flight november 23th from Palmdale/California. 4GREAT was configured to observe the \CI(1-0) line in the 4G1 channel, and the CO(8-7), CO(11-10), and CO(22-21) in the 4G2, 4G3, and 4G4 channels, respectively. In parallel, the HFA  was tuned to the \OI\ 63 $\mu$m line. Four sub-maps (each one time in horizontal and vertical scanning) were observed in the OTF total power mapping mode with the same emission-free reference position as used for \CII.   The maps have a step size of 20$''$ with 21 dumps in both direction, resulting in a map of 7$'\times$7$'$, with 0.8 sec per dump. In this paper, we focus on the \CI(1-0) data of the DR21 cloud. 
%CYG-X_NORTH_CI492_L_F643, 23.11. 2020

\subsection{JCMT observations}
With the 15 pixel HARP instrument on the JCMT telescope, a $^{12}$CO(3-2) mapping of the Cygnus-X north and south areas was carried out between 2007 and 2009 within the observing programs M08AU018 (PI N. Schneider) and M07BU019 (PI R. Simon). The observation method and reduction of this data is the same as the one described in \citet{Gottschalk2012} for the pilot study and this full dataset will be presented in more detail in a forthcoming paper. The observations have a spatial resolution of 15$^{\prime\prime}$ and a spectral resolution of 0.42 km s$^{-1}$, with a noise rms of $\sim$ 0.25 K. 

%\subsection{FCRAO observations}
%The observations of the IRAM 30m and SOFIA telescope are further complemented by $^{13}$CO(1-0) observations that were carried out with the SEQUOIA receiver on the FCRAO 14m telescope. This data is presented in \citet{Simon2006} and \citet{Schneider2006,Schneider2007,Schneider2010,Schneider2016a}. The data has an angular resolution of 45$''$ and a spectral resolution of 0.1 km s$^{-1}$. The rms noise of the spectra within this spectral resolution is around 0.4 K. The integrated $^{13}$CO map is displayed in Fig. \ref{13coMap}.

\subsection{\textit{Herschel} observations}
We employ \textit{Herschel} column density maps from the Cygnus-X region that were taken as part of the HOBYS\footnote{Herschel imaging survey of OB young stellar objects \citep{Motte2010}.} program. Low angular resolution (36$''$) column density maps are presented in \citet{Schneider2016a}. The procedure for obtaining these maps is described in \citet{Hill2011,Hill2012}. We here use column density maps at a higher spatial resolution of 18$^{\prime\prime}$ that makes use of the method described in \citet{Palmeirim2013}. The concept is to employ a multi-scale decomposition of the flux maps and assume a constant
line-of-sight temperature. The final map is then constructed from the difference maps of the convolved maps at 500 $\mu$m, 
350 $\mu$m, and 250 $\mu$m, and the temperature information from the color temperature derived from the 160 $\mu$m
to 250 $\mu$m ratio.

%\section{The DR21 ridge} \label{dr21} 
%-----------------------------------------------------------------------------------------------------------------------
\section{Results} \label{sec:dr21} 
\subsection{Spectral line integrated maps} \label{subsec:maps}
From the large variety of lines detected with the IRAM 30m telescope we selected a subset to present in this paper, i.e. the $^{13}$CO(1-0), C$^{18}$O(1-0), HCO$^{+}$(1-0) and H$^{13}$CO$^{+}$(1-0) lines, that trace both the ridge and sub-filaments. A full analysis of all detected lines with the IRAM 30m telescope is out of the scope of this paper.
Figure \ref{intIntensityMaps} displays the velocity integrated maps of these 4 lines. The overall emission distribution of the DR21 cloud is similar to what was shown in \citet{Schneider2010}, but the new maps are significantly larger. As a result, they better cover the extent of the dense sub-filaments which are best visible to the west of the DR21 ridge. Note that the coverage is not the same for all lines which is due to different setups for the observations. From Fig. \ref{intIntensityMaps} we see that the $^{13}$CO(1-0) and HCO$^{+}$(1-0) emission remains well detected outside the ridge while the C$^{18}$O(1-0) and H$^{13}$CO$^+$(1-0) lines are mostly detected towards the ridge which indicates a large concentration of mass. %The HCO$^{+}$(1-0) map also clearly reveals the extended outflow emission at the DR21 radiocontinuum position.

%This is also observed in the channel maps of $^{13}$CO in Fig. \ref{13coChanMap}. C$^{18}$O(1-0) and %H$^{13}$CO$^{+}$(1-0) are also clearly detected in the ridge, but their detection, in particular for %H$^{13}$CO$^{+}$(1-0), becomes increasingly difficult in the subfilaments. 

Figure \ref{integratedCIICI} presents the SOFIA \CII\ observations. The critical density of the \CII\ 158 $\mu$m line is typically only a few 10$^3$ cm$^{-3}$ for collisions with atomic and molecular hydrogen \citep[i.e. H \& H$_2$;][]{Goldsmith2012}, and it predominantly traces the photodissociated layers at the interface between molecular cloud and ionized phase \citep{Hollenbach1999}. It thus nicely complements the CO/HCO$^{+}$ observations as it should trace the lower-density gas. %However, \CII\ can also occur in the fully ionized phase (collisions with electrons) or in the diffuse gas further away from dense molecular cloud regions ('CO-dark gas'). 
%the outer parts of molecular clouds where UV photons still manage to penetrate 
%NICOLA: This is a little bit too short, CII can arise from various regions. And we better put a PDR paper not \citep{Beuther2014}. 
The \CII\ line integrated intensity map in Fig. \ref{integratedCIICI} reveals interesting differences compared to the IRAM molecular line data. 
Emission is detected toward the dense sub-filaments that are also seen in $^{13}$CO(1-0), HCO$^{+}$(1-0) and $^{12}$CO(3-2) (see App. \ref{sec:CIIvs12CO}) which have typical critical densities of $\sim$2$\times$10$^{3}$ cm$^{-3}$, 5$\times$10$^{4}$ cm$^{-3}$ and 3$\times$10$^{4}$ cm$^{-3}$, respectively \citep{Shirley2015}. However, \CII\ additionally also traces the lower density gas surrounding the DR21 ridge and dense sub-filaments. The \CII\ observations thus unveil a significant mass reservoir that is not located in the ridge and the sub-filaments. 
This gas has a typical \textit{Herschel} column density ($N_{\rm H_{2}}$) between 5$\times$10$^{21}$ cm$^{-2}$ and 3$\times$10$^{22}$ cm$^{-2}$ after correction for a $\sim$5$\times$10$^{21}$ cm$^{-2}$ background \citep{Schneider2016a,Schneider2022}. Even though \CII\ is detected towards these higher column densities ($N_{\rm H_{2}}$ $>$ 10$^{22}$ cm$^{-2}$), these also start to be detected in CO lines, see App. \ref{sec:13coAbundance}, which indicates that \CII\ does not trace the full column density range there. Assuming the cloud has a thickness\footnote{This is similar to the size in the plane of the sky or assumes it has a slightly flattened morphology.} of 1-3 pc indicates that \CII\ in the DR21 cloud might trace a density range between $n_{\rm H_{2}}$ = 0.5-1.6$\times$10$^{3}$~cm$^{-3}$ and $n_{\rm H_{2}}$ = 0.3-1.0$\times$10$^{4}$~cm$^{-3}$ toward the DR21 cloud. In a later paragraph we will further constrain the typical density traced by \CII. Only at the location of the DR21 radiocontinuum source there is a strong peak of \CII\ emission in the ridge. DR21 is the most evolved region in the ridge and a site of massive star formation with several compact \HII\ regions and a number of possible O-stars, including a prominent outflow source  \citep{Marston2004}. Stellar winds \citep{Cyganowski2003,Immer2014} probably dominate the dynamics of that region. As the \CII\ emission in the DR21 source is strongly affected by the local feedback from the embedded O stars, this region is mostly left out for the work presented in this paper. Lastly, we also note from Fig. \ref{integratedCIICI} that the \CII\ emission unveils filamentary gas to the east of the ridge that are not detected with Herschel or molecular line data. This is probably because of the low contrast with the significant Herschel column density background and because these filamentary structures are not visible in molecular line data.

\subsection{Spectra and channel maps} \label{subsec:spectra}
The average spectrum for the entire DR21 ridge is presented in Fig. \ref{fig:specFocus} which shows multiple velocity components in the velocity range between -10 km s$^{-1}$ and 20 km s$^{-1}$. In this paper we will exclusively focus on the emission between -8 and 1 km s$^{-1}$, highlighted in Fig. \ref{fig:specFocus}, which is the emission originating from the DR21 cloud \citep{Schneider2006,Schneider2010}. The components of the full spectrum are discussed in \citet{Schneider2023}. \\
The channel maps of $^{13}$CO(1-0) emission, shown in Fig. \ref{13coChanMap}, resolve the velocity structure in the DR21 cloud. The emission in the velocity range from -5 km s$^{-1}$ to -2.5~km~s$^{-1}$ is organized in a north-south elongated coherent structure mostly inside the $N_{\rm H_{2}}$=10$^{23}$ cm$^{-2}$ dust column density contour and forms the ridge. In this velocity range, the channel maps show a north-south velocity gradient in the ridge. At higher and lower velocities inside the ridge, the emission is more clumpy and concentrates on the DR21(OH) and DR21 locations. At velocities between -4 and -2.5 km s$^{-1}$ the sub-filament F1S becomes visible, and between -3 km s$^{-1}$ and -1 km s$^{-1}$ all the other western sub-filaments become prominent.\\
Inspecting the $^{13}$CO(1-0), C$^{18}$O(1-0), HCO$^{+}$(1-0), H$^{13}$CO$^{+}$(1-0) and \CII\ spectra over the map in Fig. \ref{iramCoverage}, we find that the spectra show skewed profiles.  This is also the case for H$^{13}$CO$^{+}$(1-0) and C$^{18}$O(1-0) which are optically thin for excitation temperatures in the DR21 ridge down to 10 K. This suggests the presence of multiple components which we will address later on. The selected positions in Fig. \ref{iramCoverage} represent typical lines in the ridge (positions 1,2,3 \& 5) and in the sub-filaments (positions 4 \& 6). Note that the HCO$^{+}$(1-0) spectra in the ridge display strong self-absorption centered on the velocities of peak emission for C$^{18}$O(1-0) and H$^{13}$CO$^+$(1-0), see \citet{Schneider2010} for more discussion. Furthermore, it becomes obvious in Fig. \ref{iramCoverage} that HCO$^{+}$(1-0) has high-velocity wing emission towards DR21 and DR21(OH), tracing the prominent molecular outflows in these regions.\\ 
The \CII\ spectra also show flat-topped and skewed line profiles in Fig. \ref{iramCoverage}. Since some bright \CII\ regions were found to experience \CII\ self-absorption and optical depth effects \citep{Graf2012,Guevara2020,Kabanovic2021}, we verified whether this could be the case for the DR21 ridge. The \CII\ spectra do not show a noteworthy dip, seen for HCO$^{+}$(1-0) which argues against self-absorption. In App. \ref{sec:opticalDepth}, the optical depth of the \CII\ line is calculated using the \dCII\ noise level which indicates a maximal optical depth of $\tau$ = 1.7. In addition, we 
performed calculations with RADEX \citep{vanderTak2007} which indicates that the optical depth for \CII\ typically is below 1 for the DR21 cloud. 
The \CII\ line shape is thus not caused by optical depth effects but rather by multiple velocity components around the DR21 ridge. This also fits with the observation in Fig. \ref{fig:specFocus} that the peak \CII\ emission is displaced from the peak molecular line emission. Therefore, we fit the spectra with multiple Gaussian velocity components (see App.~\ref{sec:TwoComps}) over the full map, and analyze the goodness of the fit, with the BTS fitting algorithm \citep{Clarke2018}. We find that the \CII\ emission in the range of $-$8 to 1 km s$^{-1}$ consist of either one or two velocity components that are typically separated by $\sim$ 2-3 km s$^{-1}$. The first component is typically around $-$4.5 km s$^{-1}$ and the second one around $-$2.5 km s$^{-1}$ (see Fig. \ref{fig:twoVelHist} of App. \ref{sec:TwoComps}). The fitting results in Fig. \ref{fig:locationsCIIfit} of App. \ref{sec:TwoComps} also show that the regions with multiple velocity components are concentrated in the close vicinity of the DR21 ridge. %However, due to the significant noise ($\sim$ 2 K) at a spatial resolution of 15$^{\prime\prime}$, it is challenging to spatially resolve the prominence of these two velocity components.

\begin{figure*}
\begin{center}
% Line fitting maps 
\includegraphics[width=0.365\hsize]{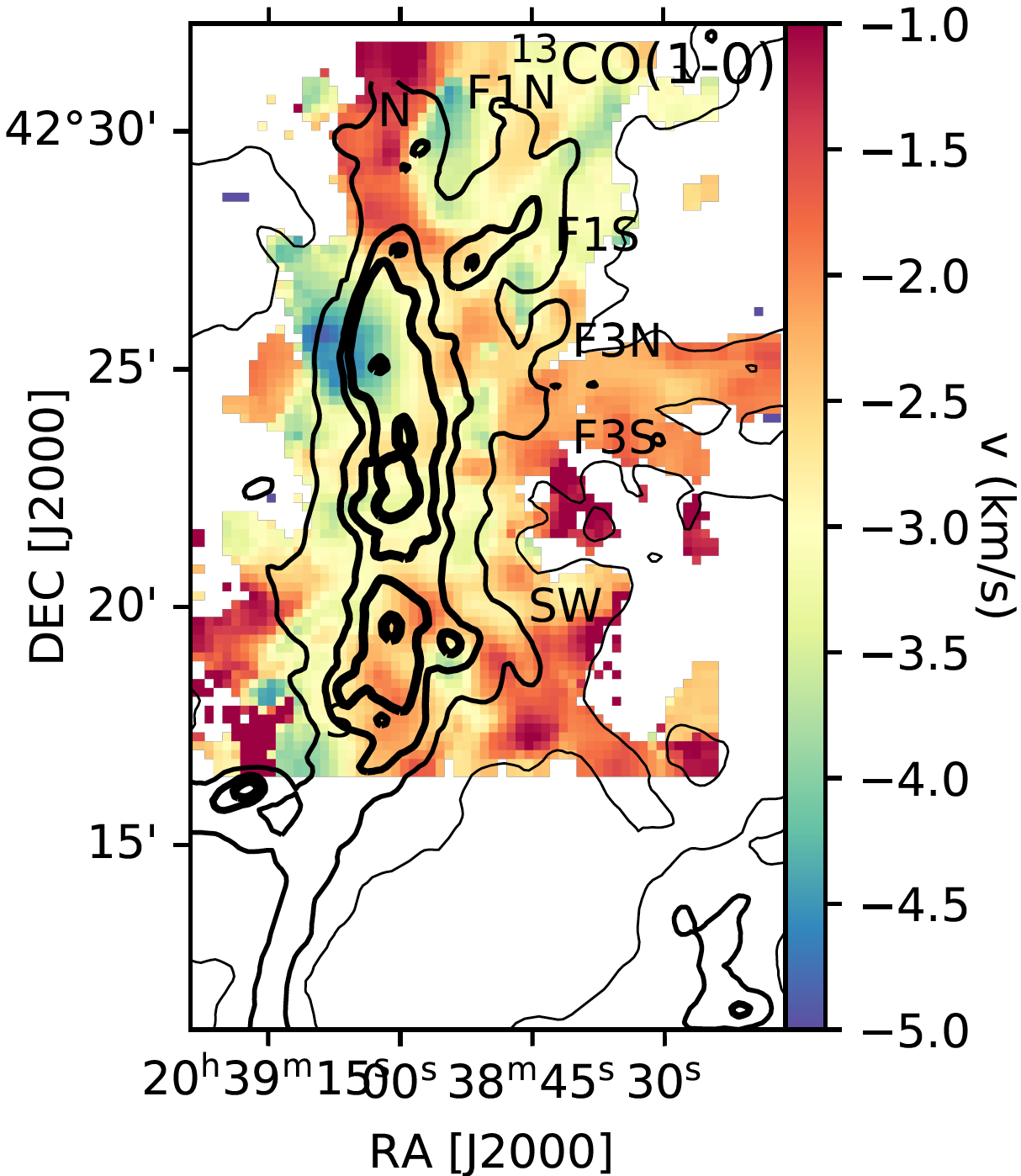}
\includegraphics[width=0.38\hsize]{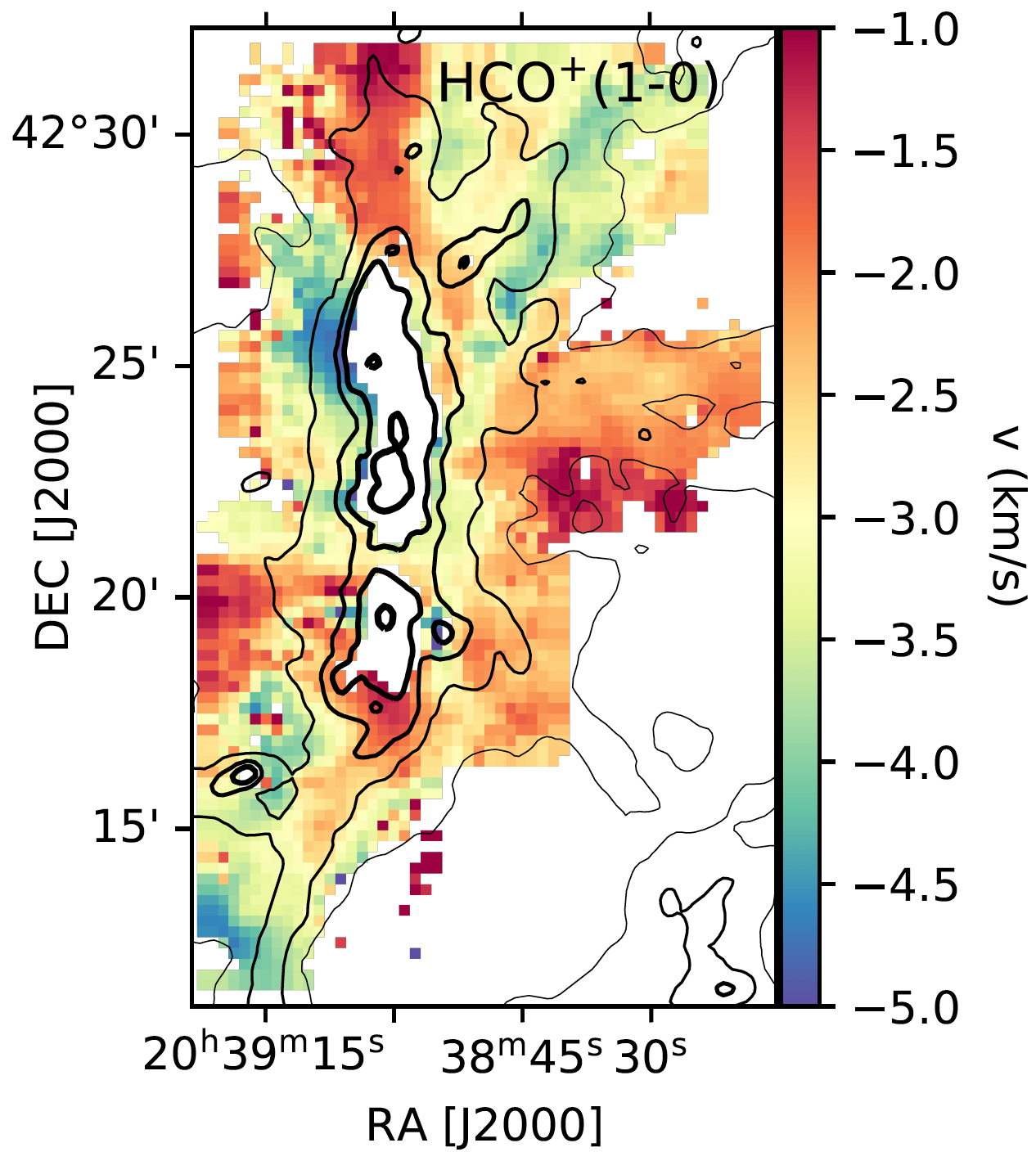}
\includegraphics[width=0.365\hsize]{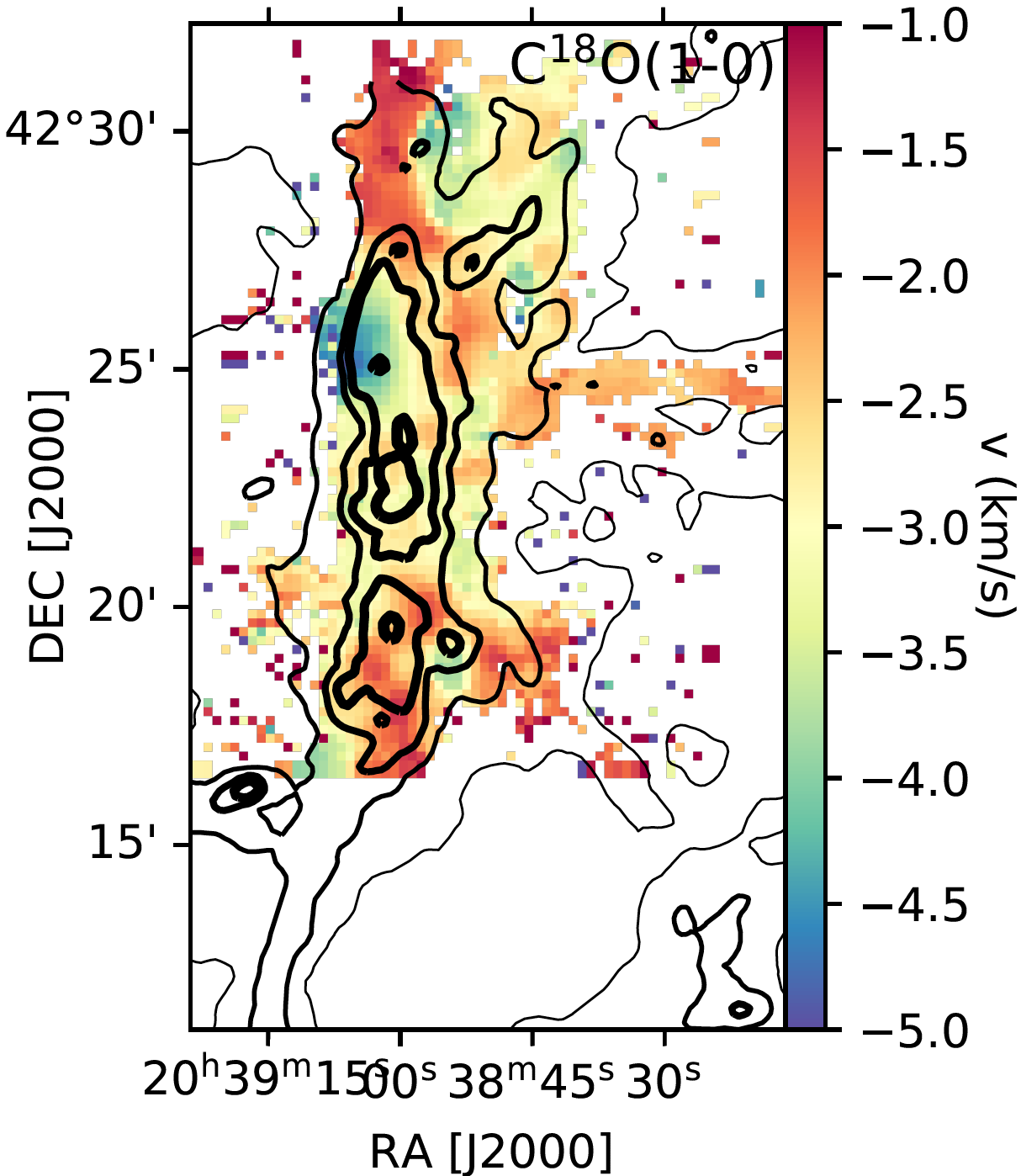}
\includegraphics[width=0.38\hsize]{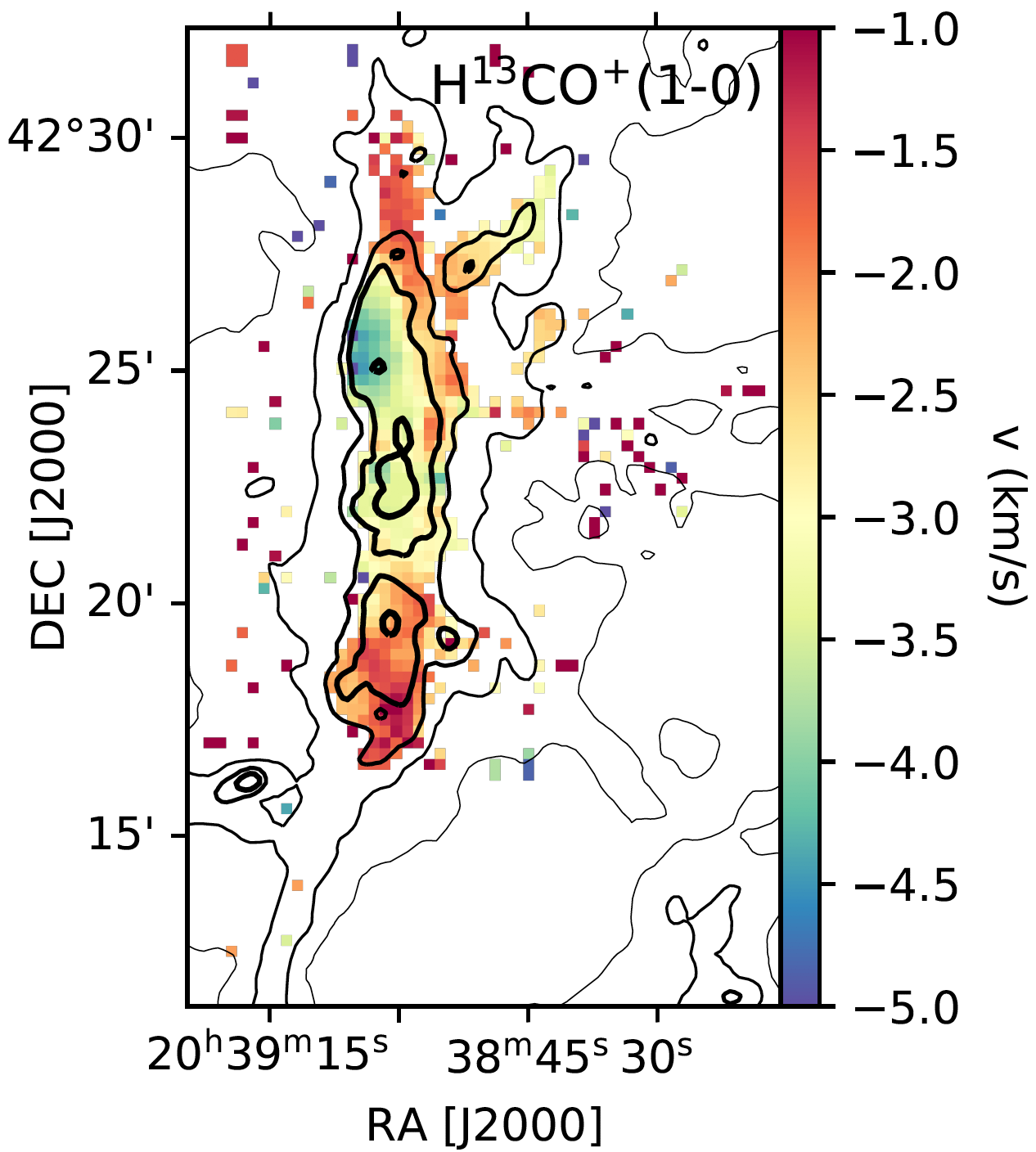}
% Moment maps 
%\includegraphics[width=0.45\hsize]{images/momentMap1_13CO.png}
%\includegraphics[width=0.38\hsize]{images/momentMap1_HCO+.png}
%\includegraphics[width=0.45\hsize]{images/momentMap1_C18O.png}
%\includegraphics[width=0.38\hsize]{images/momentMap1_H13CO+.png}
\end{center}
\caption{\textbf{Top left:} Velocity map of $^{13}$CO(1-0) over the DR21 ridge and sub-filaments, obtained from fitting a single Gaussian to the spectra. The uncertainty of the centroid velocity is below 0.15 km s$^{-1}$ over the entire map. Overplotted on the map are the \textit{Herschel} column density contours at $N_{\rm H_{2}}$ = 10$^{22}$, 2$\times$10$^{22}$, 5$\times$10$^{22}$, 10$^{23}$ and 5$\times$10$^{23}$ cm$^{-2}$ which indicate the ridge as well as the surrounding sub-filaments. The sub-filaments are indicated with their name from \citet{Hennemann2012}. \textbf{Top right:} The same for HCO$^{+}$(1-0) without fitting inside the DR21 ridge because of the HCO$^{+}$(1-0) self-absorption there. \textbf{Bottom left:} The same for C$^{18}$O(1-0). \textbf{Bottom right:} The same for H$^{13}$CO$^{+}$.}
\label{velMaps}
\end{figure*}

\begin{figure*}
\begin{center}
% Line fitting maps 
\includegraphics[width=0.49\hsize]{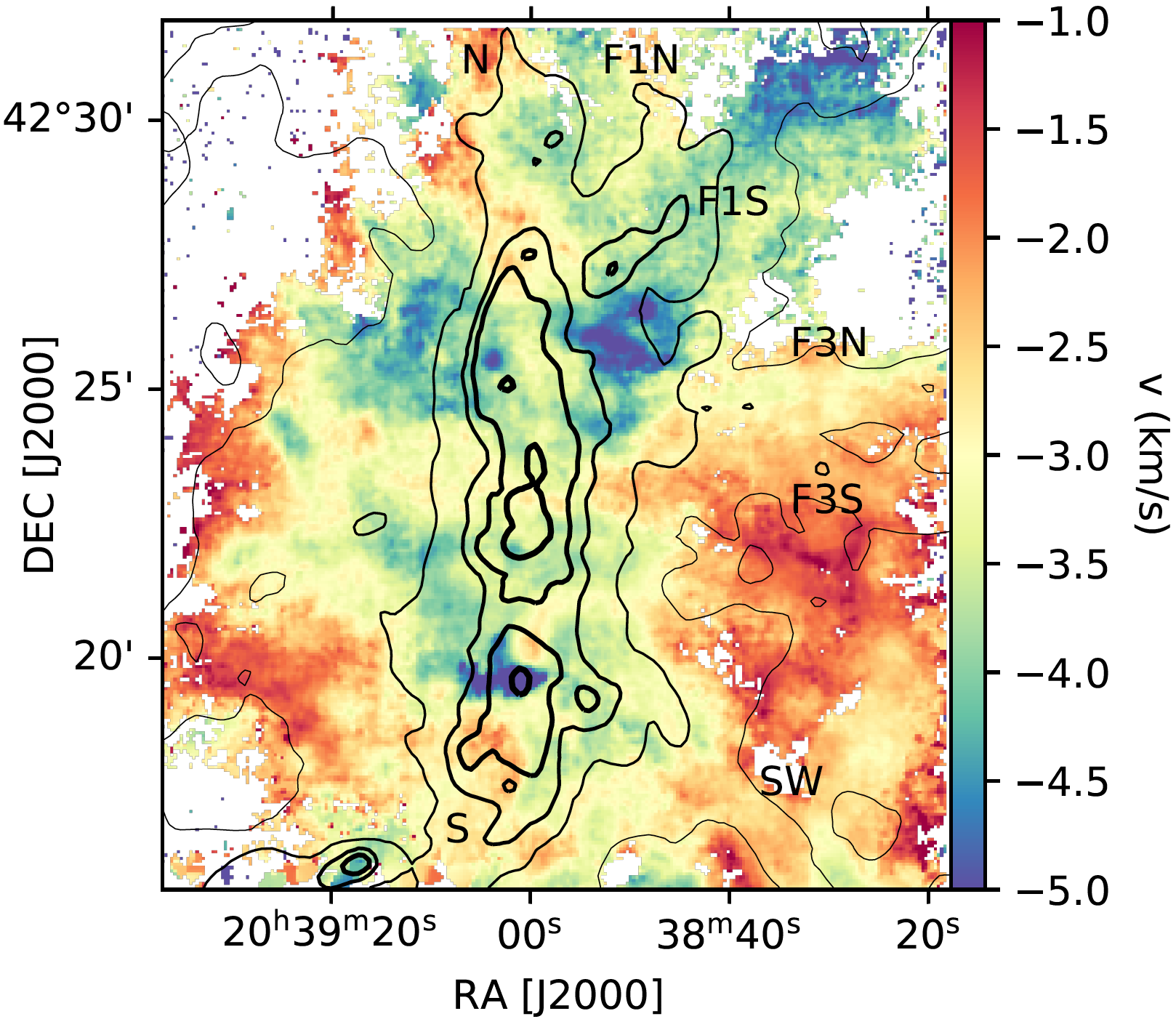}
\includegraphics[width=0.49\hsize]{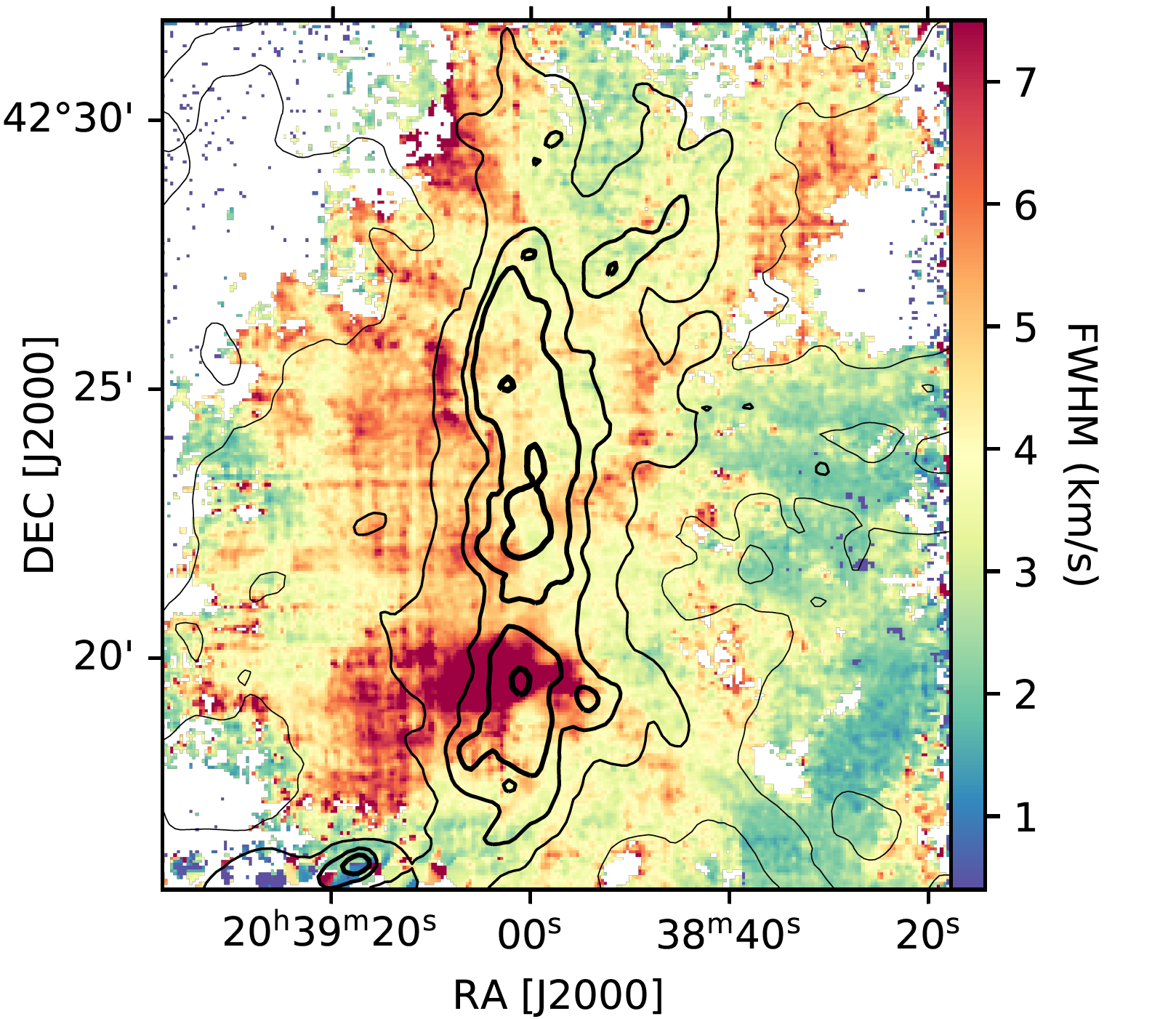}
% Moment maps 
%\includegraphics[width=0.49\hsize]{images/moment1MapCII-DR21_zoom_1.png}
%\includegraphics[width=0.49\hsize]{images/moment2MapCII-DR21_zoom_1.png}
\end{center}
\caption{\textbf{Left:} Velocity field observed with \CII. The black contours indicate the \textit{Herschel} column density contours at $N_{\rm {H_{2}}}$ = 10$^{22}$, 2$\times$10$^{22}$, 5$\times$10$^{22}$, 10$^{23}$ and 5$\times$10$^{23}$ cm$^{-2}$. The sub-filaments are indicated with their name. \textbf{Right:} The same for the \CII\ FWHM.}
\label{velFieldCII}
\end{figure*}

\subsection{The velocity field of the DR21 cloud} \label{subsec:velocity}
Even though the spectra are slightly more complex than a single Gaussian in the -8 to 1 km s$^{-1}$ velocity range, we performed a single Gaussian line fitting on the data sets to obtain information on the global velocity distribution and linewidth for the different tracers. A fit was accepted when the brightness was higher than 3$\times$ the noise rms in the different tracers. Note that moment maps might be better suited to give a view on the global dynamics in the region. However, in large regions of the observed map the signal-to-noise (S/N) of the data is not abundant while excellent S/N is required for producing trustworthy moment maps \citep[e.g.][]{Teague2019}. As a result, the produced first and second moment maps have large uncertainties in the subfilaments and do not provide a good visualization of the velocity field there.\\ %In contrast to the \CII\ observations (Sect.~\ref{subsec:spectra}), the optically thin molecular lines  show basically a single velocity component around $-$3 km s$^{-1}$ (Fig.~\ref{iramCoverage}), except of HCO$^+$ that suffers from a prominent self-absorption dip at that velocity in the ridge (this is why we excluded this area for the fit). The other lines show some broadening so that we may see the same overlap of several velocity components as for \CII. However, the angular resolution of $\sim$23-29$''$ of the data does not allow to draw a firm conclusion. \\
The resulting velocity maps %\footnote{We note that moment maps look very similar to the maps done with line fitting which is a confirmation that the fitting provides a good representation of the global kinematics.} 
for the molecular lines are presented in Fig. \ref{velMaps} and show similar velocity gradients over the ridge as the ones presented in \cite{Schneider2010} using only N$_{2}$H$^{+}$(1-0). We confirm that there is not a single gradient perpendicular to the ridge, but that the velocity pattern over the ridge shows organised velocity gradients with altering orientation. These altering velocity gradients are mixed with a clear north-south velocity gradient from -4.5 to -2.5 km s$^{-1}$. Remarkably, the velocities of the sub-filaments (N, F1N, F3N, F3S and SW) in all lines is redshifted (v$_{LSR}\,\sim$-2.5 to -1 km s$^{-1}$) with respect to the ridge (bulk emission around -3.5 km s$^{-1}$). Only the F1S sub-filament at velocities around -3.5 km s$^{-1}$ is not significantly redshifted with respect to the ridge but is not blueshifted either. Lastly, the southern sub-filament appears slightly redshifted close to the ridge, but  more to the south this filament becomes less redshifted with respect to the ridge. This observation of dominantly redshifted sub-filaments and a complete lack of a truly blueshifted subfilament with respect to the ridge is noteworthy and confirms the observations in the channel maps of $^{13}$CO(1-0) in Fig. \ref{13coChanMap}.\\
The \CII\ observations trace the lower density gas kinematics in the cloud, and thus provide a complementary view on the surrounding gas kinematics in the cloud off the dense filamentary structures. %Because the \CII\ emission is more complex with at least two velocity components (Sect.~\ref{subsec:spectra}), we here performed moment maps. 
The velocity field is shown in Fig. \ref{velFieldCII}. There is, similar to the molecular line observations, mostly systematically redshifted emission (v$\sim$-3 to -1 km s$^{-1}$) in the cloud around the ridge. This redshifted gas becomes increasingly prominent further away from the ridge. In addition, Apps. \ref{sec:TwoComps} \& \ref{sec:TwoCompsMol} show that there is an excellent correspondence in velocity of the subfilaments and this redshifted \CII\ emission, indicating that the subfilaments are directly embedded in this mass reservoir seen with \CII. However, the \CII\ velocity field also shows additional features. In particular, blueshifted velocities down to v=-5 km s$^{-1}$ are found in localized regions directly east and north-west of the ridge. No blueshifted gas is found further away from the ridge and the F1S filament. This fits with the observation in Fig. \ref{fig:specFocus} that the average \CII\ emission has an offset from the molecular line emission and that it is associated with the presence of two \CII\ velocity components (at -4.5 and -2.5 km s$^{-1}$) established in the previous section. Lastly, from Fig. \ref{fig:velMapAndNum} in App. \ref{sec:TwoComps} we also note that these regions with more blueshifted velocities overlap with the regions that have two fitted velocity components. In these regions the blueshifted velocity component is thus the dominant source of emission which affects the observed velocity field. %At these specific locations the more blueshifted component of \CII\ might become dominant, which appears to happen in particular close to the ridge. %In this scenario of multiple velocity components, the strong velocity gradients over the ridge are thus likely associated with local convergent flows that intersect at the position of the ridge.

\begin{figure*}
\begin{center}
\includegraphics[width=0.45\hsize]{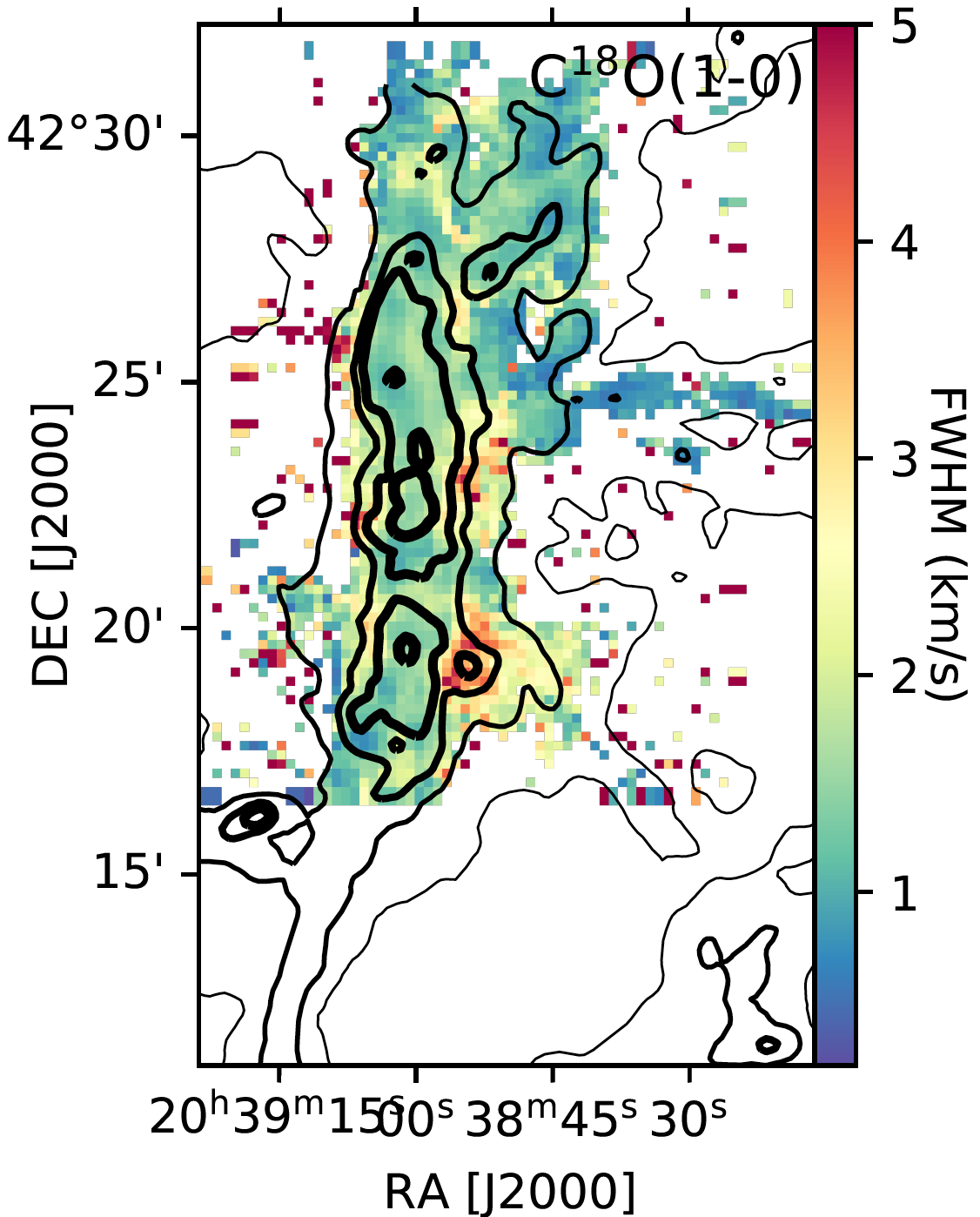}
\includegraphics[width=0.45\hsize]{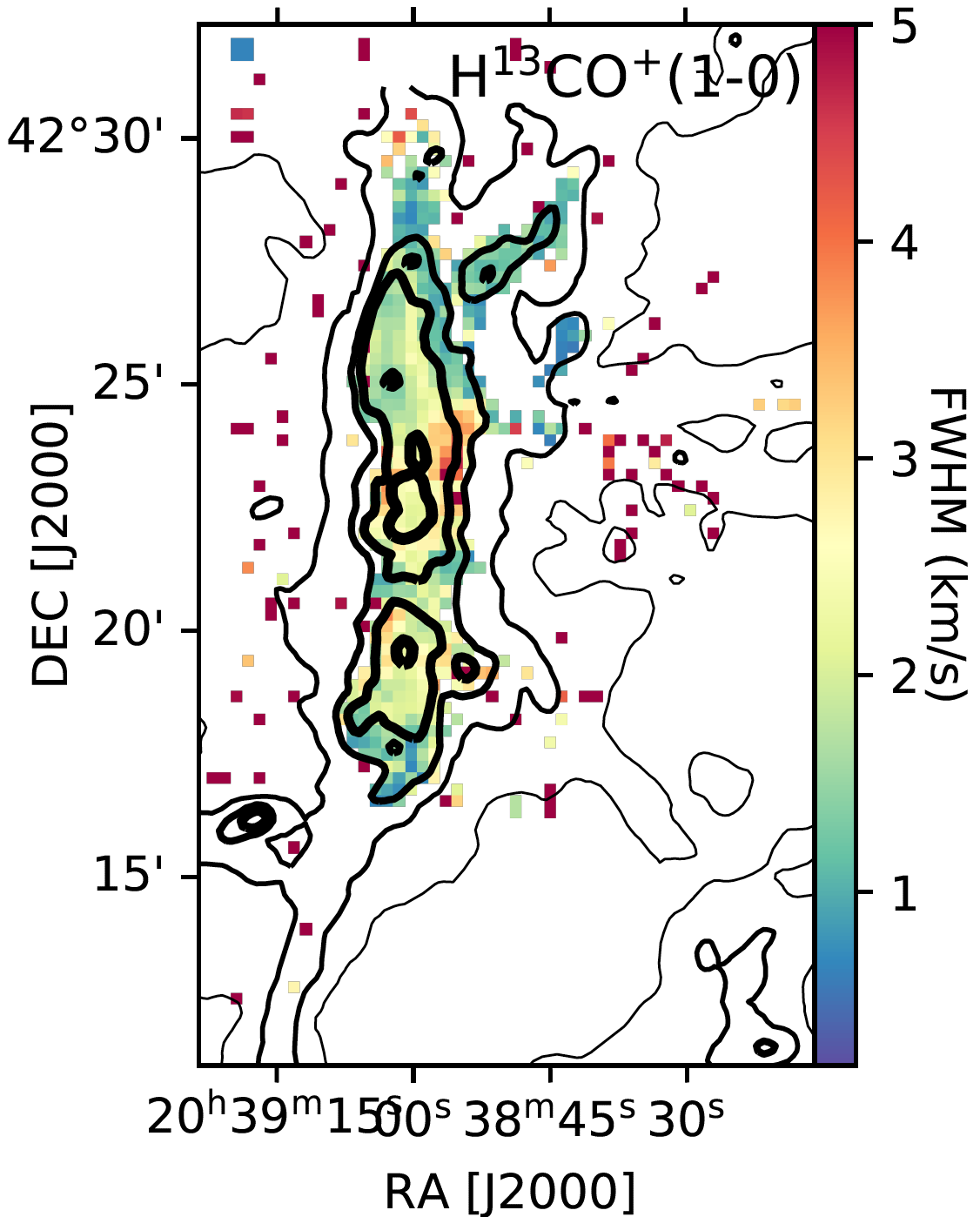}
\end{center}
\caption{\textbf{Left:} FWHM map of C$^{18}$O(1-0) over the DR21 ridge and sub-filaments. The uncertainty of the FWHM is generally below 0.1 km s$^{-1}$ and always below 0.4 km s$^{-1}$ over the entire map. Overplotted on the map are the \textit{Herschel} column density contours at $N_{\rm {H_{2}}}$ = 10$^{22}$, 2$\times$10$^{22}$, 5$\times$10$^{22}$, 10$^{23}$ and 5$\times$10$^{23}$ cm$^{-2}$. %Note that we used a different color scale compared to Fig. \ref{velFieldCII} to highlight the linewidth changes for the narrower molecular lines. 
\textbf{Right:} The same for H$^{13}$CO$^{+}$(1-0). }
\label{widMaps}
\end{figure*}

\subsection{Linewidths} \label{subsec:linewidth}
The Gaussian line fitting also provides maps of the FWHM for the spectra. The results for the optically thinner lines C$^{18}$O(1-0) and H$^{13}$CO$^{+}$(1-0) are presented in Fig.~\ref{widMaps}. %Note that the FWHM is related to the velocity dispersion $\sigma$ with $\sigma = {\rm FWHM}/\sqrt{8 \, \ln(2))}$. 
The C$^{18}$O(1-0) map indicates, where detected, that the sub-filaments have a lower FWHM than the ridge, typically $\lesssim$ 1 km s$^{-1}$. In the ridge, a FWHM between 1 and 2.5 km s$^{-1}$ is observed (which typically has two velocity components, see App. \ref{sec:TwoCompsMol}). Noteworthy is the strong local increase of the FWHM up to 4 km s$^{-1}$ at locations where the sub-filaments connect to the ridge. This behaviour is also observed in the velocity dispersion maps of N$_{2}$H$^+$(1-0) \citep{Schneider2010} and NH$_{3}$ \citep{Keown2019}. One would expect that this is the result of overlapping velocity components, associated to the sub-filament and the ridge, at this connection. However, App. \ref{sec:TwoCompsMol} shows that these regions are best fitted with a single Gaussian component. This suggests either that the overlapping velocity components are closely blended together or that there is a rapid change in the line-of-sight velocity field at the edge of the ridge for example due to accretion shocks.\\ 
Comparing the FWHM maps for the molecular lines in Fig. \ref{widMaps} with those of the \CII\ line, we observe three clear differences. Firstly, the \CII\ line has a significantly larger linewidth (typically $>$ 4-5 km s$^{-1}$) than the molecular lines over the full map. 
%Nicola: I added that for the DR21 outflow. 
Note that the largest FWHMs are associated with the DR21 outflow which is not the focus of our study. 
Secondly, the highest \CII\ FWHM values, excluding the DR21 outflow, are found outside of the dense DR21 ridge, mostly in the eastern region without molecular dense sub-filaments. This east-west asymmetry with respect to the ridge for the \CII\ linewidth is remarkable and seems to be correlated with the lack of dense sub-filaments east of the ridge. Thirdly, examining the map in more detail, it is also observed that the \CII\ emission directly surrounding the sub-filaments has a significantly larger linewidth than the emission towards the sub-filaments. 
%Nicola: see my comment, I would be careful with this statement. We should talk about that.. 
%These differences, compared to the molecular line FWHM maps, are not necessarily surprising. 
%The \CII\ line traces the lower density, UV-excited outer gas while the molecular lines trace the dense, molecular ridge and sub-filaments. 
In Fig. \ref{iramCoverage}, it is observed that this broader \CII\ linewidth is the result of the flat-topped spectra which are due to more blueshifted \CII\ emission with respect to the molecular lines. This more blueshifted \CII\ emission is the result of the prominent blueshifted \CII\ velocity component found in App. \ref{sec:TwoComps} and thus is not necessarily the result of higher turbulent support.
This demonstrates that the molecular lines miss an important part of the picture to understand the full DR21 molecular cloud dynamics and evolution.

%-----------------------------------------------------------------------------------------------------------------------
\section{Analysis}\label{sec:analysis}

\subsection{The C$^{+}$ column density and excitation}\label{sec:CIIexcitation}
\begin{figure}
    \centering
    \includegraphics[width=0.97\hsize]{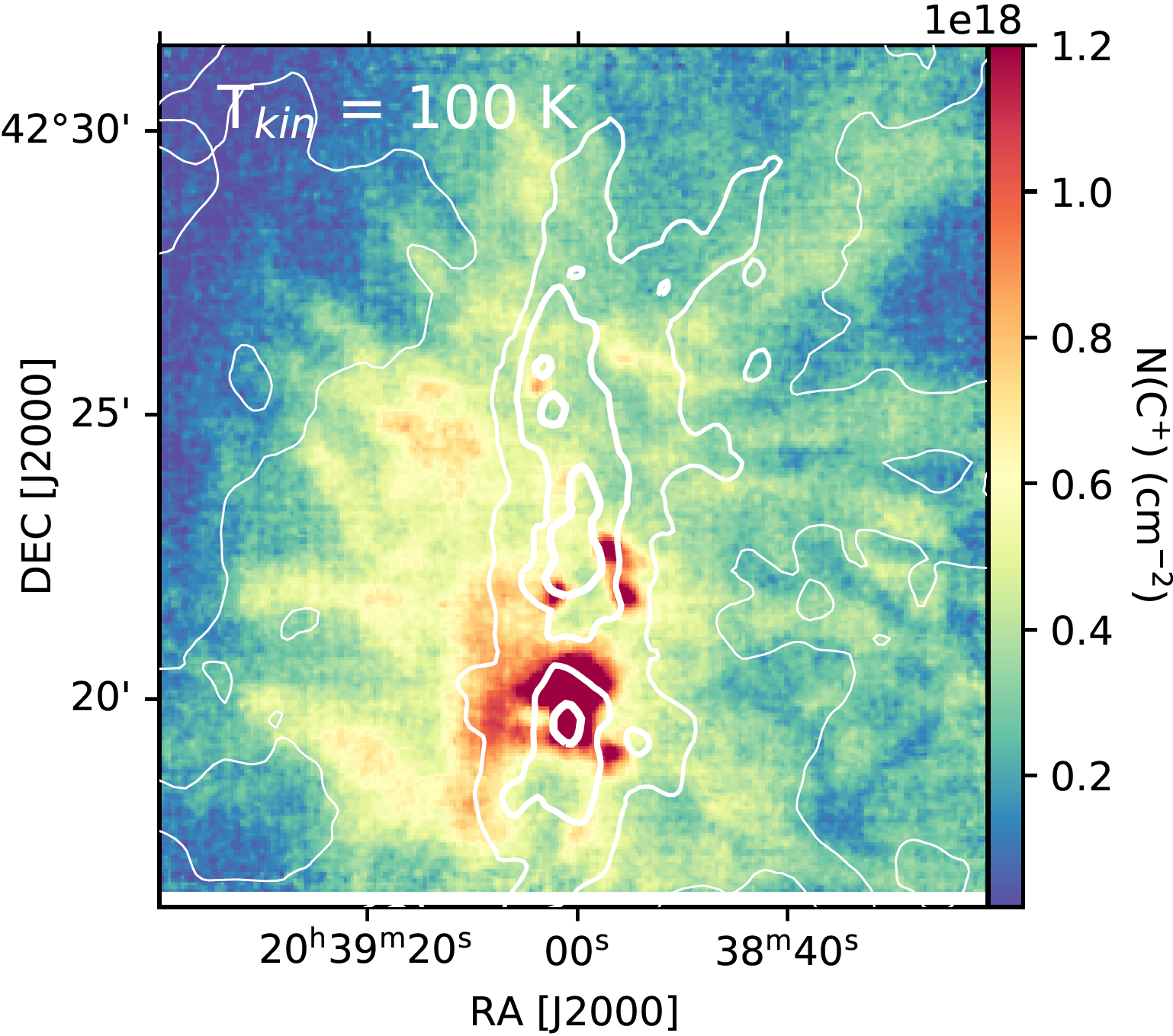}
    \includegraphics[width=\hsize]{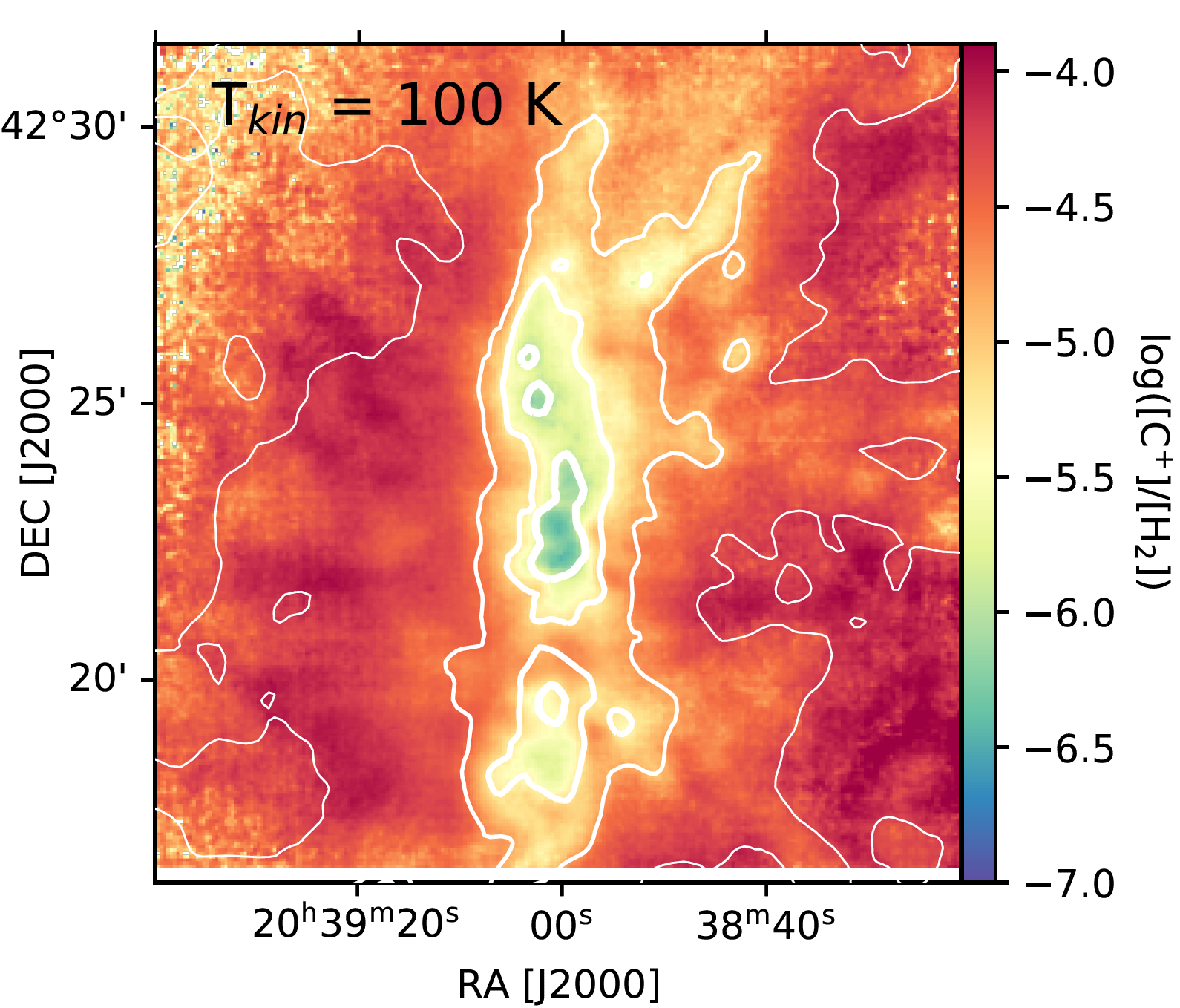}
    \caption{{\bf Top}: C$^{+}$ column density map between -8 and 1~km~s$^{-1}$ assuming a kinetic temperature of 100~K and $n_{\rm H_{2}}$~=~5$\times$10$^{3}$ cm$^{-3}$ for the excitation conditions. Note that these excitation conditions, which should be representative for most of the map, are most likely not valid towards the peak around the DR21 radiocontinuum region in the south of the map as this region experiences strong internal heating. {\bf Bottom}: The resulting [C$^{+}$]/[H$_{2}$] abundance ratio deduced from the \textit{Herschel} column density map over the DR21 cloud on a log scale.}
    \label{fig:Nc+map}
\end{figure}

In Sect. \ref{subsec:maps}, it was estimated that \CII\ traces gas between 5$\times$10$^{2}$ and 10$^{4}$ cm$^{-3}$. Here, we will explore the excitation conditions and regions traced by the \CII\ emission in more detail. To determine the typical temperature of the C$^{+}$ gas, we consider the heating from FUV photons. From the FUV field map calculated in \citet{Schneider2016b}, we find that the DR21 cloud is located in an FUV field up to G$_{0}$ = 200-400 Habing due to heating from local UV source in the region \citep[see also][]{Schneider2023}. This fits with the typical line integrated \CII\ intensity towards the surrounding cloud of $\sim$ 50 K km s$^{-1}$ which is expected for an FUV field of G$_{0}$ = 200 Habing in the PDR Toolbox \citep{Kaufman2006,Pound2008,Pound2023} at densities between 5$\times$10$^{2}$ and 10$^{4}$ cm$^{-3}$. For this FUV-field strength and density range, the PDR Toolbox predicts temperatures of 100-200 K for the \CII\ emitting gas of the photo-dissociation region (PDR).\\
With these excitation conditions, it is possible to produce the C$^{+}$ column density map over the full extent of the FEEDBACK map using the equation from \citet{Goldsmith2012}
\begin{equation}
    {\scriptstyle\Delta {\rm T_{A}} = 3.43\times10^{-16} \times\left[ 1 + 0.5{\rm e}^{91.25/{\rm T_{kin}}} \left( 1 + \frac{2.4\times10^{-6}}{{\rm C_{ul}}} \right) \right]^{-1} \frac{{\rm N(C^{+})}}{\delta {\rm v}}}
\end{equation}
with $\Delta$T$_{A}$ the brightness temperature for a uniform source that fills the beam, T$_{kin}$ the kinetic temperature, C$_{ul}$ the collisional de-excitation rate, N(C$^{+}$) the C$^{+}$ column density and $\delta$v the linewidth. C$_{ul}$ is given by 
\begin{equation}
    {\rm C_{ul}} = {\rm n} \times {\rm R_{ul}}
\end{equation}
where R$_{ul}$ is the de-excitation rate coefficient given by
\begin{equation}
    {\rm R_{ul}} = 3.8 \times 10^{-10}\,{\rm cm}^{3}\,{\rm s}^{-1}\,({\rm T_{kin}}/100)^{0.14}
\end{equation}
Using a T$_{kin}$ = 100 K and $n_{\rm H_{2}}$ = 5$\times$10$^{3}$ cm$^{-3}$ then gives C$^{+}$ column densities N(C$^{+}$) $\approx$ 0.4-1.0$\times$10$^{18}$ cm$^{-2}$ over the extent of the cloud which is shown in Fig. \ref{fig:Nc+map}. However, note that these assumed excitation conditions are not valid for the DR21 radiocontinuum region as it is a dense and strongly irradiated PDR region. Comparing this C$^{+}$ column density map with the deduced H$_{2}$ column density map from the \textit{Herschel} data also allows to estimate the [C$^{+}$]/[H$_{2}$] abundance ratio over the map. This shows that we find a C$^{+}$ abundance around 10$^{-4}$ in the outer parts of the cloud around the ridge and sub-filaments, which is of the order of the elemental abundance ratio of carbon ($\chi_{C}$) in the ISM \citep[e.g.][]{Glover2012}. Basically all carbon is thus found in the ionized state (C$^{+}$) in these regions. Towards the sub-filaments, we find under these conditions that $\sim$ 10\% of the gas is traced by \CII\ and towards the ridge this drops even lower to values between 0.1 and 1\%. We do however note that the density and temperature are an important assumption. A higher temperature up to 200 K can reduce the C$^{+}$ column density with $\sim$ 50\% while a lower temperature can increase the C$^{+}$ column density by a factor 2. A lower typical density of 10$^{3}$ cm$^{-3}$ can increase the column density with an additional factor 2. However, such a N(C$^{+}$) increase due to a low density would result in a C$^{+}$ abundance above the elemental abundance. This suggest that a significant part of the \CII\ emission originates from regions with a typical density of $n_{\rm H_{2}} \sim$ 5$\times$10$^{3}$ cm$^{-3}$. Combining this density with the Herschel column density for the surrounding clouds results in a maximal line-of-sight depth of 2 pc. This implies that the surrounding cloud of the DR21 ridge has a sheet-like morphology.\\ 
The same analysis is done for $^{13}$CO(1-0) in App.~\ref{sec:13coAbundance}. This analysis indicates that the $^{13}$CO abundance peaks in some sub-filaments and at the edges of the DR21 ridge. The abundance then decreases in the ridge and in the outer cloud. Both in the ridge and the outer part of the cloud, $^{13}$CO(1-0) appears to trace only a fraction of the total gas mass. Combined, this provides compelling evidence that \CII\ indeed traces most of the surrounding gas in the DR21 cloud. The regions surrounding the ridge and sub-filaments are thus CO poor gas that lights up in \CII. In combination with the molecular line data, \CII\ thus provides a global view of the cloud.

%filament with all the subfilaments of the cloud.
%Because \CII\ traces more extended gas, it also provides a view on the linewidth east of the ridge. This clearly %shows that the \CII\ emission east of the ridge has a significantly larger linewidth than the region west of the %filament with all the subfilaments of the cloud. Visually inspecting the fitted spectra, confirms indeed an %increasing linewidth to the east of the ridge. Examining the map in more detail, it is also observed that the \CII\ %emission directly surrounding the subfilaments has a significantly larger linewidth than the \CII\ emission towards %the subfilaments.%, this is indicates that there is a relatively strong transition from the subfilament velocity to %the velocity of the ridge at the location where it was proposed that the subfilaments are accreted %\citep{Schneider2010}. With the current data, it is not possible to confidently contrain the physical length of this %transition, because it might not be resolved by the beamsize. This implies a transition length $\leqslant$ 0.2 pc.

\begin{figure*}
\includegraphics[width=0.5\hsize]{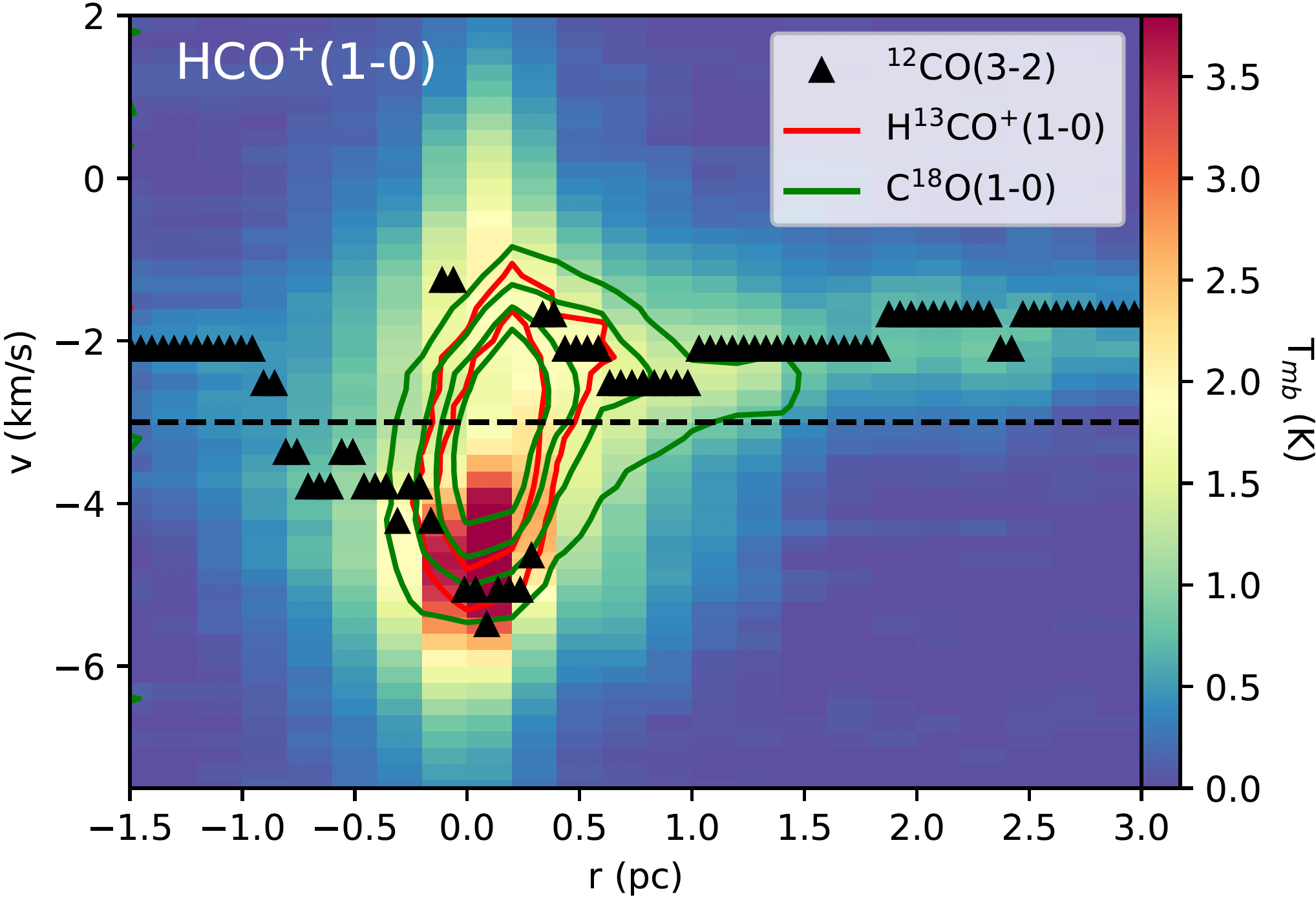}
\includegraphics[width=0.5\hsize]{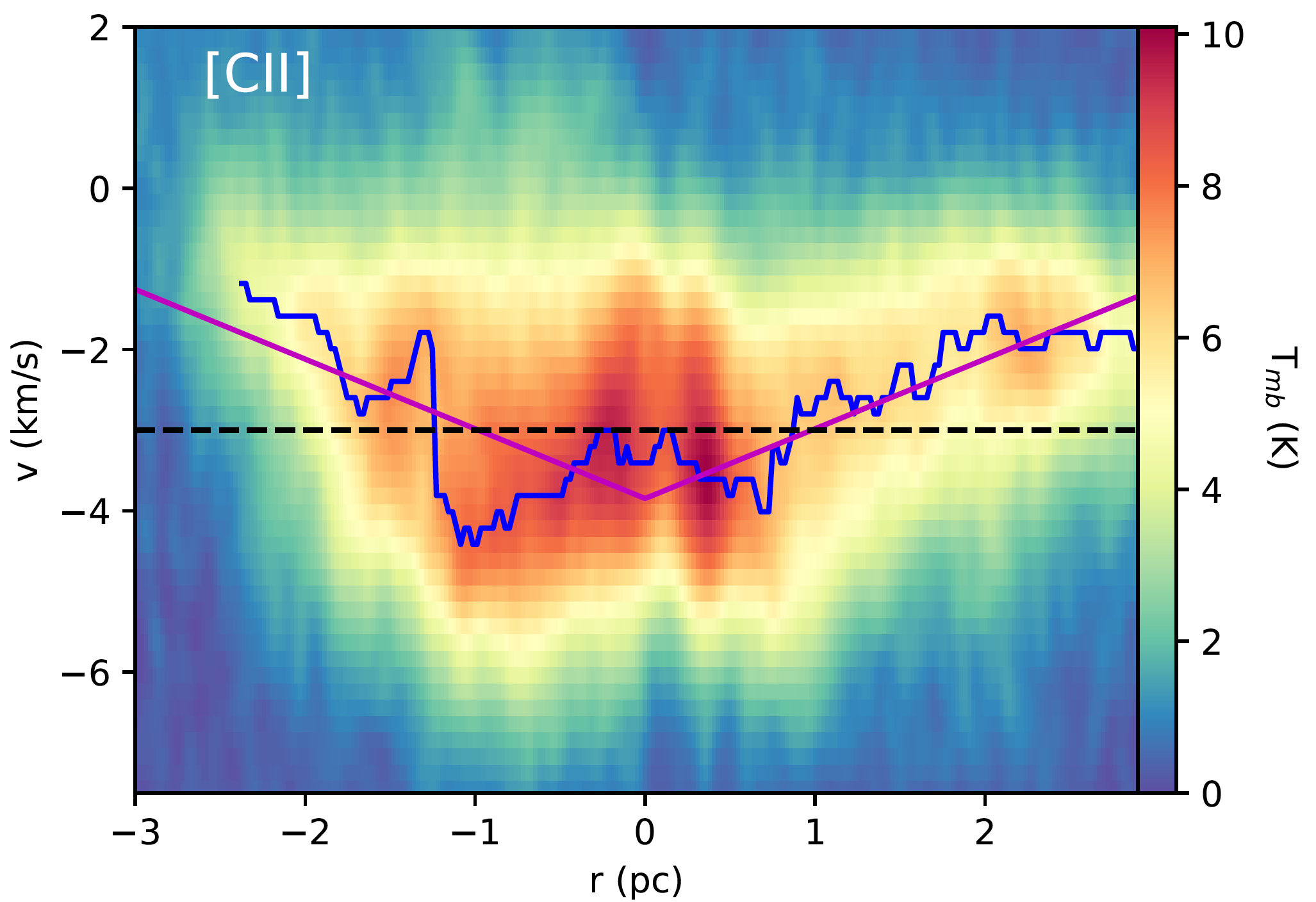}
\caption{{\bf Left}: HCO$^{+}$(1-0) PV diagram with the green and red contours indicating the emission from C$^{18}$O(1-0) and H$^{13}$CO$^{+}$(1-0), respectively. The PV diagram was constructed perpendicular to the filament from the region indicated by the red box in Fig. \ref{integratedCIICI} by averaging the intensities along declination axis in this box. The black triangles follow the peak brightness velocity of $^{12}$CO(3-2) as a function of the distance (r) from the center of the ridge. This indicates that the gas is redshifted at both sides of the ridge which is at a velocity of $\sim$ -3 km s$^{-1}$ (indicated by the dashed horizontal line). At the ridge, there is a large increase in linewidth, associated with the velocity gradients observed over the ridge in Fig. \ref{velMaps}. This fits with the observed HCO$^{+}$(1-0) blue asymmetry in the PV diagram due to self-absorption in a collapsing cloud. %For $^{12}$CO(3-2) a high opacity has to  be taken into account and that at high velocities (v$_{lsr} < $ -6 km s$^{-1}$ and v$_{lsr} > $ 0 km s$^{-1}$) it also traces protostellar outflows.
%This fits with the observation that the H$^{13}$CO$^{+}$(1-0) C$^{18}$O(1-0) contours show a dispersion that fits very well with self-absorbed profile in HCO$^{+}$ that indicates collapse. 
{\bf Right}: The \CII\ PV diagram perpendicular to the DR21 ridge over the same region with the blue line following the peak brightness velocity. %of the ambient gas traced by \CII\ which is redshifted with respect to the ridge. The blue line indicates the peak brightness evolution of \CI, and the black triangles indicate the $^{12}$CO(3-2) evolution. 
On large scales ($|$r$|>$ 1 pc), the \CII\ emission shows a redshifted V-shape followed with a sharp transition to more blueshifted emission at ($|$r$|<$ 1 pc) which makes the velocity profile deviate from a pure V-shape. To highlight this, we fitted a V-shape to the PV diagram which is indicated in magenta. We interpret this to be due to a second more spatially concentrated velocity component that is associated with the formation of the DR21 ridge. The Pearson coefficients for the central velocity as a function of radius are -0.76 and 0.89 at the east and west of the ridge, respectively. This points to a relatively well organized, but not perfect correlation, for the central velocity with radius.
}
\label{pvDiagrams}
\end{figure*}

\subsection{Position-velocity diagram of the ridge} \label{subsec:pv}
The position-velocity (PV) diagram perpendicular to the DR21 ridge gives an additional view on the cloud kinematics with respect to the ridge. Figure~\ref{pvDiagrams} shows PV diagrams for the optically thick -HCO$^{+}$(1-0) \& $^{12}$CO(3-2)- and thin -H$^{13}$CO$^{+}$(1-0) and C$^{18}$O(1-0)- molecular lines observed with the IRAM 30m and JCMT (see the red box in Fig. \ref{integratedCIICI} for the covered region). It confirms that the emission from subfilaments outside the ridge is at more redshifted velocities with respect to the densest gas in the ridge. This gives rise to a V-like shape in the PV diagram, similar to what was observed for the Musca filament \citep{Bonne2020b} which has about the same size as the DR21 ridge but has a two orders of magnitude lower mass. For the Musca filament this was proposed to be the result of magnetic field bending by the interaction of an overdensity with a more diffuse region in the colliding \HI\ cloud.  %which we also call a $^{\prime}$soft$^{\prime}$ collision (Schneider et al. in prep.). 
As for this previous study of the low-mass Musca filament, the DR21 ridge is located at the apex of this V-like shape. However, the V-shape for DR21 in molecular lines is not as clear as the ones observed in Musca. This can be related to the rapid linewidth increase within 1 pc of the DR21 ridge, while the Musca filament is trans-sonic \citep{Hacar2016,Bonne2020b}. Such a rapid linewidth increase in the densest regions is predicted to be the result of pc scale gravitational acceleration in massive clouds \citep[e.g.][]{Peretto2006,Peretto2007,Hartmann2007,Gomez2014,Watkins2019} and could also be the result of accretion driven turbulence \citep{Klessen2010}. %The gravitationally driven linewidth is further supported by the fact that the linewidth of 1-2 km s$^{-1}$, see Figs. \ref{widMaps} and \ref{pvDiagrams}, corresponds to the typical velocity fields perpendicular to the ridge, see Fig.~\ref{velMaps} and \citet{Schneider2010}. 
A gravitational collapse scenario is also supported by the strong blue asymmetry self-absorption of HCO$^{+}$(1-0) that covers the same velocity interval as the perpendicular velocity gradients over the ridge \citep{Schneider2010}.\\% It should also be noted that the presence of protostellar outflows in the ridge, traced with $^{12}$CO, provides further contamination to the PV diagram perpendicular to the ridge. However this does not affect the interpretation of the PV diagram as this occurs at higher velocities. NICOLA: This last sentence is useful, why out?\\
Figure \ref{pvDiagrams} also presents the \CII\ PV diagram which traces the surrounding gas in the DR21 cloud. The PV diagram shows that the surrounding gas forms a redshifted V-shape perpendicular to the DR21 ridge on large scales ($|$r$| >$ 1 pc). The Pearson correlation coefficients for the central \CII\ velocity as a function of radius are -0.76 and 0.89 on the east and west side of the ridge, respectively. With bootstrapping we find that these coefficients are far out of the 95\% intervals for the null hypothesis that there is no correlation, i.e. [-0.19, 0.20] (east) and [-0.18, 0.17] (west), and thus demonstrate a clearly organized velocity field perpendicular to the ridge. As the cloud appears to be organized in a sheet, this points at a curved sheet. However, in \CII\ it is also observed that there are deviations from a pure V-shape due to blueshifted emission in the vicinity of the ridge at $|$r$|$ $\sim$ 1 pc. This deviation from the V-shape is particularly visualized in the PV diagram by the sharp peak velocity transition at $|$r$|$ $\sim$ 1 pc and might explain why the Pearson correlation coefficients are not equal to one. We propose that this traces two distinct but converging velocity components at -4.5 and -2.5 km s$^{-1}$ in the DR21 cloud that drive continuous mass inflow to the DR21 ridge which is located at the intersection of these flows. 
%Note that here we are not talking of the proposed high-velocity ($>$ 10 km s$^{-1}$) cloud-cloud collision between the 
%DR21 cloud and a cloud at 9 km s$^{-1}$ \citep{Dickel1978,Dobashi2019}, but rather of organized accretion flows on the DR21 ridge in the cloud formed by the soft collision.
% Nicola: Better Dobashi et al., our scenario is again different as the one of Dickel and Dobashi.  
%Note that here we are not talking of the proposed high-velocity ($>$ 10 km s$^{-1}$) collision in the DR21 cloud \citep[][ Schneider et al. in prep.]{Dickel1978}, but rather the organized accretion flows on the DR21 ridge in the cloud formed by the soft collision.

%\begin{figure}
%    \centering
%    \includegraphics[width=\hsize]{images/pvDiagramRaDR21_zoom_y340-460+CI.png}
%    \caption{The PV diagram of the \CII\ line perpendicular to the DR21 ridge (indicated in Fig.~\ref{integratedCIICI}). The red triangles indicate the velocity of maximal \CII\ brightness at each radius. The blue solid line indicates the same for the \CI\ emission.}
%    \label{pvDiagramCII}
%\end{figure}

\subsection{Convergence of blue- and redshifted gas in the DR21 cloud} \label{subsec:convergence}
\begin{figure*}
\centering
\includegraphics[width=0.325\hsize]{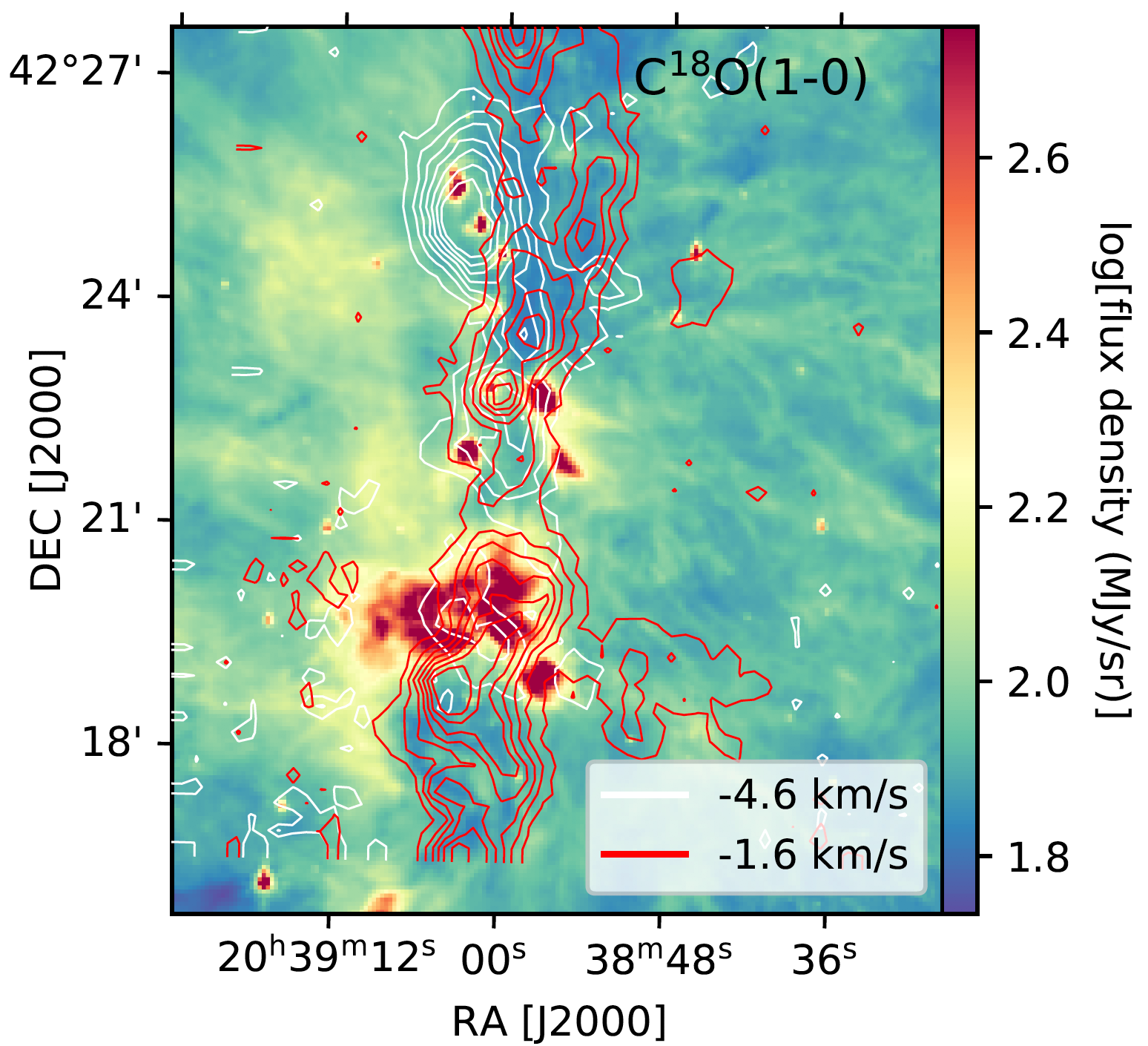}
\includegraphics[width=0.32\hsize]{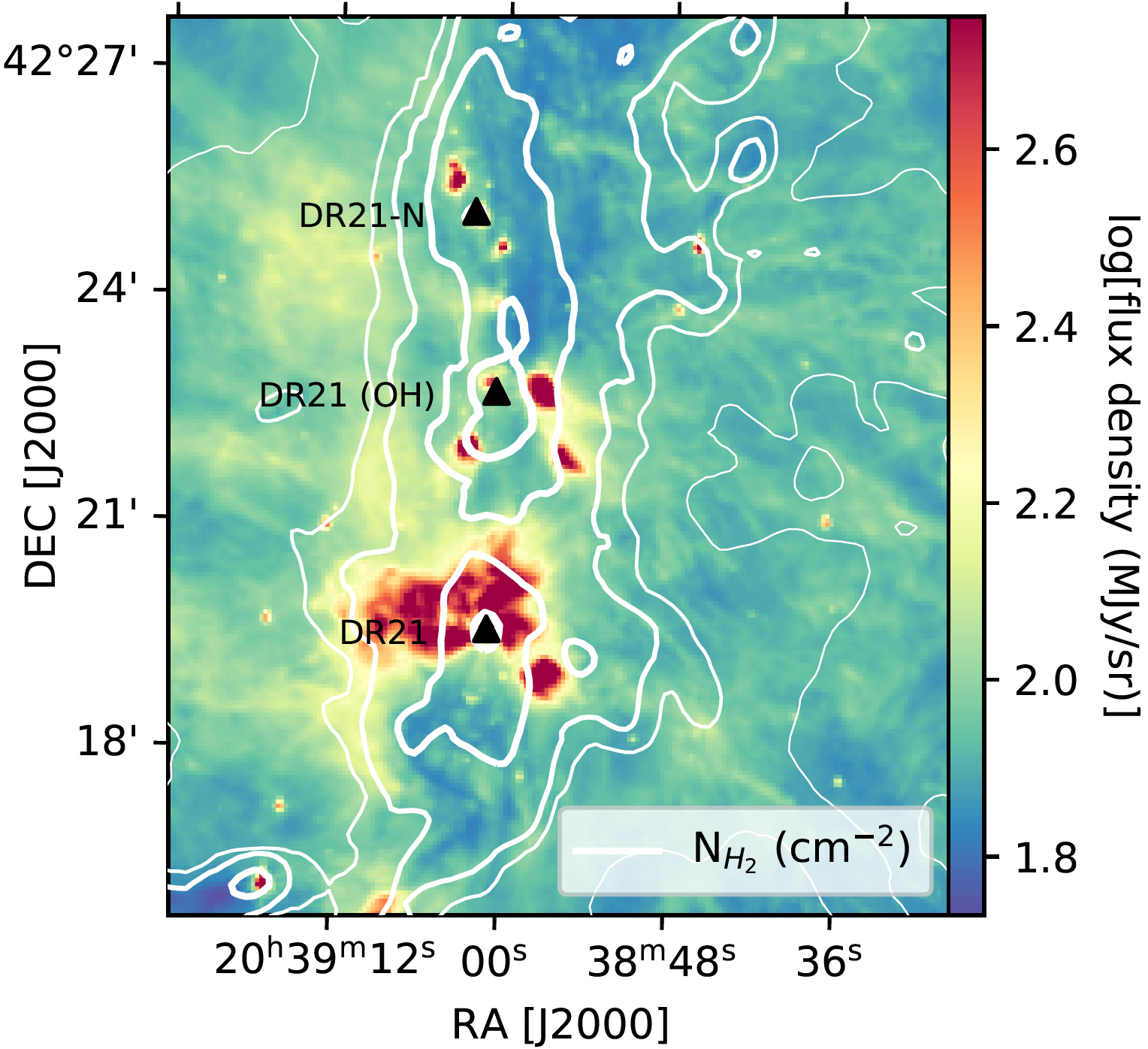}
\includegraphics[width=0.31\hsize]{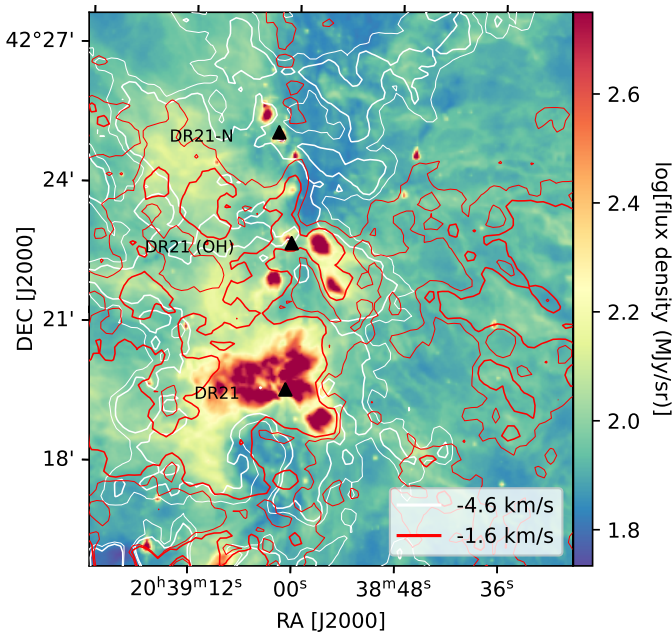}
\caption{\textit{Spitzer} 8 $\mu$m image of the DR21 cloud overlaid with velocity contours of C$^{18}$O(1-0) (left) and \CII\ (right) at -1.6 km s$^{-1}$ and -4.6 km s$^{-1}$, and overlaid with the \textit{Herschel} column density (middle). The northern and southern part of the ridge shows 8 $\mu$m extinction, i.e. is 'IR-dark', while in the middle there is strong emission at the locations of DR21 and DR21(OH). Some protostellar sources show up as localized emission spots. Outside of the ridge, the 8 $\mu$m emission is extended, rather diffuse, with some feather-like features. The contours of C$^{18}$O(1-0) indicate that the more blueshifted emission is located towards the region with IR emission in the ridge, while the redshifted emission is concentrated towards the strong 8 $\mu$m extinction. The structure of the \CII\ emission is more complex as it more diffuse and also affected by the feedback in DR21 and local protostellar sources (e.g. around DR21(OH)). However, there  is overall a correspondance on the large scales of the blueshifted gas to bright Spitzer 8 $\mu$m regions and redshifted gas to 8 $\mu$m extinction. The figure in the middle indicates the location of DR21, DR21 (OH) and DR21-N for reference of their location.}
\label{ExtinctionContours}
\end{figure*}

The molecular line observations demonstrated the existence of an organized velocity field over the ridge, similar to the results presented in \cite{Schneider2010}. This internal velocity field is proposed to be connected to the inflowing sub-filaments in the surrounding cloud, and is typically considered to be tracing inflow. Using 8~$\mu$m observations from the \textit{Spitzer} Space Telescope \citep{Hora2009}, we attempt to portray a 3D configuration of the ridge and sub-filaments. Inspecting the 8~$\mu$m map of the DR21 cloud (Fig. \ref{ExtinctionContours}), it becomes obvious that heavily reduced 8~$\mu$m emission ('IR-dark') traces regions inside the ridge, except for the bright compact sources at DR21 and DR21(OH) which shows internal regions heated by stellar/YSO feedback. Consequently, this indicates that the gas associated with strong 8~$\mu$m extinction is located at the front side of the DR21 ridge with $N_{\rm H_{2}}\,>$ 10$^{23}$ cm$^{-2}$, as this is a foreground layer absorbing 8~$\mu$m emission. The region of the ridge with more extended 8~$\mu$m emission is then located towards the back of the DR21 ridge. Towards the sub-filaments similar, but lower contrast, 8~$\mu$m extinction is observed. The overlay of the 8~$\mu$m map with contours of molecular and \CII\ line emission at v = -4.6 km s$^{-1}$ and -1.6 km s$^{-1}$ indicates that the most blueshifted molecular line emission corresponds to the extended 8~$\mu$m emission features, while the redshifted line emission corresponds to the darkest 8~$\mu$m regions. This clearly indicates that the redshifted gas is located in front of the blueshifted gas from our point of view, which confirms that the flows in the ridge and cloud are indeed converging. The red- and blueshifted velocity component that make the V-shape in the PV diagram are thus accreted on the ridge. Note that there are several contaminants (DR21, protostellar objects,...), a lower contrast and a more complex contour distribution that make the \CII\ map less straightforward to evaluate. Nonetheless, there is a clear association at several locations between the distribution of the redshifted gas and 8~$\mu$m extinction in the surrounding cloud that is traced by this line.

\subsection{Virial analysis of the DR21 cloud} \label{subsec:virSec}

\begin{figure}
    \centering
    \includegraphics[width=\hsize]{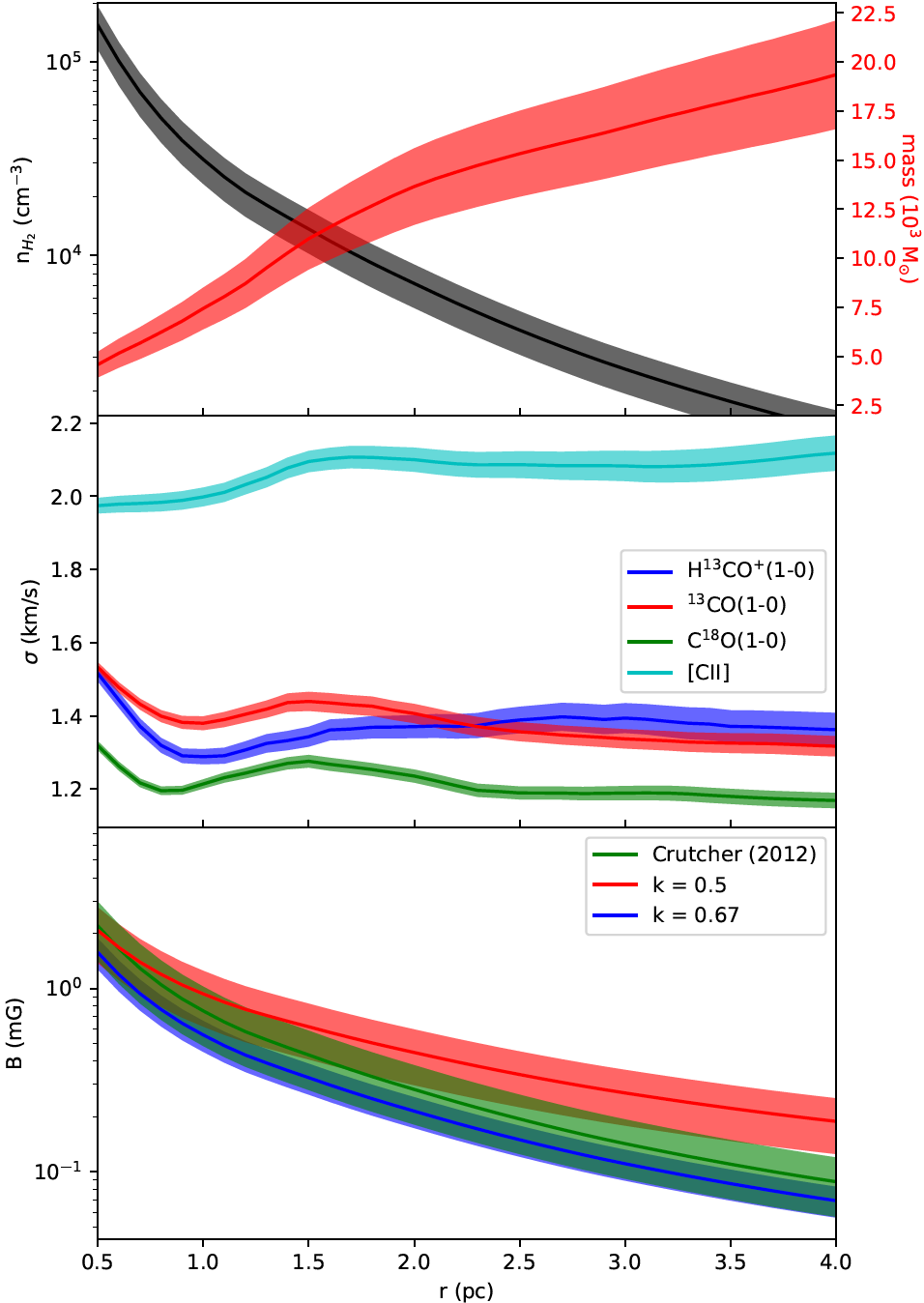}
    \caption{{\bf Top}: Radial profile of the average density in the DR21 cloud (in black). The average density within each radius was estimated based on the mass within the considered radius. The mass within the radius r, is shown at the axis on the right (in red). {\bf Middle:} The same for the velocity dispersion of H$^{13}$CO$^{+}$(1-0) (in blue), $^{13}$CO(1-0) (in red),  C$^{18}$O(1-0) (in green) and \CII\ (in cyan). The molecular lines show very similar behaviour while the \CII\ line has a significantly higher linewidth. However, note that at the outer radii (r $\sim$ 3-4 pc) the cloud is not fully covered by the IRAM 30m map  (Figs.~\ref{iramCoverage} and ~\ref{intIntensityMaps}). {\bf Bottom}: Estimated characteristic magnetic field strength for the the gas within a radius for the three different considered profiles. The green curve shows the predictions by the \citet{Crutcher2010} relation. The red and blue curves use a magnetic field strength of 0.94 mG at the density of the ridge and k = 0.5 and 0.67, respectively, in B $\propto$ n$_{{\rm H}}^{k}$.}
    \label{fig:densMassSigmaBfieldVirial}
\end{figure}

\begin{figure*}
\includegraphics[width=0.49\hsize]{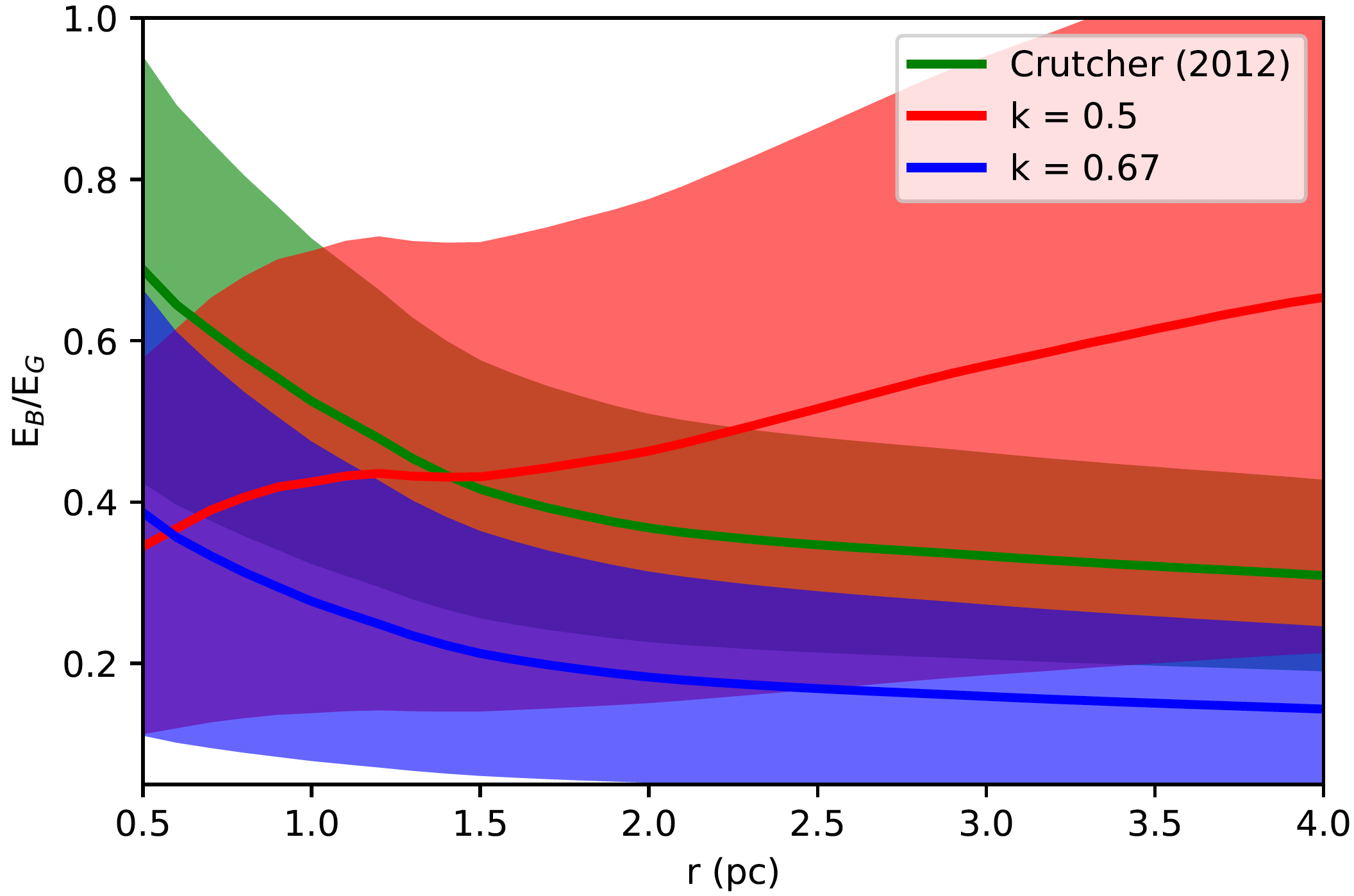}
\includegraphics[width=0.49\hsize]{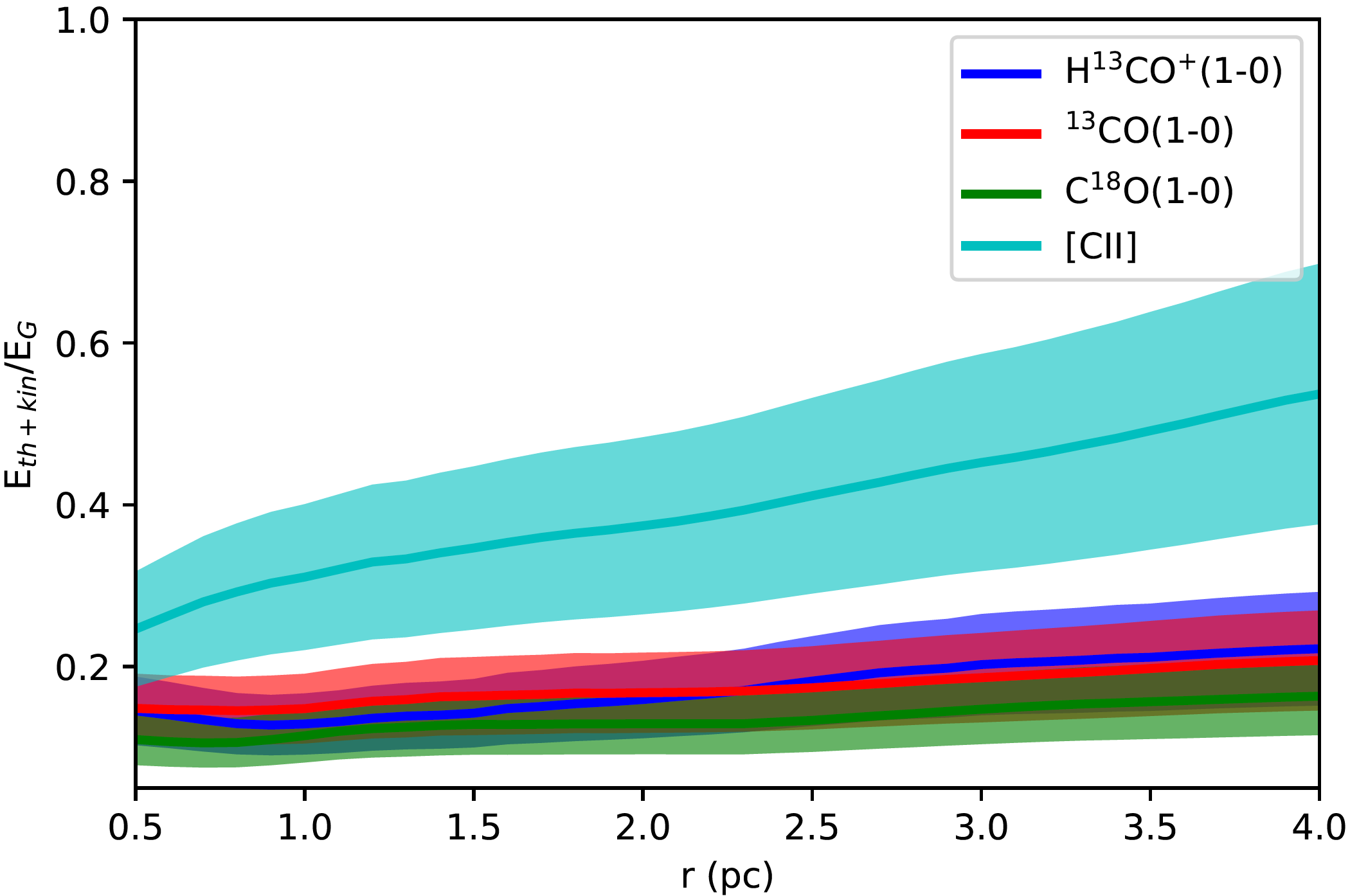}
\caption{Ratio of the magnetic (left) and thermal + kinetic energy (right) over the gravitational energy as a function of the radius in the DR21 cloud. The radius is centered on DR21(OH). For the thermal + kinetic energy, the linewidth of H$^{13}$CO$^{+}$(1-0), $^{13}$CO(1-0), C$^{18}$O(1-0) and \CII\ are used. It appears that the regions of cloud traced by \CII\ experience more support against gravitational collapse than the molecular regions. However, the presence of multiple components has to be kept in mind which makes this an upper limit. The three approaches to estimate the magnetic field support are displayed on the left. It shows that the magnetic support is always smaller than the gravitational energy, but that the importance strongly depends on the assumptions for the magnetic field strength evolution.}
\label{supportRatios}
\end{figure*}

\begin{figure}
    \centering
    \includegraphics[width=\hsize]{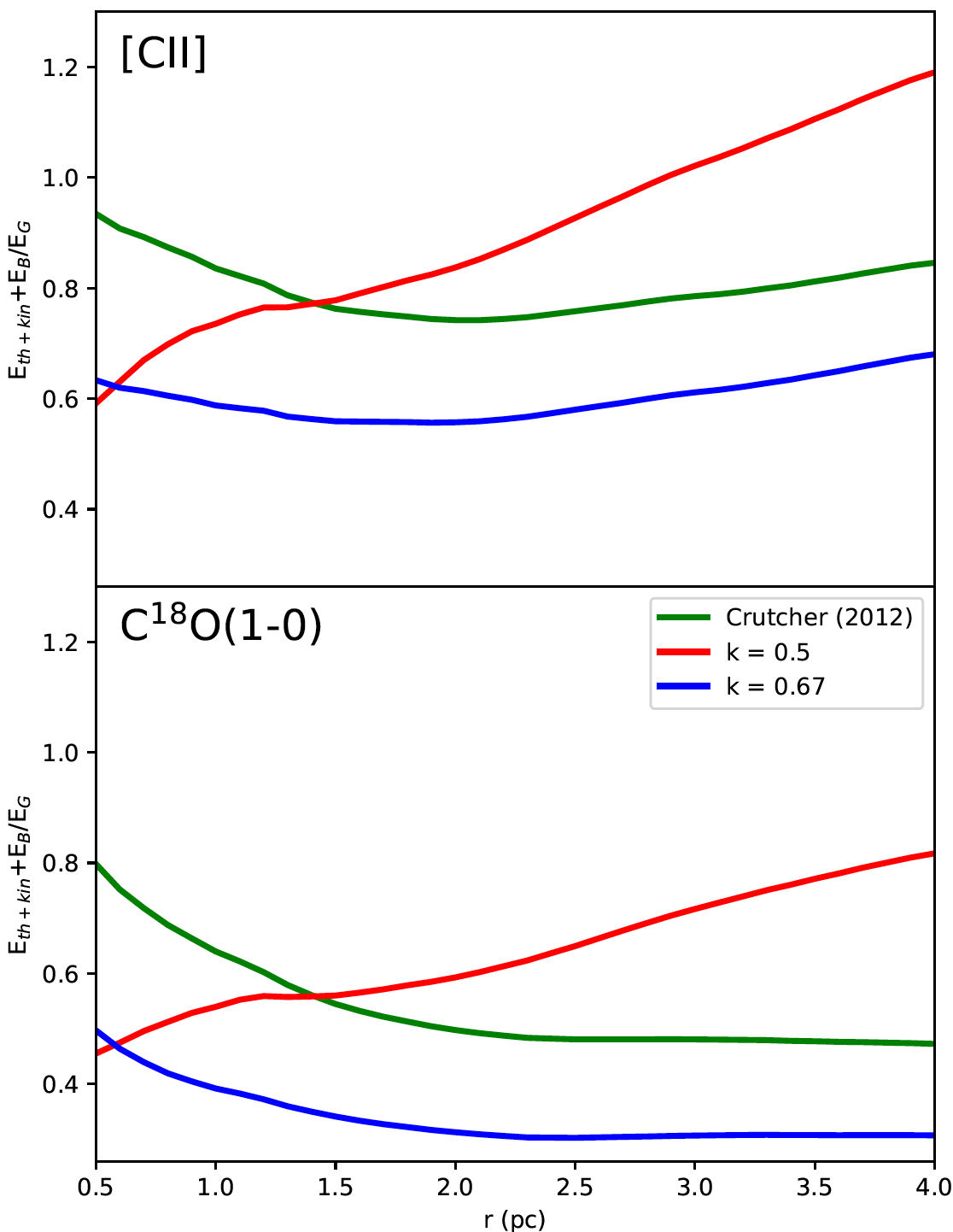}
    \caption{ The ratio of the energy terms that provide support, i.e. E$_{th+kin}$ and E$_{B}$, over the gravitational energy as a function of radius in the DR21 cloud. This is shown for \CII\ (top), which has the largest linewidth, and C$^{18}$O(1-0) (bottom), which has the smallest linewidth. Only when considering \CII\ and a k=0.5 evolution of the magnetic field do the support terms get larger than the gravitational energy at r $>$ 3 pc.}
    \label{fig:totEnergyProfs}
\end{figure}

In \citet{Schneider2010} it was found that the DR21 ridge is gravitationally collapsing. With the newly obtained data sets, we can further investigate the global stability of the molecular cloud as a function of its radius, centered on the column density peak in DR21(OH). This is done with the different tracers by estimating the gravitational potential energy, the turbulent energy and the magnetic energy. In this analysis, the density profile of the cloud might affect the results that use the simple equations below which correspond to a sphere with uniform density. The same is valid for the more sheet-like morphology that we propose for the surrounding DR21 cloud to maintain a plausible C$^{+}$ abundance. However, the impact of these differences from a uniform sphere are predicted to be relatively small \citep[$\lesssim$ 50\% for typical clump density profiles and a not too flattened ellipsoid cloud based on][]{Bertoldi1992}. Therefore, considering the uncertainties, we think the analysis below is reasonable.\\
%E.g. using the equations from \citet{Bertoldi1992} and a density profile $\rho \propto$ r$^{-2}$ would increase the gravitational energy with 60 \% and more shallow density profiles would only reduce the difference with a uniform density distribution\\
The thermal and turbulent energy is calculated using 
\begin{equation}
{\rm E_{T}} = \frac{3}{2} {\rm M} \, \sigma^{2}
\end{equation}
with M the mass in the region determined from the \textit{Herschel} dust column density map and $\sigma$ the velocity dispersion of the studied region (which we approximate here with the observed linewidth as there is no direct measure of the kinetic temperature). The gravitational energy is determined by 
\begin{equation}
{\rm E_{G}} = -\frac{3}{5}\frac{{\rm GM^{2}}}{{\rm R}}
\end{equation}
where G is the gravitational constant and R the radius. The magnetic energy is calculated using
\begin{equation}
{\rm E_{mag}} = \frac{1}{2}{\rm M \,V_{A}^2}
\end{equation}
with V$_{\rm A}$ the Alfv\'en speed, which is given by V$_{\rm A}$ = $\frac{{\rm B}}{\sqrt{\mu_{0}\rho}}$ with $\mu_{0}$ the vacuum permeability,  $\rho$ the density, and B the magnetic field strength.\\
The evolution of mass as a function of radius from DR21(OH) is shown in Fig. \ref{fig:densMassSigmaBfieldVirial}. The increase in mass flattens as a function of radius at r $>$ 1.5 pc which is expected with mass concentration in the ridge. For the error on the mass, we assume an uncertainty of 15\% for the derived column density as this corresponds to typical differences when creating the column density map with different methods \citep[e.g.][]{Peretto2016}. To estimate the thermal and turbulent internal support of the cloud, we assume that the linewidths of the observed lines, i.e. $^{13}$CO(1-0), C$^{18}$O(1-0), H$^{13}$CO$^{+}$(1-0) and \CII, are a good proxy. This turns out to be a good approximation as their linewidth is dominated by non-thermal motion. Note, that this approach ignores that a significant part of the linewidth might be associated with organized convergent flows driven by gravitational collapse or inertial motion instead of internal support \citep{Traficante2018c,Traficante2018b,Traficante2020} and that we have demonstrated the presence of multiple velocity components associated with inflow. The linewidth as a function of radius for the three considered molecules is presented in Fig. \ref{fig:densMassSigmaBfieldVirial}. It shows an increasing linewidth towards small radii and a fairly constant linewidth at r $>$ 1.5 pc. The increasing linewidth towards the ridge can fit with gravitationally driven inflow. This would also fit with predictions by \citet{Traficante2020} since most of the mass in the DR21 cloud has column densities $\gg$ 0.1 g cm$^{-2}$ (i.e. $N_{\rm H_2}\gg$ 2.6$\times$10$^{22}$ cm$^{-2}$). The estimated kinetic support based on the linewidth is thus an upper limit on the turbulent support, and might significantly overestimate it. We do have to note that the current IRAM 30m do not entirely cover the studied radii, but the \CII\ data showed that these regions are highly CO-poor.\\\\
Determining the magnetic field in the cloud as a function of its size is the most uncertain quantity because there are only a few dust polarization magnetic field observations in the region. Based on these observations, it was proposed in \citet{Ching2017} that the magnetic field strength in the DR21 ridge is 0.94 mG. %and they also find that the magnetic field strength in the massive dense cores (MDCs) of DR21 follows B $\propto$ n$_{H}^{0.54 \pm 0.30}$. 
Therefore, we used two approaches to estimate the magnetic field strength evolution in the cloud. First, we make use of the magnetic field strength relation from \citet{Crutcher2010} which is given by
\begin{equation}
{\rm B} = \left\{
    \begin{array}{ll}
        0.01\: {\rm mG\;\;\;\;\;\;\;\;\;\;\;\;\;\;\;\; (n}_{\rm H} < 300 {\rm \:cm}^{-3})\\
        0.01\:(\frac{{n}_{\rm H}}{300})^{0.65}\: {\rm mG\;\;\;\; (n}_{\rm H} > 300 {\rm \:cm}^{-3})
    \end{array}
\right.
\end{equation}
Secondly, we use B $\propto$ n$_{\text{H}}^{\text{k}}$ and start from the constraint that the magnetic field strength is 0.94 mG at the density of the ridge. To extrapolate the relation, we use two values: k = 0.5 and k = 0.67. These exponents represent two asymptotic cases: k = 0.67 is generally considered to describe the case where the magnetic field is dominated by gravitational collapse, and k = 0.5 the case where the magnetic field plays an important role in support against gravitational collapse \citep[e.g.][]{Basu1997,Hennebelle2011}. The resulting magnetic field strengths as a function of radius for the different estimates are shown in Fig. \ref{fig:densMassSigmaBfieldVirial}. Over the full cloud there is an uncertainty up to a factor 4 for the magnetic field strength based on the different relations used. This will thus have a significant impact on the magnetic energy as will be shown in the next paragraph.\\
The calculated energy terms are shown in Fig. \ref{supportRatios} \& Fig. \ref{fig:totEnergyProfs}, compared to the gravitational energy. The thermal and kinetic energy, estimated from the velocity dispersion, is very similar for the different molecules and has values $<$ 20\% of the gravitational energy. In addition, it was pointed out that these values likely are an upper limit. The relation for \CII, tracing the lower density gas, is different as it reaches up to 40-60\% of the gravitational energy. This seems to suggest that in the lower density gas, which is particularly found at r $>$ 2-3 pc, there might be significant thermal or turbulent support. However, it has to be taken into account that we found multiple velocity components which do not necessarily contribute to support against collapse in the region. The magnetic field energy depends on its assumed field strength evolution. Assuming k = 0.5 indicates that the cloud can experience some support from the magnetic field on large scales ($\sim$ 3-4 pc), but gradually this magnetic support decreases leading to an increasing importance of gravitational collapse when approaching the ridge (r $<$ 3 pc). Assuming k = 0.67, the magnetic field provides little support at low densities while the higher density regions towards the ridge (r $<$ 1 pc) experience an increased importance of the magnetic field. Yet, in both cases the values still indicate that the gravitational energy dominates over the cloud. This becomes particularly evident in Fig. \ref{fig:totEnergyProfs} where the total support terms over the gravitational terms are plotted. This shows that only considering k = 0.5 and the velocity dispersion of the \CII\ emission allow for support against gravitational collapse, but this ignores that the \CII\ velocity dispersion might not be associated with mass inflow rather than turbulent support.

\section{Discussion} \label{sec:discuss} 

Combining the results and analysis from Sects.~\ref{sec:dr21} and \ref{sec:analysis}, we will now discuss the cloud evolution leading to high-mass star formation in the DR21 ridge. 

\subsection{Magnetic field bending followed by gravitational collapse in the DR21 ridge} \label{subsec:collapse} 
From the fitted velocity maps and PV diagrams we found indications of an accelerating velocity field, based on the rapid linewidth increase in the ridge and a blueshifted asymmetry in the self-absorbed molecular lines towards the ridge. Additionally, the observations showed that basically all sub-filaments and the surrounding cloud appear to be organized in a flattened structure that gives rise to a redshifted V-shape perpendicular to the ridge. \CII\ also displays a second blueshifted velocity component that is localized in the vicinity of the ridge and looks like a more localized blueshifted V-shape.\\
The accelerating velocity field can be the result of the proposed gravitationally driven inflow \citep[e.g.][]{Peretto2006,Hartmann2007}. However, to explain the observed redshifted V-like velocity field in the surrounding cloud, as well as the systematic redshift of the sub-filaments, we have to consider the apparent flattened geometry of the molecular cloud. If the flattened geometry is curved, this can provide a straightforward explanation for the observed velocity field if this curved sheet geometry is associated with the magnetic field bending in a collision as predicted by \citep{Hartmann2001,Inoue2013,Vaidya2013,Inoue2018,Abe2020}. The presence of a strong magnetic field in the compressed sheet might then even initially prevent gravitational acceleration and significantly affect the velocity field until the massive ridge is reached. We note that this is the same mechanism that was also proposed to explain the observations toward the Musca cloud that has a very similar kinematic V-shape perpendicular to the filament \citep{Bonne2020a}. There, it was proposed that Musca formed at the apex of a bent magnetic field as the result of an overdensity interacting with diffuse \HI\ gas in a colliding flow. %traces magnetic field bending around the filament due to a soft \HI\ cloud collision (v$_{col} \sim$ 7-10 km s$^{-1}$) that formed the Musca filament \citep{Bonne2020a}.\\ %We talk of a soft collision because the Musca filament did not form by a head-on collision of two fully molecular clouds, e.g. presented in \citep{Haworth2015}, but rather by the compression of a $\sim$ 5 pc size overdensity during interaction with the more diffuse regions in the 50 pc size colliding clouds/flows. Bending of the magnetic field around an overdensity in a collision, giving rise to the observed kinematic signature, was predicted based on simulations carried out by \citet[e.g.][]{Inoue2013,Vaidya2013,Abe2020}.\\
%This scenario would provide a straightforward explanation for the observed redshifted V-shape of the DR21 region, where the sub-filaments are substructures that formed in the inflowing mass reservoir. 
In fact, the DR21 region is proposed to be in a high-velocity (v$_{col} \gtrsim$  20 km s$^{-1}$) collision with the diffuse, mostly atomic, regions of the other velocity component between v$_{LSR}$ = 9 and 15 km s$^{-1}$ in the Cygnus-X region \citep{Schneider2023}. This points to the same formation mechanism for the DR21 cloud as for the Musca cloud. However, Cygnus-X region is more than an order of magnitude more massive than the Chamaeleon-Musca region \citep[e.g.][]{Reipurth2008} and the collision velocity is higher, which can explain why the DR21 ridge is forming massive stars \citep[e.g.][]{Dobbs2020,Liow2020}. %The proposed high-velocity collision can also fit with the observed high CH$^{+}$ abundance towards the DR21 cloud over the full velocity range associated with the collision (i.e. v$_{LSR}$ = -5 to 15 km s$^{-1}$) \citep{Godard2012}. This high CH$^{+}$ abundance was proposed to be the result of a high-velocity shocks in the region. 
The second \CII\ velocity component we found in the vicinity of the DR21 ridge at v$_{LSR}$ $\sim$ -4.5 km s$^{-1}$ is then likely gas associated with the $\gtrsim$20 km s$^{-1}$ collision, responsible for the formation of the DR21 ridge, that leads to accretion on the ridge from the blueshifted side.\\
One could consider the scenario that the DR21 cloud is a swept-up shell created by expansion from a local bubble. However, there is no evident massive cluster candidate near the DR21 cloud that could drive the expansion of such a massive region \citep{Maia2016} and the study of the multiple velocity components by \citet{Schneider2023} does not unveil a typical expanding shell morphology in the region.\\\\ %Furthermore, looking into the longitudinal velocity field along the axis of the ridge using e.g. C$^{18}$O(1-0), H$^{13}$CO$^{+}$(1-0) and \CII\ also seems to challenge this interpretation. If the ridge fragmented in a local expanding shell, a curved velocity field, similar to the one seen perpendicular to the ridge, would be expected along the axis of the DR21 ridge. However, the velocity fields in Figs. \ref{velMaps} and \ref{velFieldCII} rather show a blue- to redshifted velocity gradient from north to south along the ridge, not a curved velocity field.\\\\
In summary, we propose that the DR21 ridge was formed as the result of a high-velocity collision in the Cygnus-X region which bends the magnetic field around the ridge. This bending is at the origin of the inflow seen in the form of a kinematic V-shape in the flattened surrounding cloud. However, due to the high-density in the compressed cloud, gravity increasingly starts to dominate the curved kinematics driven by the magnetic field bending. This explains the observed accelerated motion in the ridge and the results from the virial analysis. The proposed scenario thus provides a comprehensive explanation for the rich set of observations that trace the density range of the DR21 cloud. Observations that trace the line-of-sight magnetic field morphology, and thus the proposed bending, would be invaluable to further test this scenario.\\
Lastly, we note that gravity did not take over the cloud kinematics in Musca \citep{Bonne2020b}. This is an important difference because the increasing importance of gravity on larger scales in the DR21 cloud allows more effective mass provision to the center(s) of collapse, e.g. the DR21(OH) and DR21-N clumps \citep{Schneider2010}.

\subsection{Mass accretion on the DR21 ridge} \label{subsec:accretion} 
The mass inflow towards the ridge guided by the sub-filaments can be estimated using $\dot{\rm M}_{acc}$ = N $\pi$R$^{2}$v$_{inf}$n \citep{Schneider2010,Peretto2013}, with N the number of inflowing filaments, R the radius, n the density and v$_{inf}$ the velocity of the infalling filaments. Working with 7 molecular inflowing subfilaments with n = 10$^{5}$ cm$^{-3}$, R = 0.1 pc \citep{Hennemann2012} and an inflow velocity of 1-2 km s$^{-1}$, we get an estimated mass accretion rate of 1.3-2.6$\times$10$^{-3}$ M$_{\odot}$~yr$^{-1}$ similar to the estimate of \citet{Schneider2010}. In \cite{Schneider2010}, it was also shown that the F3 sub-filament continues its flow towards the DR21(OH) clump in the ridge and may provide inflow to multiple MDCs. However, this mass inflow rate is roughly an order of magnitude lower than the required mass accretion rate ($\sim$ 10$^{-2}$ M$_{\odot}$ yr$^{-1}$) to replenish the DR21 ridge within 1 Myr.\\
Taking into account that the mass inflow probably also happen off the sub-filaments, we can calculate this contribution of mass inflow to the ridge. We consider inflow over the full surface of the ridge for which we assume a cylindrical geometry with R = 0.36 pc and L = 4.15 pc \citep{Hennemann2012}. For the inflow, we work with a velocity of $\sim$ 1-2 km s$^{-1}$ and n = 10$^{4}$ cm$^{-3}$, which is justified by $N_{\rm {H_{2}}} \sim$ 3$\times$10$^{22}$ cm$^{-2}$ within the 1 pc vicinity of the ridge. This density is at the high range of densities in the ambient gas, which can be expected if the gas is gravitationally concentrated when approaching the ridge. These values give an estimated mass accretion rate of 0.56-1.1$\times$10$^{-2}$ M$_{\odot}$ yr$^{-1}$, which can replenish the DR21 ridge in 1-2 Myr. Note that this might be an overestimate up to a factor $\sim$2 as the inflow is predominantly sideways in the sheet and not uniform. 1-2 Myr is an expected lifetime for the DR21 ridge when taking the expression from \cite{Clarke2015} for the timescale of a free-fall longitudinal collapse for a filament
\begin{equation}
\tau_{{\rm ff}} = (0.49 + 0.26 \,{\rm A})({\rm G} \,\rho)^{-\frac{1}{2}}
\label{longitudinalCollapseTimescale}
\end{equation}
where A (= 5.8) is the initial aspect ratio of the filament half length over the filament radius and $\rho$ (= 3.9$\times$10$^{-19}$ g cm$^{-3}$) the average density of the ridge. This results in a $\tau_{ff}$ of 0.39 Myr. Considering support against longitudinal collapse from the magnetic field, which is perpendicular to the DR21 ridge at the center \citep{Vallee2006,Matthews2009,Ching2022}, 1-2 Myr likely is an appropriate time scale to replenish the DR21 ridge. We thus propose that the DR21 ridge can be continuously replenished by mass inflow from the surrounding cloud.\\
%Note that in contrast to the sub-filaments it is not expected that all of this off-filament inflow will directly provide mass to the MDCs. This inflow might rather encounter the DR21 ridge as a discontinuity, such that it will only provide mass accretion on the ridge. However, the gravitational collapse of the ridge, seen e.g. with HCO$^{+}$, will then provide more mass to the MDCs to maintain very active high-mass star formation over the full ridge. %We thus propose a scenario where filamentary inflow directly provides mass to the MDCs in the ridge because of gravitational attraction by the mass of the ridge ($\sim$ 10$^{5}$ M$_{\odot}$). 
Only when feedback from the newly formed massive stars disperses the dense ridge and prevents further mass accretion it will end active high-mass star formation. This dispersal of the ridge by stellar feedback, which is proposed to happen on relatively short timescales after the first massive stars form \citep[][]{Hollyhead2015,Luisi2021,Chevance2022,Bonne2022b} (i.e. $\lesssim$ 3 Myr), might thus be important to maintain a low star formation effiency (SFE). Even though  stellar feedback might halt mass provision to the ridge, it is observed around the DR21 radiocontinuum source that regions of dense gas a bit further out, in particular the nearby sub-filaments (S \& SW), remain present. This might be able to maintain continued lower mass star formation while stellar feedback is already dispersing the ridge. For example, low-mass cores were detected in the sub-filament F3 \citep{Hennemann2012}. 

\subsection{The link with high-mass star formation}
The above scenario implies that high-mass star formation might only occur over an interval of the total star formation time in a cloud. In the DR21 cloud, the formation of high-mass stars would be directly related to the lifetime of the massive ridge. This potentially shorter phase of high-mass star formation might fit with the observed shallow slope of the core mass function (CMF) in high-mass star forming regions \citep{Motte2018b,Liu2018,Pouteau2022} compared to the power law tail of the initial mass function, as well as the slightly shallower initial mass functions (IMF) in young clusters ($<$ 4 Myr) of Cygnus-X \citep{Maia2016}. However, further work with e.g. the ALMA-IMF sample \citep{Motte2022}, will be needed to better understand the evolution of these top-heavy CMFs in high-mass star forming regions and how it connects with the IMF of young stellar clusters as well as the evolution of high-mass star forming clouds and filaments \citep{Pouteau2022}.
%Lastly, we note that the magnitude of the velocity field inside the DR21 cloud, $\sim$2-3 km s$^{-1}$ (see Fig. \ref{pvDiagrams}), is not that different from lower mass regions. This suggests that it is the orders of magnitude higher density as a result of the different c, which appears to favor an increasing importance for gravity on the cloud scale, between low- and high-mass star formation.

%\begin{figure}
%    \centering
%    \includegraphics[width=\hsize]{images/sketchDR21scenario.png}
%    \caption{Schematic representation of the proposed collision scenario responsible for the formation of the DR21 ridge. The red and light blue clouds represent the mostly atomic colliding clouds presented in Schneider et al. (in prep.). The curved molecular cloud traced with the \CII\ emission, subfilaments and DR21 ridge is indicated with orange, green and dark blue, respectively. Note the presence of the \CII\ emission behind the DR21 ridge responsible for the formation of the DR21 ridge. The dashed black lines indicate the curved magnetic field and the full curved arrows indicate the associated curved velocity field.}
%    \label{fig:sketchScenario}
%\end{figure}

%-----------------------------------------------------------------------------------------------------------------------

\section{Conclusion} \label{sec:conclusions} 
We have presented new molecular line data of $^{13}$CO(1-0), C$^{18}$O(1-0), HCO$^{+}$(1-0) and H$^{13}$CO$^{+}$(1-0) from the IRAM 30m telescope and a \CII\ map from the FEEDBACK SOFIA legacy survey of the DR21 cloud. %, and combined the data with available \textit{Herschel} and JCMT observations of the full Cygnus-X region. 
In this study, we exclusively focus on the emission between -8 km s$^{-1}$ and 1 km s$^{-1}$, which is associated with the DR21 cloud, for all lines.\\
The molecular lines trace the dense ridge ($N_{\rm H_{2}} > $10$^{23}$ cm$^{-2}$)  and sub-filaments while \CII\ traces the surrounding molecular cloud (i.e. $N_{\rm H_{2}} \lesssim $ 10$^{22}$ cm$^{-2}$) that embeds these filamentary structures. We demonstrate that this surrounding cloud seen with \CII\ is mostly CO poor. Both in the molecular and \CII\ spectra we find indications of more than one velocity component within the studied velocity range. We also note there is an offset for the peak velocity in the cloud between the \CII\ and molecular lines, confirming they trace different regions of the cloud. Inside the ridge, the molecular lines confirm the results from \citet{Schneider2010} while the observations of the sub-filaments show they are systematically redshifted with respect to the ridge. The \CII\ kinematics show a redshifted V-like shape, between -3 and -1 km s$^{-1}$, perpendicular to the ridge that intersects with a more blueshifted velocity component (at v$_{LSR}$ $\sim$ -5 km s$^{-1}$) at the location of the ridge. We confirm that these two intersecting velocity components converge, and that there is continuous mass accretion on the DR21 ridge.  The V-shape velocity field embeds all the redshifted sub-filaments, and density requirements to explain the observed \CII\ emission indicate that the V-shape has a flattened morphology. %We therefore propose that the sub-filaments are substructures forming in the inflowing mass reservoir due to dissipation of kinetic energy which explains the lower linewidth associated with dense filamentary structures. 
Tracing the CO poor gas, seen in \CII\ towards the DR21 cloud, is thus essential for a comprehensive view on cloud evolution and filament formation.\\
Performing a virial analysis, including estimates for the magnetic field support, we find that the cloud should be gravitationally collapsing on pc scales. The surrounding gas traced by \CII\ appears to have more support against gravitational collapse, but this does not take into account the organised velocity field and presence of two velocity components that are not associated with turbulent support.\\\\
We propose that there is continuous mass accretion onto the DR21 ridge. This accretion is initiated by the bending of the magnetic field around the ridge due to a large-scale collision. In this collision, an overdensity is compressed into a dense filament/ridge during the interaction with diffuse atomic regions of the colliding cloud. However, because of the high density in the DR21 cloud, gravity is taking over the cloud dynamics which facilitates rapid mass provision to a (few) center(s) of collapse where massive clusters are forming. We found that the continuous mass inflow, visible thanks to \CII, can replenish the ridge. Therefore we propose that the DR21 ridge will remain present until stellar feedback will disperse it and halt massive star formation\\
Lastly, we note that this scenario is the same as proposed for the low-mass Musca cloud \citep{Bonne2020b}, except for the increasing importance of gravitational collapse in the massive DR21 cloud. This hints that the same scenario might be at the origin of filament formation in both a low- and high-mass star forming region. Furthermore, recent magnetic field and velocity observations of some nearby low- and high-mass star forming regions \citep[e.g.][]{Tahani2022b,Arzoumanian2022} appear to have a straightforward explanation in this scenario. This leads us to tentatively propose that these types of collisions are a widespread mechanism in the Milky Way to initiate a wide variety of star formation activity in the Milky Way.


\begin{acknowledgments}
We thank W. Lim and F. Comeron for insightful comments on an early draft of the paper. Based in part on observations made with the NASA/DLR Stratospheric
Observatory for Infrared Astronomy (SOFIA). SOFIA is jointly operated
by the Universities Space Research Association, Inc. (USRA), under
NASA contract NAS2-97001, and the Deutsches SOFIA Institut (DSI) under
DLR contract 50 OK 0901 to the University of Stuttgart. This work was supported by the Agence National de Recherche (ANR/France) and the Deutsche Forschungsgemeinschaft (DFG/Germany)
through the project "GENESIS" (ANR-16-CE92-0035-01/DFG1591/2-1) and 
the DFG project number SFB 956. The FEEDBACK project is supported by the Federal Ministry of Economics and Energy (BMWI) via DLR, Projekt Number 50 OR 2217 (FEEDBACK-plus). Financial support for the SOFIA Legacy Program, FEEDBACK, at the University of Maryland was provided by NASA through award SOFO070077 issued by USRA.
\end{acknowledgments}

\vspace{5mm}
\facilities{SOFIA (upGREAT), IRAM 30m, Herschel, JCMT}

\software{astropy \citep{Astropy2013} 
          %Cloudy \citep{2013RMxAA..49..137F}, 
          %SExtractor \citep{1996A&AS..117..393B}
          }

\appendix

\section{\CII\ and JCMT $^{12}$CO(3-2) observations}\label{sec:CIIvs12CO}
\begin{figure}
    \centering
    \includegraphics[width=0.6\hsize]{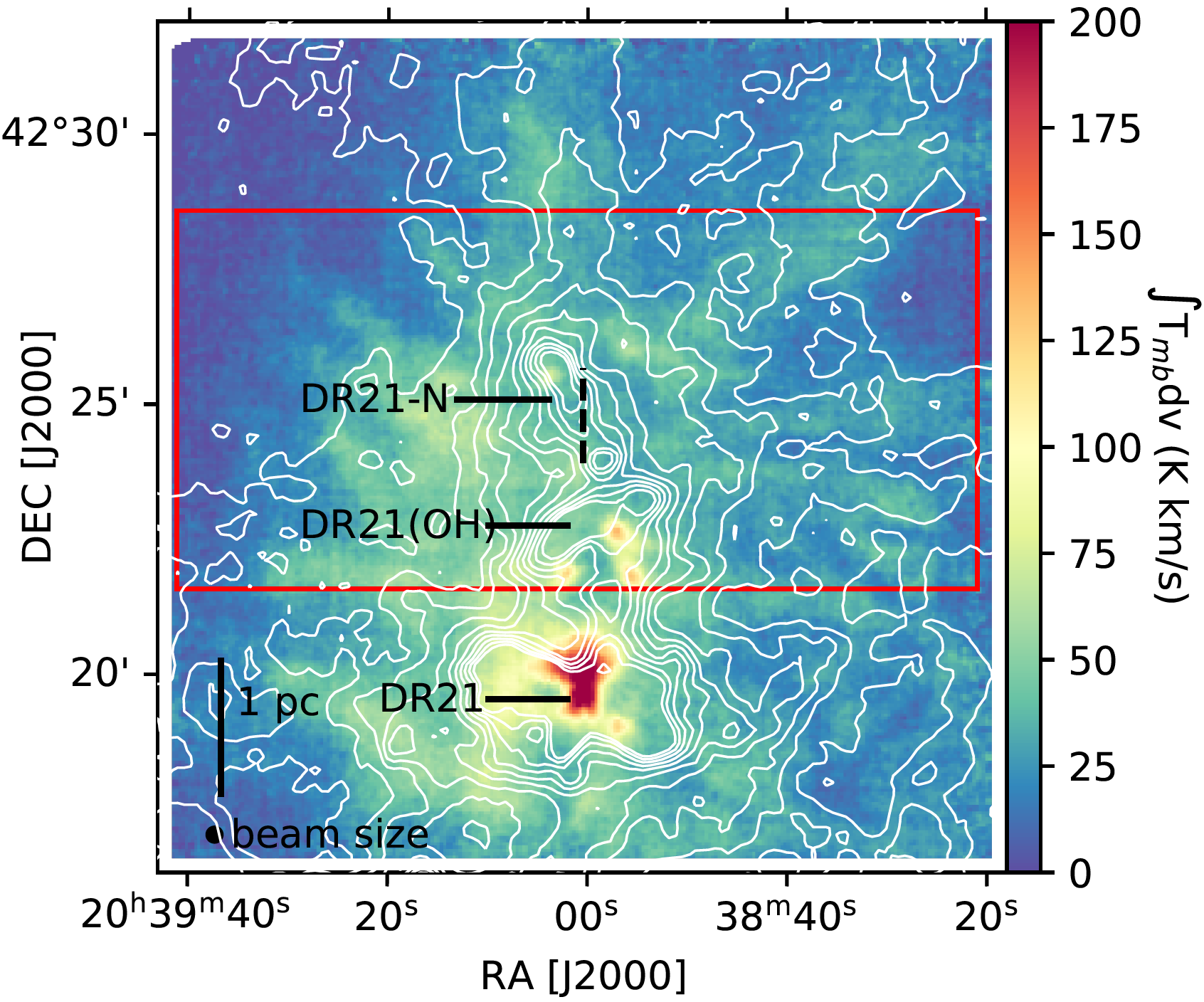}
    \caption{The \CII\ integrated intensity map from -8 to 1 km s$^{-1}$ as in Fig. \ref{integratedCIICI}.  Here, the contours indicate the JCMT $^{12}$CO(3-2) emission starting at 20 K km s$^{-1}$ with incremental steps of 20 K km s$^{-1}$.}
    \label{fig:CIIvsCO}
\end{figure}

In Fig. \ref{fig:CIIvsCO} we present a comparison of the \CII\ integrated intensity and JCMT $^{12}$CO(3-2) integrated intensity. This JCMT data over the full Cygnus-X region will be the topic of a forthcoming paper. Because of the larger column density than for $^{13}$CO, the $^{12}$CO emission is detected further into the cloud compared to $^{13}$CO (see Fig. \ref{intIntensityMaps}). However, $^{12}$CO still is particularly sensitive to the massive ridge and surrounding subfilaments and has a clearly distinct morphology compared to \CII. This confirms that \CII\ provides a unique view on the lower-column density gas that is not accessible with molecular lines.

\section{The $^{13}$CO abundance}\label{sec:13coAbundance}
\begin{figure}
    \centering
    \includegraphics[width=0.6\hsize]{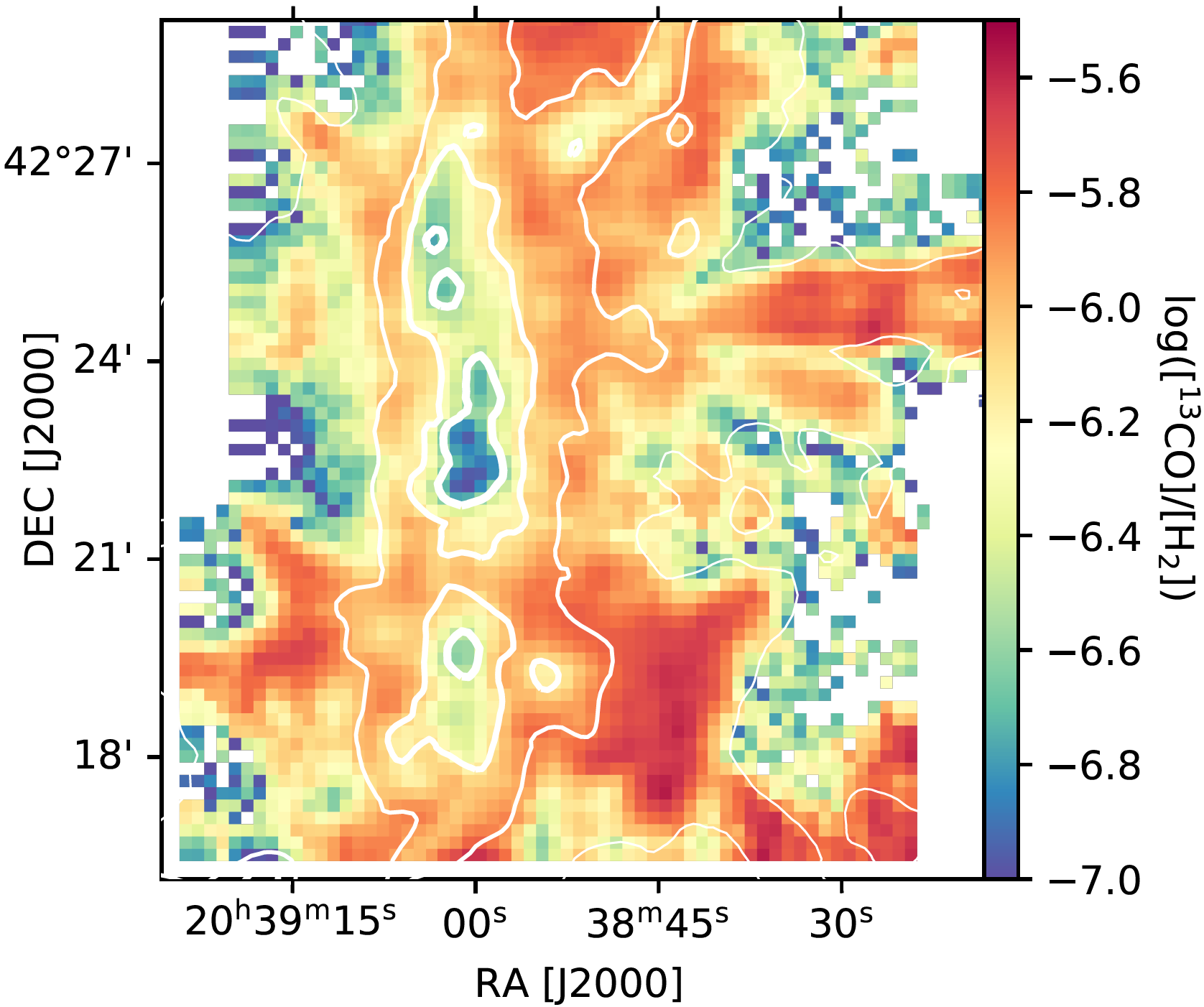}
    \caption{Map of the $^{13}$CO abundance on a log-scale over the DR21 cloud. This is deduced from the $^{13}$CO(1-0) intensity, a kinetic temperature of 18 K and the \textit{Herschel} column density map. The white contours indicate the \textit{Herschel} column density map at N = 10$^{22}$ cm$^{-2}$, 3$\times$10$^{22}$ cm$^{-2}$, 10$^{23}$ cm$^{-2}$ and 4$\times$10$^{23}$ cm$^{-2}$.}
    \label{fig:13coAbundanceMap}
\end{figure}

\begin{figure}
    \centering
    \includegraphics[width=0.6\hsize]{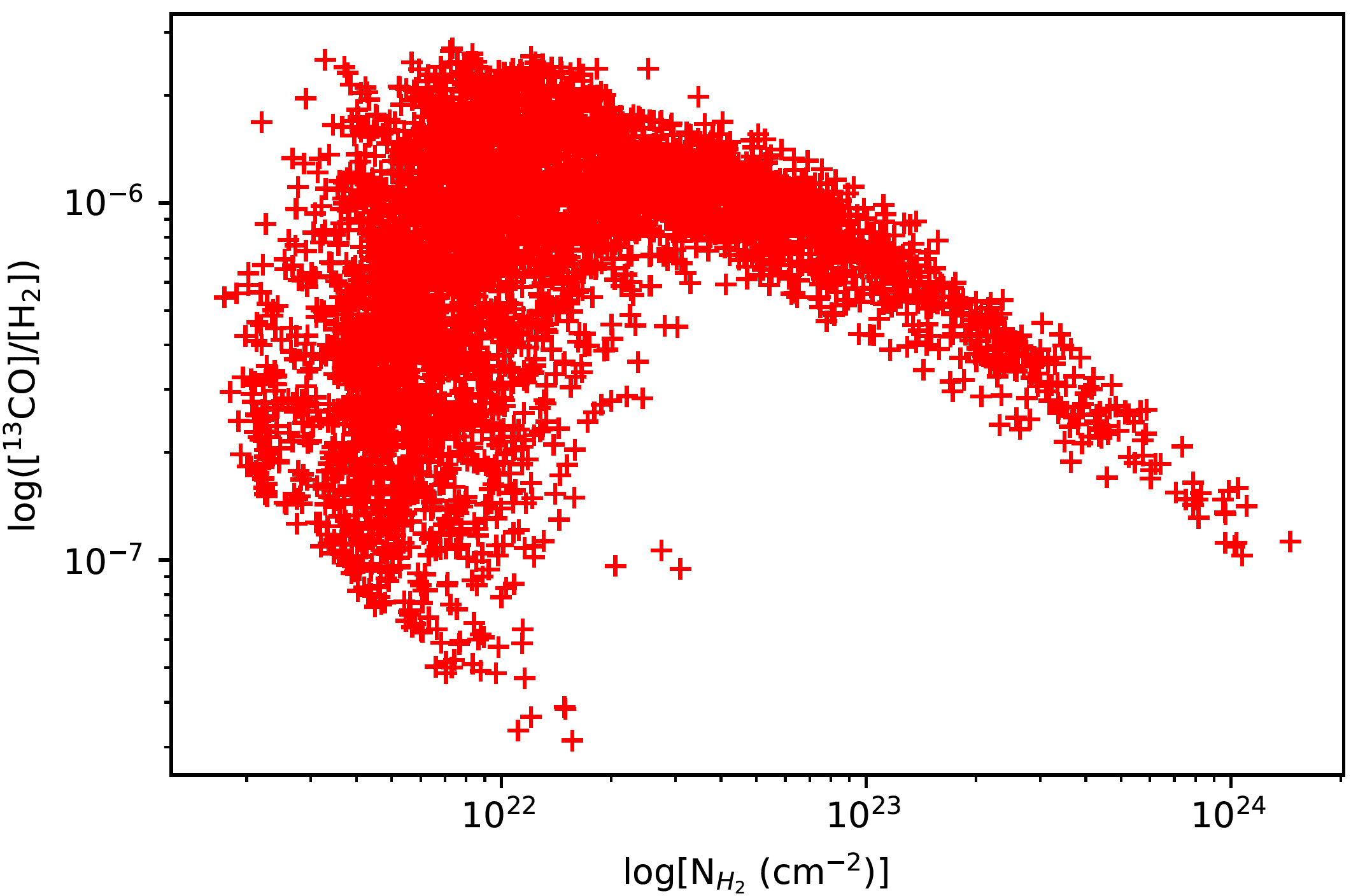}
    \includegraphics[width=0.6\hsize]{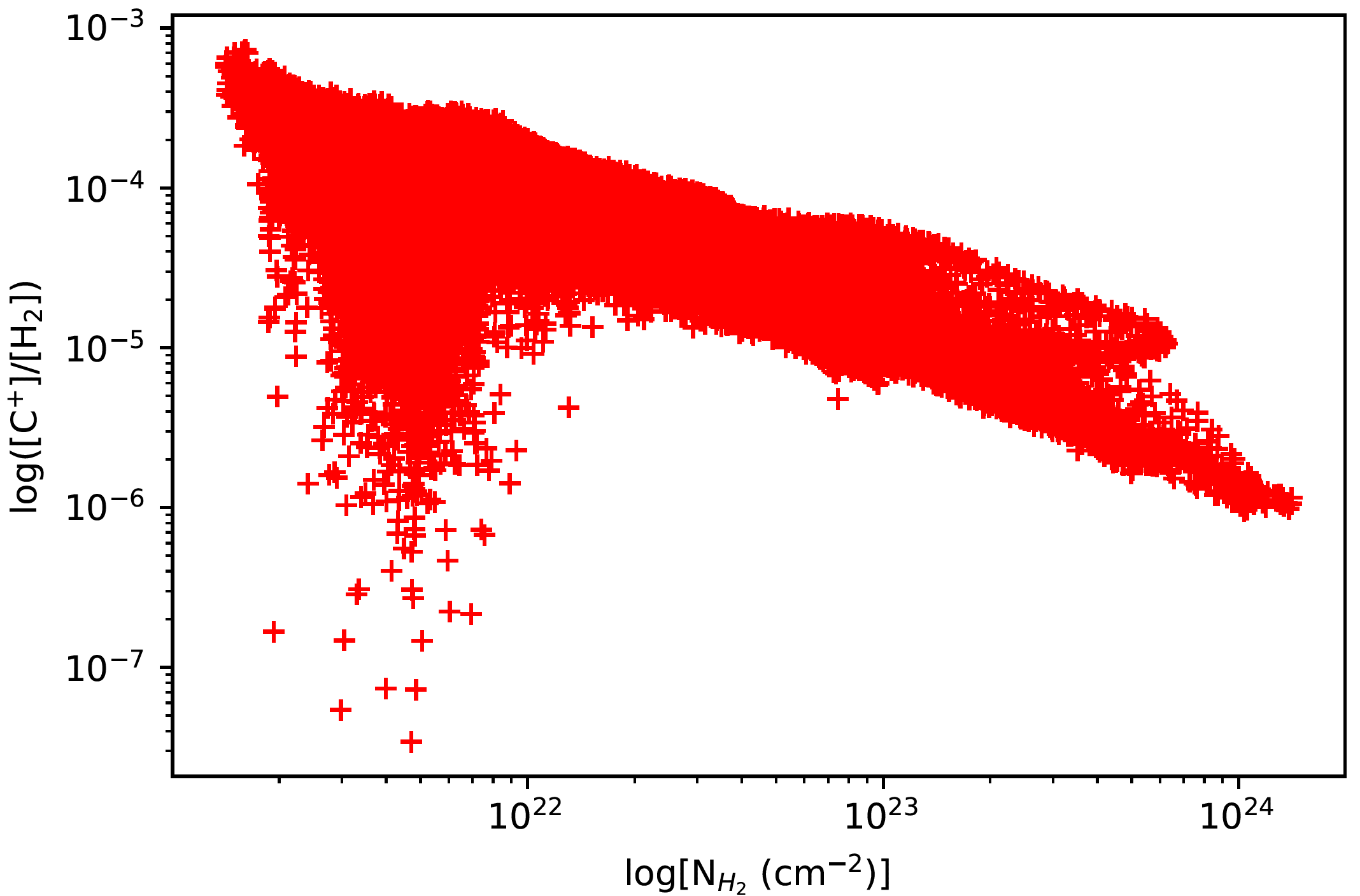}
    \caption{{\bf Top}: $^{13}$CO abundance as a function of the \textit{Herschel} column density for all pixels in the map. {\bf Bottom}: The same for the C$^{+}$ abundance.}
    \label{fig:abundanceCorrelations}
\end{figure}

The calculation of the $^{13}$CO column density map allows to estimate $^{13}$CO abundance based on the background subtracted \textit{Herschel} column density map \citep{Schneider2016a}. This then allows to explore which regions of the cloud are traced by the observed $^{13}$CO(1-0) transition. The calculation of the $^{13}$CO column density follows the calculations presented in \citep{Mangum2015}. Below we recapitulate the equations and values used. Note that these equations assume that the molecule is optically thin. This assumption is not necessarily valid for $^{13}$CO(1-0) over the full map, in particular at the ridge. However, due to the high density in the ridge, which will result in freeze-out of the molecule, the opacity should not be excessive even in these regions of very high column density. When studying the outer regions of the ridge and the ambient gas, C$^{18}$O is barely detected which implies that the $^{13}$CO opacity will not be significantly above 1 for standard isotopologue ratios. Nonetheless, it has to be kept in mind this assumption is a limitation of the calculations below.\\
The total column density is then calculated from
\begin{equation}
    {\rm N}[{\rm cm}^{-2}] = {\rm f(T_{ex})}\int {\rm T_{mb}dv}
\end{equation}
Here T$_{mb}$ is the main beam brightness in Kelvin, v the velocity in km s$^{-1}$ and f(T$_{ex}$) is given by
\begin{equation}
    {\rm f(T_{ex})} = \frac{3{\rm hZ}}{8\pi^{3}\mu^{2}{\rm J_{t}}}\frac{{\rm exp(h}\nu /{\rm kT_{ex})}}{[1-{\rm exp(-h}\nu /{\rm kT_{ex})}]({\rm J(T_{ex})-J(T_{BG})})}
\end{equation}
In this equation, Z is the partition function
\begin{equation}
    {\rm Z} = \frac{2{\rm kT_{ex}}}{{\rm h}\nu} + 1/3
\end{equation}
with k the Boltzmann constant, h the Planck constant, $\nu$ the transition frequency (for $^{13}$CO(1-0) this is 110.201 GHz) and T$_{ex}$ the excitation temperature for which we assume a value of 18 K, see also \citet{Schneider2010}.\\
Further in the equation of f(T$_{ex}$), $\mu$ is the dipole moment (for $^{13}$CO(1-0) this is 0.112 Debye), J$_{t}$ is the upper value of the rotational quantum number (i.e. 1) and J(T$_{ex}$) is given by
\begin{equation}
    {\rm J(T_{ex})} = \frac{{\rm h}\nu}{{\rm k[exp(h}\nu /{\rm kT_{ex}})-1]}
\end{equation}
For the background (BG), a temperature of 2.73 K is assumed based on the cosmic microwave background (CMB).\\\\
The estimated $^{13}$CO abundance map, based on the Herschel column density map, is presented in Fig. \ref{fig:13coAbundanceMap}. Using the assumptions above, $^{13}$CO is most abundant at the edges of the ridge and in the sub-filaments (F3N, F3S and SW in particular). Inside the ridge, there is a significant decrease with increasing column density, see also Fig. \ref{fig:abundanceCorrelations}, which is expected because of CO depletion at densities $n_{\rm H_{2}} >$ 10$^{4}$ cm$^{-3}$. To trace the DR21 ridge kinematics, one thus needs to rely on tracers like HCO$^{+}$, N$_{2}$H$^{+}$ and deuterated molecules to probe the very dense gas of the ridge. In the surrounding cloud, at $N_{\rm H_{2}} <$ 1.5$\times$10$^{22}$ cm$^{-2}$ where $^{13}$CO is detected, there is also an observed lower $^{13}$CO abundance. Indeed, Fig. \ref{fig:abundanceCorrelations} shows that the $^{13}$CO abundance starts decreasing again towards lower column densities. This is the result of an increasing CO dark gas fraction in the DR21 cloud below these column densities. From Fig. \ref{fig:abundanceCorrelations}, it is also observed that at $N_{\rm H_{2}} <$ 1.5$\times$10$^{22}$ cm$^{-2}$ the \CII\ abundance continues to increase up to the typical carbon abundance in the ISM. This indicates that \CII\ allows to trace the CO dark gas in the surrounding DR21 cloud.

\section{The \CII\ optical depth}\label{sec:opticalDepth}
\begin{figure*}
    \centering
    \includegraphics[width=0.325\hsize]{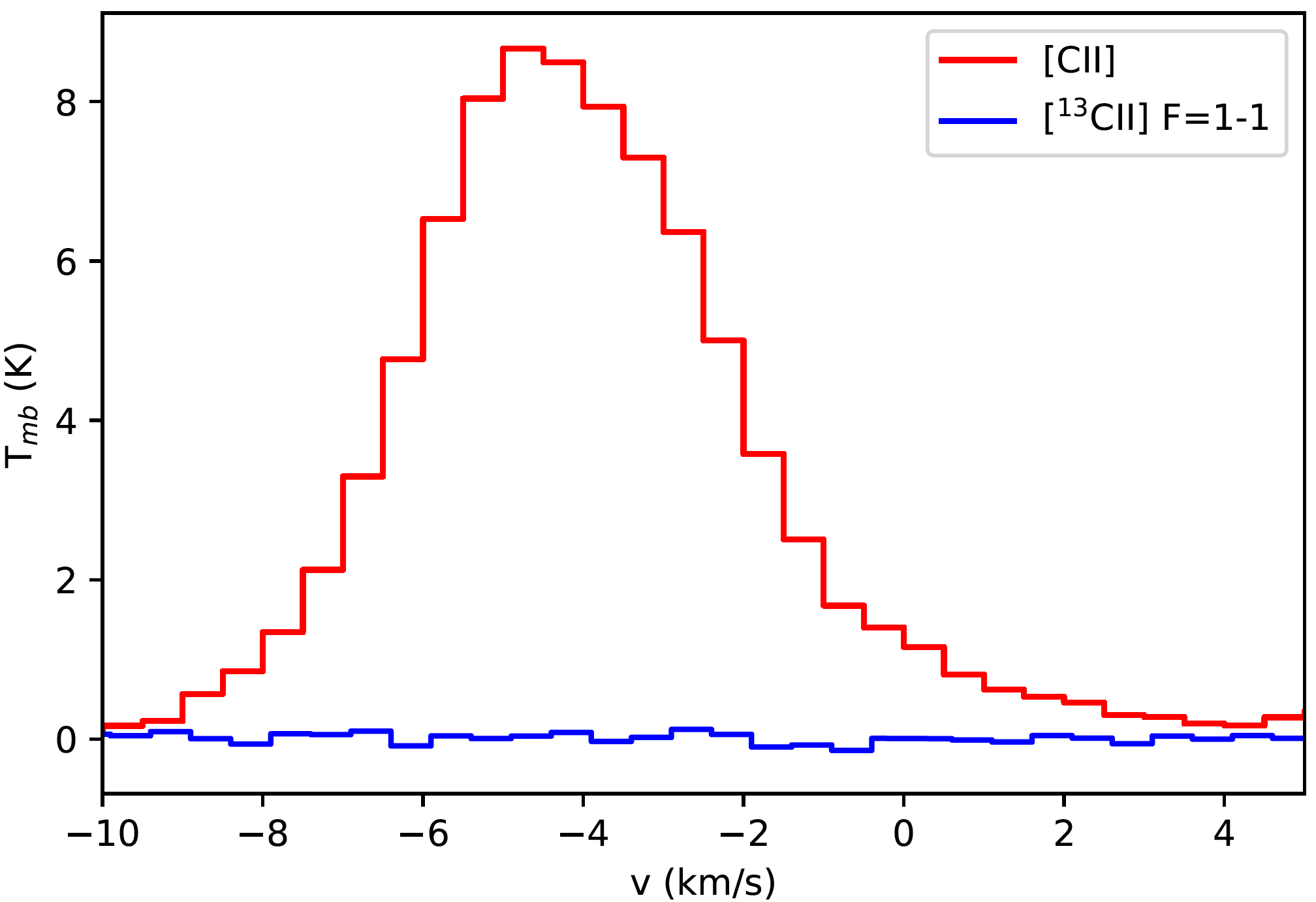}
    \includegraphics[width=0.325\hsize]{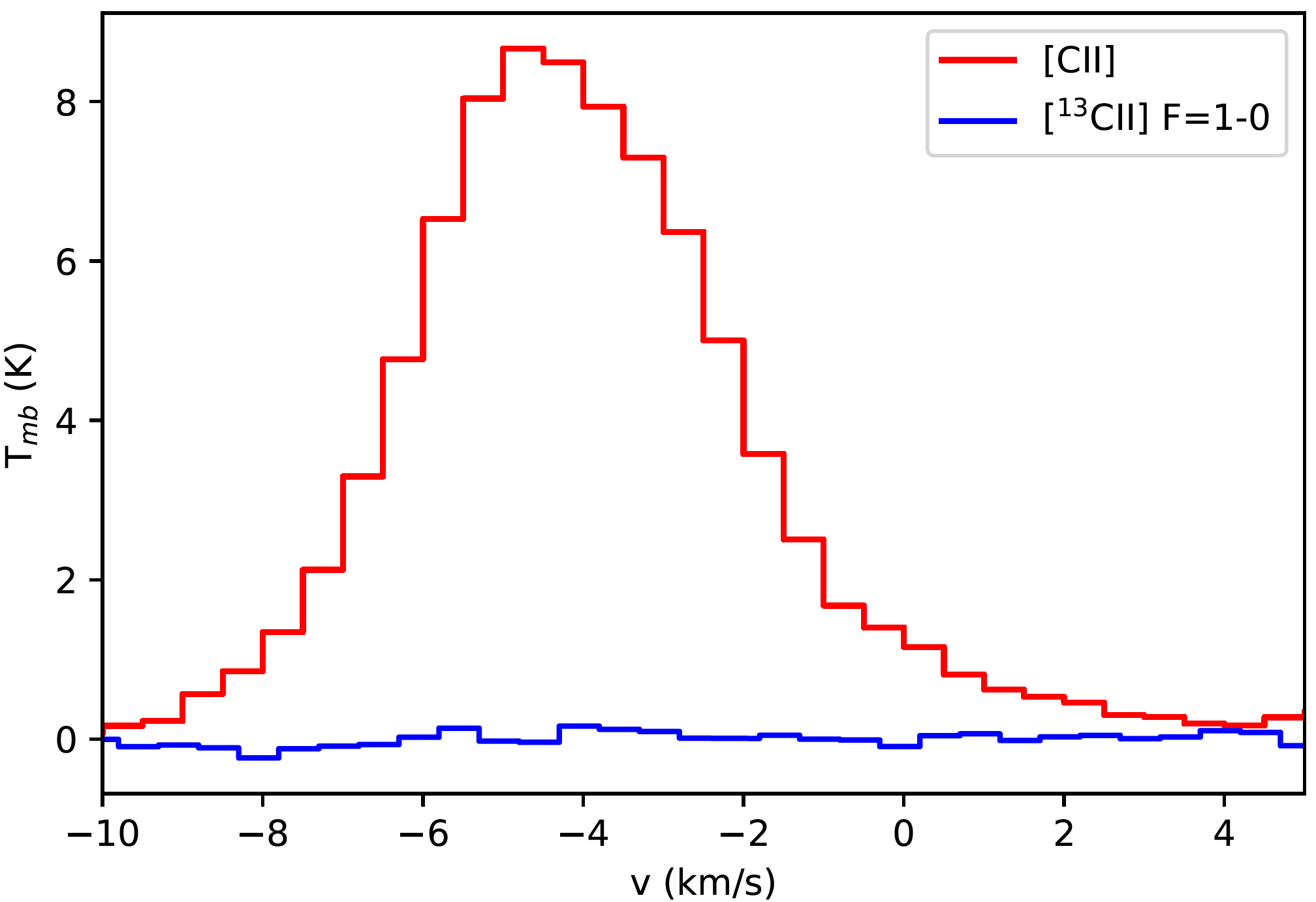}
    \includegraphics[width=0.325\hsize]{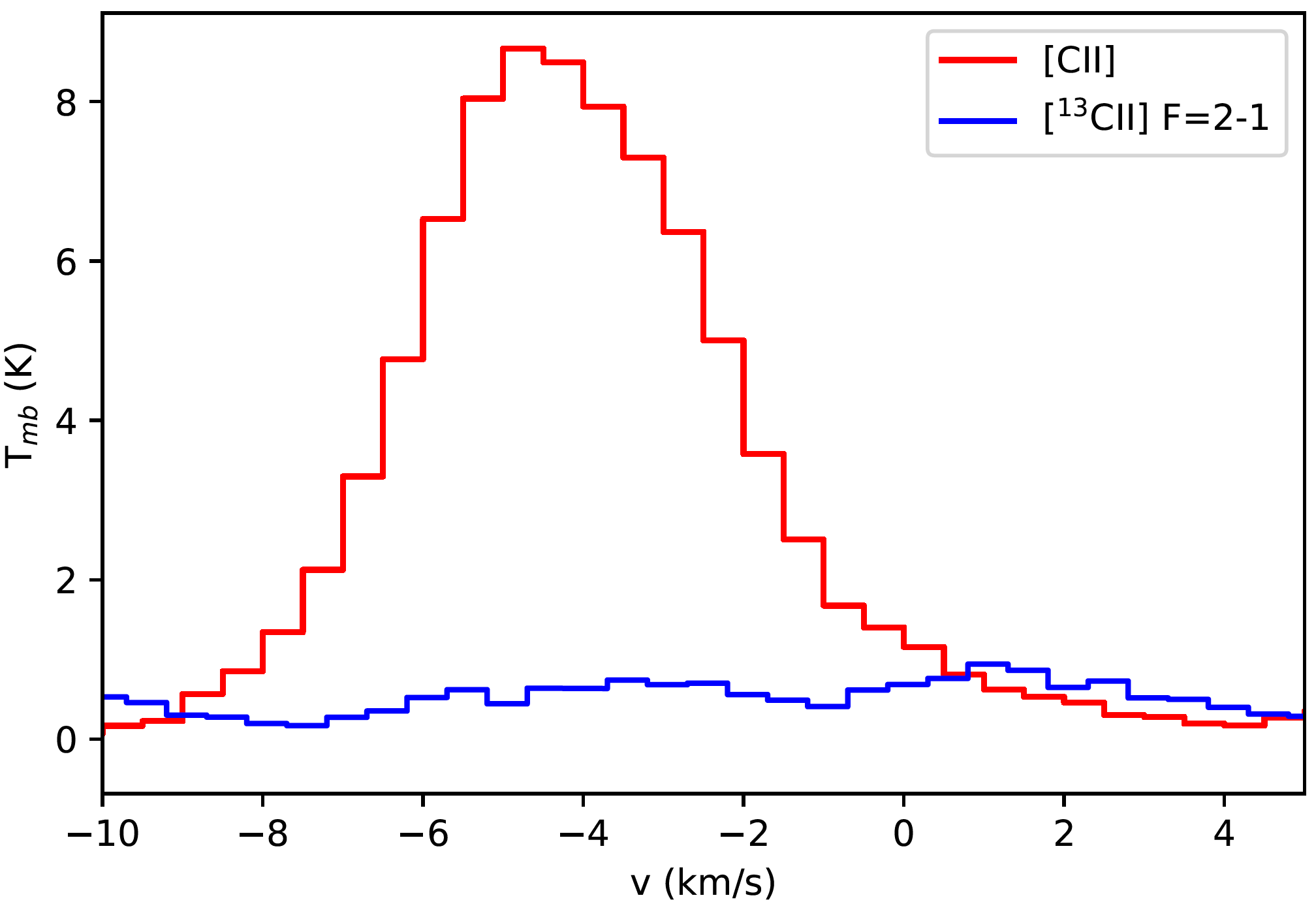}
    \caption{{\bf Left}: \dCII\ F = 1-0 transition shifted to the same velocity as the \CII\ line. There is no detected counterpart for this \dCII\ transition at the velocities of the DR21 ridge component. {\bf Middle}: The same for the \dCII\ F = 1-1 transition. {\bf Right}: The same for the \dCII\ F = 2-1 transition. For this transition, there is confusion from the higher velocity components \citep{Schneider2023}. Therefore, we do not use this transition here.}
    \label{fig:13CIIspectra}
\end{figure*}

In order to constrain the \CII\ optical depth towards the DR21 cloud, and in particular the ridge, we work with the average spectrum over the ridge in order to reduce the noise rms. This decrease in noise rms is important in order to try and detect the \dCII\ hyperfine structure lines that  are located at -65.2 km s$^{-1}$ (F = 1-0), 11.7 km s$^{-1}$ (F = 2-1) and 62.4 km s$^{-1}$ (F = 1-1) with respect to the \CII\ line. The hyperfine structure lines have a relative strength of 0.25 (F = 1-0), 0.625 (F = 2-1) and 0.125 (F = 1-1). Ideally one thus works with the F = 2-1 transition, but this line is very close to the actual \CII\ emission and contaminated by the observed bridging with a higher velocity component \citep{Schneider2023}. At the location of the F = 2-1 transition no indication of additional \dCII\ emission is found, see Fig. \ref{fig:13CIIspectra} but we will not work with this transition because of the confusion from the actual \CII\ emission. Therefore we focus on the second brightest transition (F = 1-0) at -65.2 km s$^{-1}$. Fig. \ref{fig:13CIIspectra} shows that this transition is not detected at a noise rms of 7.6$\times$10$^{-1}$ K within 0.5 km s$^{-1}$ for the averaged spectrum. Although not detected, we still try to obtain an idea of the optical depth associated with this noise level. This can be done using the equation
\begin{equation}
    \frac{\rm T_{mb,12}}{\rm T_{mb,13}} = \frac{1 - {\rm e}^{-\tau}}{\tau}\alpha
\end{equation}
With $\tau$ the optical depth and and $\alpha$ the local $^{12}$C/$^{13}$C abundace for which we take a value of 60 \citep{Wilson1994} and a correction factor of 0.25 for the relative strength of the F = 1-0 transition. Then using the noise rms value (7.6$\times$10$^{-1}$ K) results in an optical depth of $\tau =$ 1.7. Since the F = 1-0 transition is not detected, this indicates $\tau =$ 1.7 rather is an upper limit which indicates that the \CII\ line is optically thin or marginally optically thick at best towards the DR21 ridge.\\
To further investigate the potential optical depth of the \CII\ line, we also run RADEX simulations \citep{vanderTak2007} using input values obtained in this paper for the \CII\ emitting region towards the DR21 ridge. These calculations are presented in Tab. \ref{tab:radexOpacity} which shows that the predicted optical depth typically is below unity. Note that the highest obtained optical depths, which are still only of the order of unity and occur at T$_{kin} \approx$ 30 K, have a too low corresponding line brightness such that they likely are not representative for the observed emission. This is in agreement with the results from Sec. \ref{sec:CIIexcitation} which indicates that the kinetic temperature in the \CII\ emitting gas is likely of the order of 100 K (for which we obtain lower optical depths).

\begin{table*}[]
    \centering
    \begin{tabular}{cccccc}
    \hline
    \hline
         N(C$^{+}$) (10$^{17}$ cm$^{-2}$) & T$_{kin}$ (K) & $n_{\rm H_{2}}$ (cm$^{-3}$) & FWHM (km s$^{-1}$) & $\tau$ & T$_{R}$ (K)\\
         \hline
         5.0 & 30 & 10$^{4}$ & 4.0 & 7.7$\times$10$^{-1}$ & 1.6\\
         5.0 & 50 & 10$^{4}$ & 4.0 & 6.1$\times$10$^{-1}$ & 5.2\\
         5.0 & 75 & 10$^{4}$ & 4.0 & 4.7$\times$10$^{-1}$ & 9.0\\
         5.0 & 100 & 10$^{4}$ & 4.0 & 3.9$\times$10$^{-1}$ & 1.2$\times$10$^{1}$\\
         8.0 & 30 & 10$^{4}$ & 4.0 & 1.2 & 2.2\\
         8.0 & 50 & 10$^{4}$ & 4.0 & 9.7$\times$10$^{-1}$ & 7.4\\
         5.0 & 30 & 10$^{3}$ & 4.0 & 8.2$\times$10$^{-1}$ & 4.3$\times$10$^{-1}$\\
         \hline
    \end{tabular}
    \caption{Results produced with RADEX for conditions (N(C$^{+}$), $n_{\rm H_{2}}$, T$_{kin}$, FWHM) that could be representative of the \CII\ emitting gas. T$_{R}$ is the radiation temperature, which is equivalent to the main beam temperature under the assumption that the beam filling is unity (this should be a reasonable assumption seeing the extended \CII\ emission). The optical depth ($\tau$) predictions are basically always below unity which suggests relatively optically thin \CII\ emission. }
    \label{tab:radexOpacity}
\end{table*}

\section{\CII\ multi-component fit}\label{sec:TwoComps}
\begin{figure}
    \centering
    \includegraphics[width=0.49\hsize]{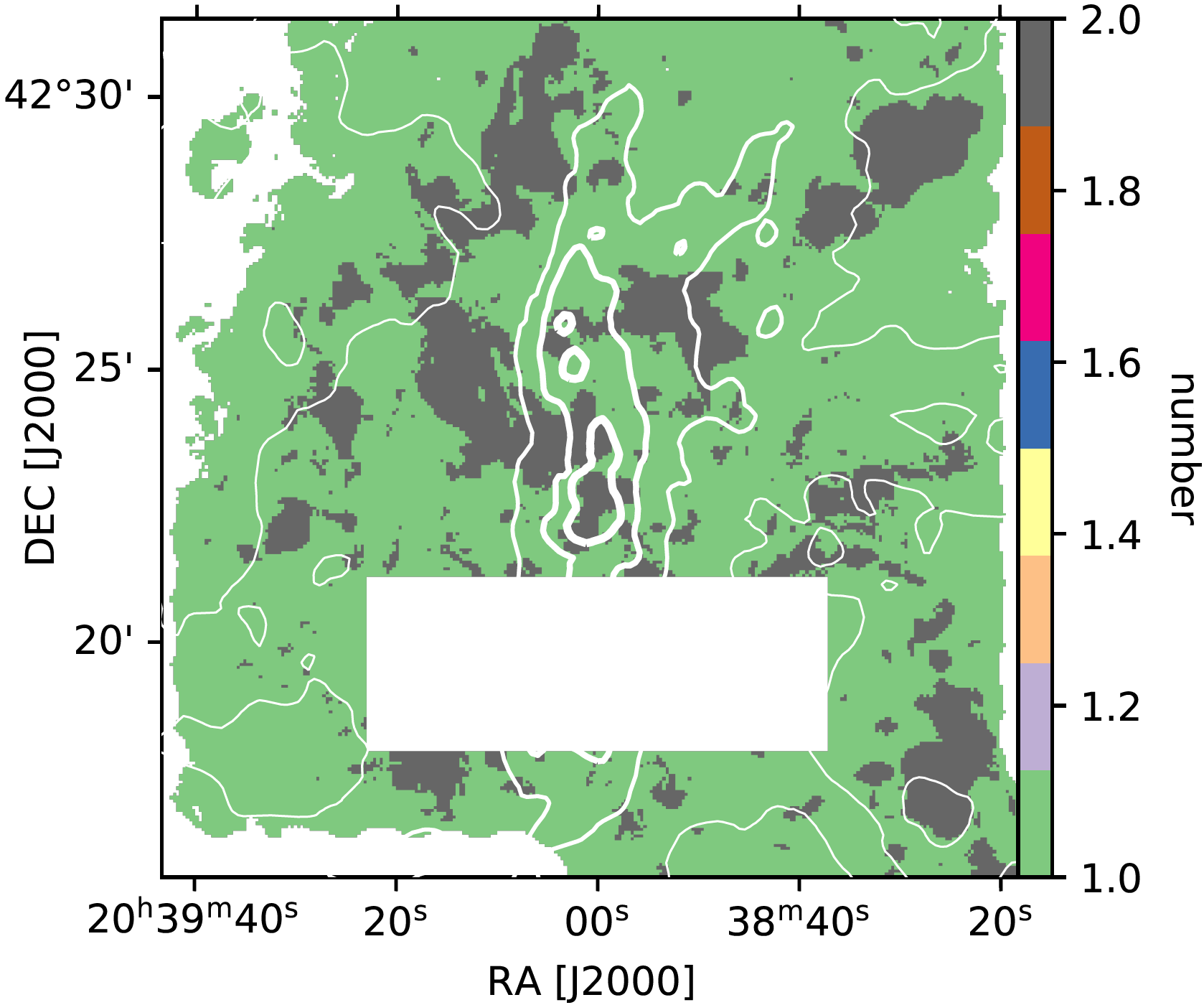}
    \caption{Number of fitted velocity components within the -8 to 1 km s$^{-1}$ velocity range. The white box indicates the region around the DR21 radio continuum source and its associated outflow which artificially increases the number of fitted velocity components. The white contours indicate the Herschel column density at $N_{\rm {H_{2}}}$ = 10$^{22}$, 3$\times$10$^{22}$, 10$^{23}$ and 4$\times$10$^{23}$ cm$^{-2}$.}
    \label{fig:locationsCIIfit}
\end{figure}

\begin{figure}
    \centering
    \includegraphics[width=0.495\hsize]{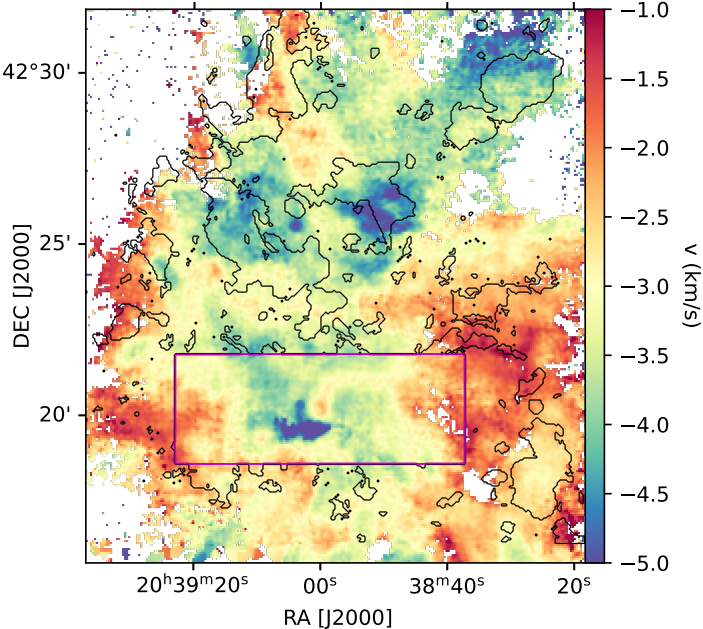}
    \includegraphics[width=0.47\hsize]{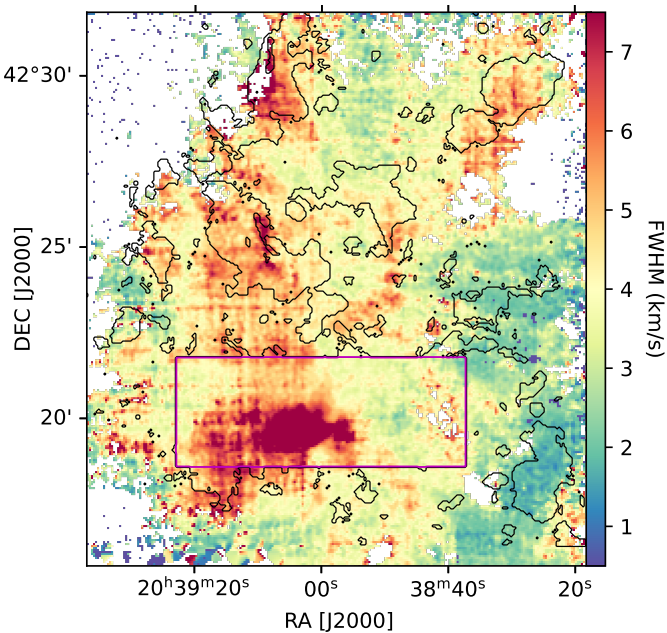}
    \caption{{\bf Left}: \CII\ velocity field from Fig. \ref{velFieldCII} with the black contours indicating the regions that are fitted with two velocity components between v$_{LSR}$ -8 and 1 km s$^{-1}$. {\bf Right}: The same for the \CII\ FWHM map.}
    \label{fig:velMapAndNum}
\end{figure}

\begin{figure}
    \centering
    \includegraphics[width=0.5\hsize]{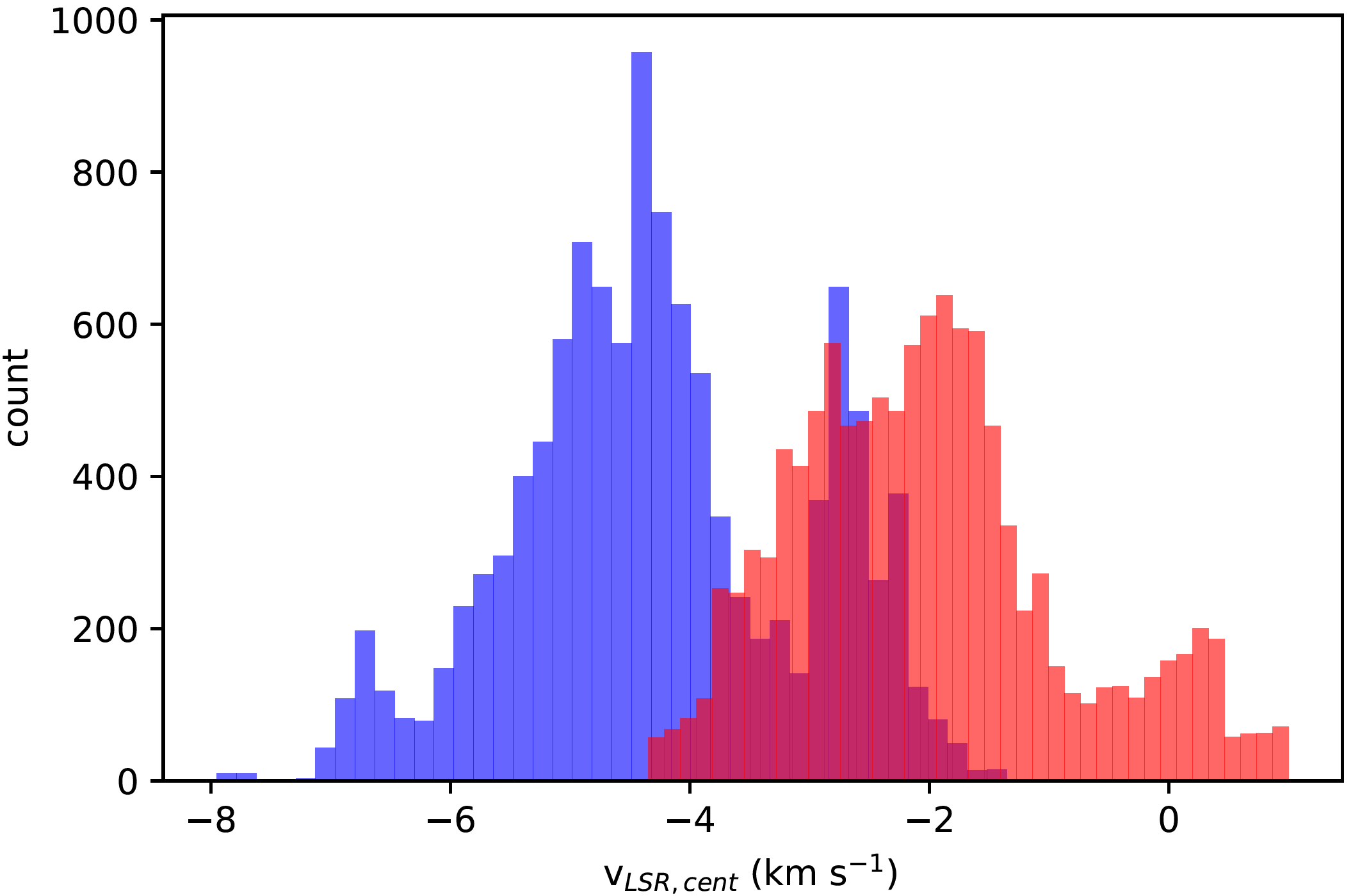}
    \caption{Peak velocity distribution of the blueshifted (blue) and redshifted (red) velocity components of the double Gaussian fit to the data.}
    \label{fig:twoVelHist}
\end{figure}

\begin{figure}
    \centering
    \includegraphics[width=\hsize]{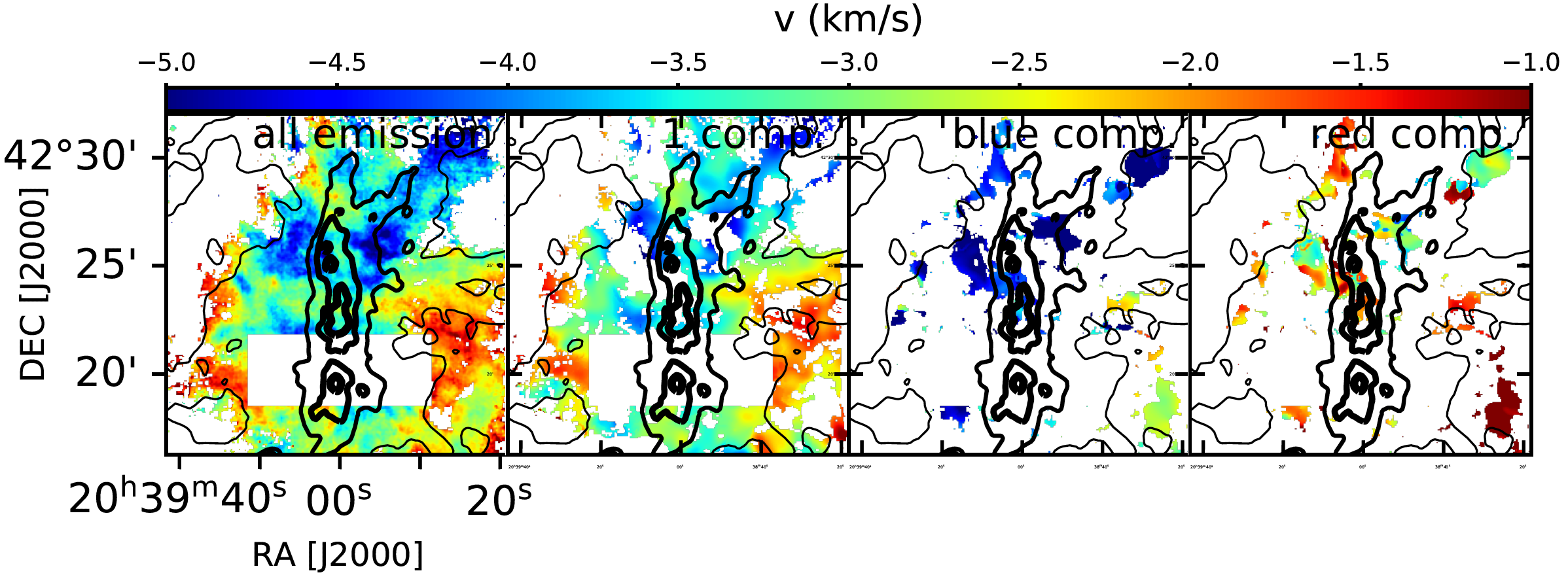}
    \includegraphics[width=\hsize]{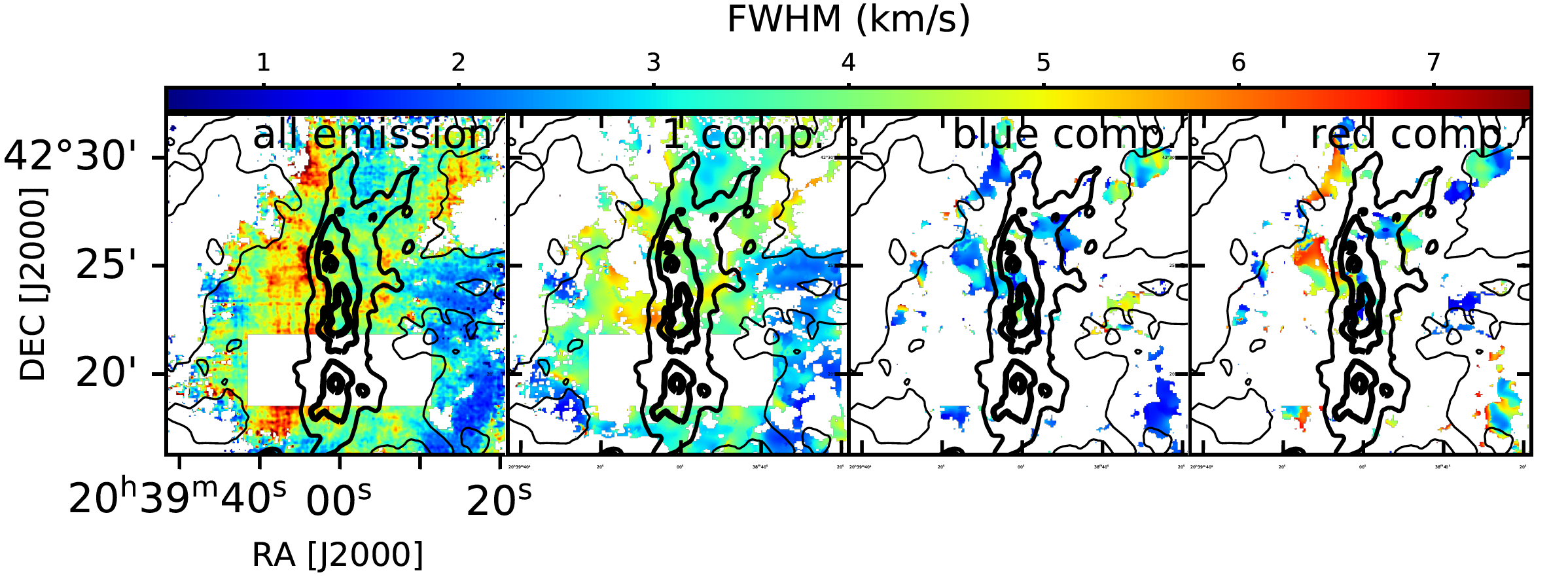}
    \caption{{\bf Top}: (left) The \CII\ velocity field of the DR21 cloud when taking into account all emission between -8 and 1 km s$^{-1}$. The black contours indicate the Herschel column density at $N_{\rm {H_{2}}}$ = 10$^{22}$, 3$\times$10$^{22}$, 10$^{23}$ and 4$\times$10$^{23}$ cm$^{-2}$. (middle left) The \CII\ velocity field in the pixels that have only one velocity component. (middle right) The \CII\ velocity field  of the blueshifted component in the pixels that were fitted with two Gaussian components. (right) The \CII\ velocity field of the redshifted component in the pixels that were fitted with two Gaussian components. Bottom: The same as above for the FWHM.}
    \label{fig:cii4figs}
\end{figure}

As the \CII\ emission likely is not optically thick, we examine here whether more than one Gaussian component is required to explain the observed \CII\ line spectral line profiles. To fit all spectra over the full map within the -8 to 1 km s$^{-1}$ velocity range, we employ the Beyond The Spectrum (BTS) fitting tool which is updated with the Akaike Information Criterion (AIC) \citep[][in prep.]{Clarke2018,Clarke2019}. However, we do exclude the DR21 radiocontinuum source and associated outflow from the analysis. As the \CII\ noise rms is still relatively high at the native resolution, which can mask multiple velocity components, we smoothed the data to a spatial resolution of 30$^{\prime\prime}$. This value was taken to balance significantly reducing the noise rms while avoiding the inclusion of large velocity gradients in a single beam which would give rise to additional velocity components. Note that the 30$^{\prime\prime}$ data is also used in \citet{Schneider2023}. In the fitting iteration procedure, when a fit has been done with n velocity peaks the fit will be repeated with n-1 and n+1 velocity peaks and the corrected AIC is calculated for each new fit. Then the fit with the smallest number of velocity peaks is chosen unless a higher number of velocity peaks is smaller by $\Delta_{AIC} >$ 10 (see e.g. \citet{Clarke2019} for the equations). This process is iterated until the number of peaks does not change anymore. The defined $\Delta_{AIC}$ parameter was found to be robust not to overfit noisy spectra, in particular for large spectral cubes (Clarke et al. in prep.).\\
The resulting number of velocity components over the map in the -8 to 1 km s$^{-1}$ velocity range are presented in Fig. \ref{fig:locationsCIIfit}. This demonstrates that the regions with two velocity components are in particular located at the edge of the DR21 ridge. To test this result, we also fitted the \CII\ data cube between -8 and 1 km s$^{-1}$ at the 20$^{\prime\prime}$ resolution. This demonstrated a similar spatial distribution of the two velocity components. Only, because of the higher noise rms, the procedure found less pixels with two velocity components. Comparing the number of velocity components with the \CII\ velocity field in Fig. \ref{fig:velMapAndNum}, we find that regions with two velocity components have a close spatial correlation with the regions that have a blueshifted velocity in \CII. This confirms that the location of the DR21 ridge is associated with the intersection of the proposed V-shape accretion flow with an additional velocity component. Lastly, Fig. \ref{fig:twoVelHist} shows the velocity distribution of the two fitted velocity components, while Fig. \ref{fig:cii4figs} shows the associated maps for the velocity and linewidth. The more blueshifted velocity component is typically found within the range of -6 km s$^{-1}$ and -4 km s$^{-1}$ while the redshifted component varies between -4 and -1 km s$^{-1}$. The typical velocity difference of the two converging components is between 2 and 3 km s$^{-1}$.

\section{C$^{18}$O(1-0) and H$^{13}$CO$^{+}$(1-0) multi-component fit}\label{sec:TwoCompsMol}
\begin{figure}
    \centering
    \includegraphics[width=0.32\hsize]{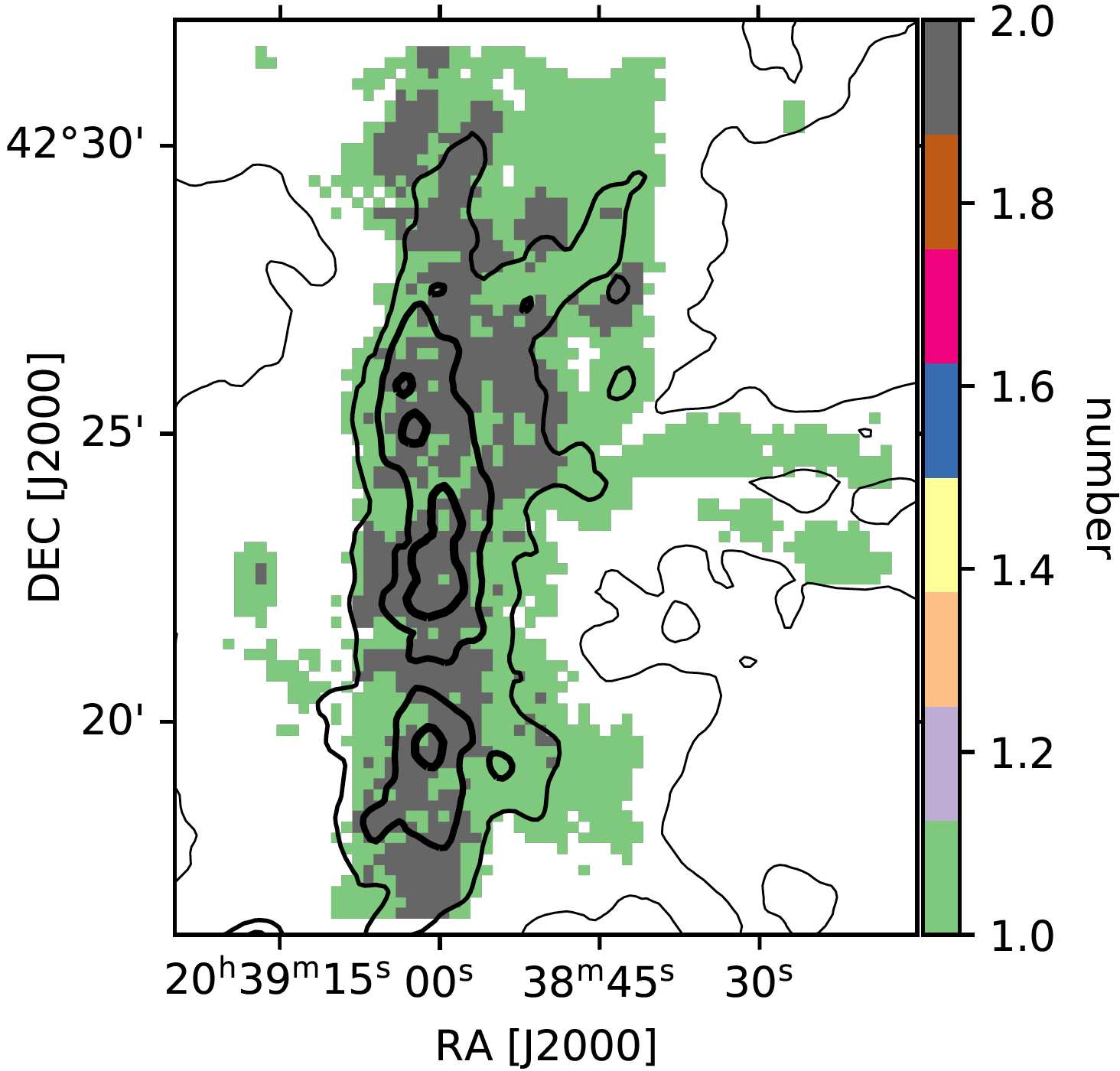}
    \includegraphics[width=0.31\hsize]{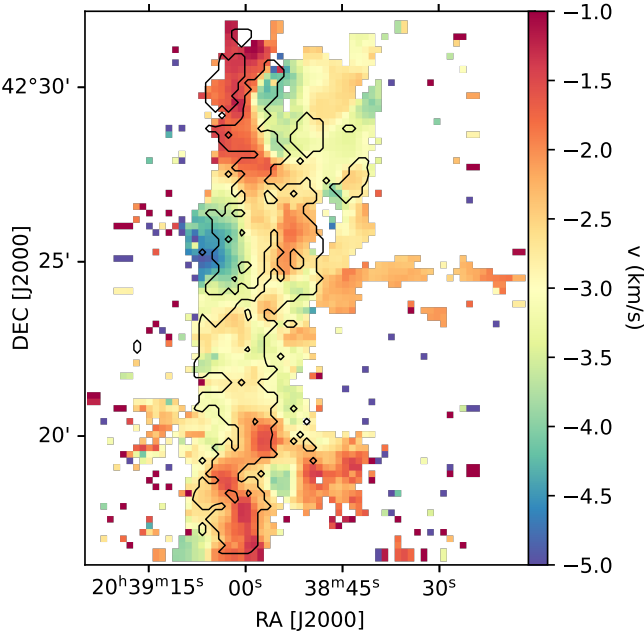}
    \includegraphics[width=0.31\hsize]{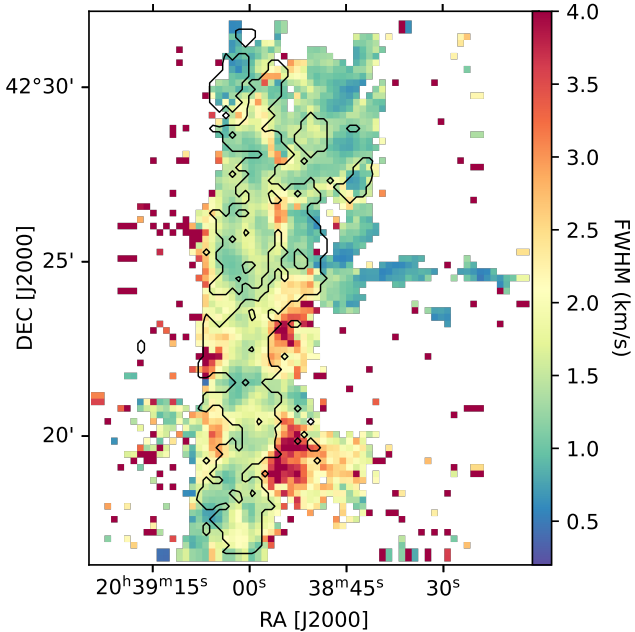}
    \caption{{\bf Left}: The same as Fig. \ref{fig:locationsCIIfit} for C$^{18}$O(1-0) with the column density contours in black. Middle \& right: Same as Fig. \ref{fig:velMapAndNum} for C$^{18}$O(1-0).}
    \label{fig:locationsC18Ofit}
\end{figure}

\begin{figure}
    \centering
    \includegraphics[width=0.5\hsize]{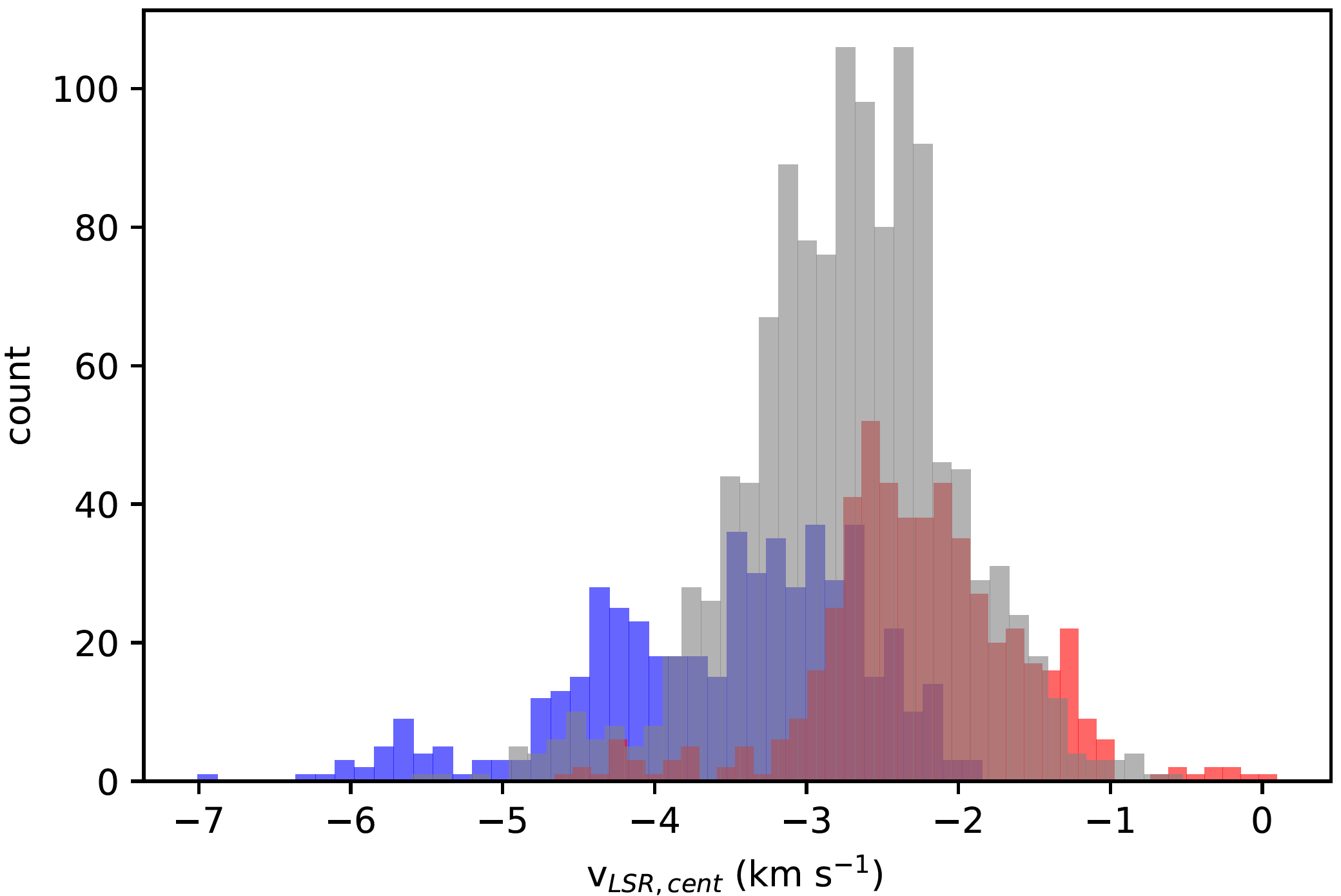}
    \caption{The same as Fig. \ref{fig:twoVelHist} for C$^{18}$O(1-0). The grey histogram indicates the peak velocity distribution for the pixels with a single fitted velocity component.}
    \label{fig:velHistC18O}
\end{figure}

\begin{figure}
    \centering
    \includegraphics[width=\hsize]{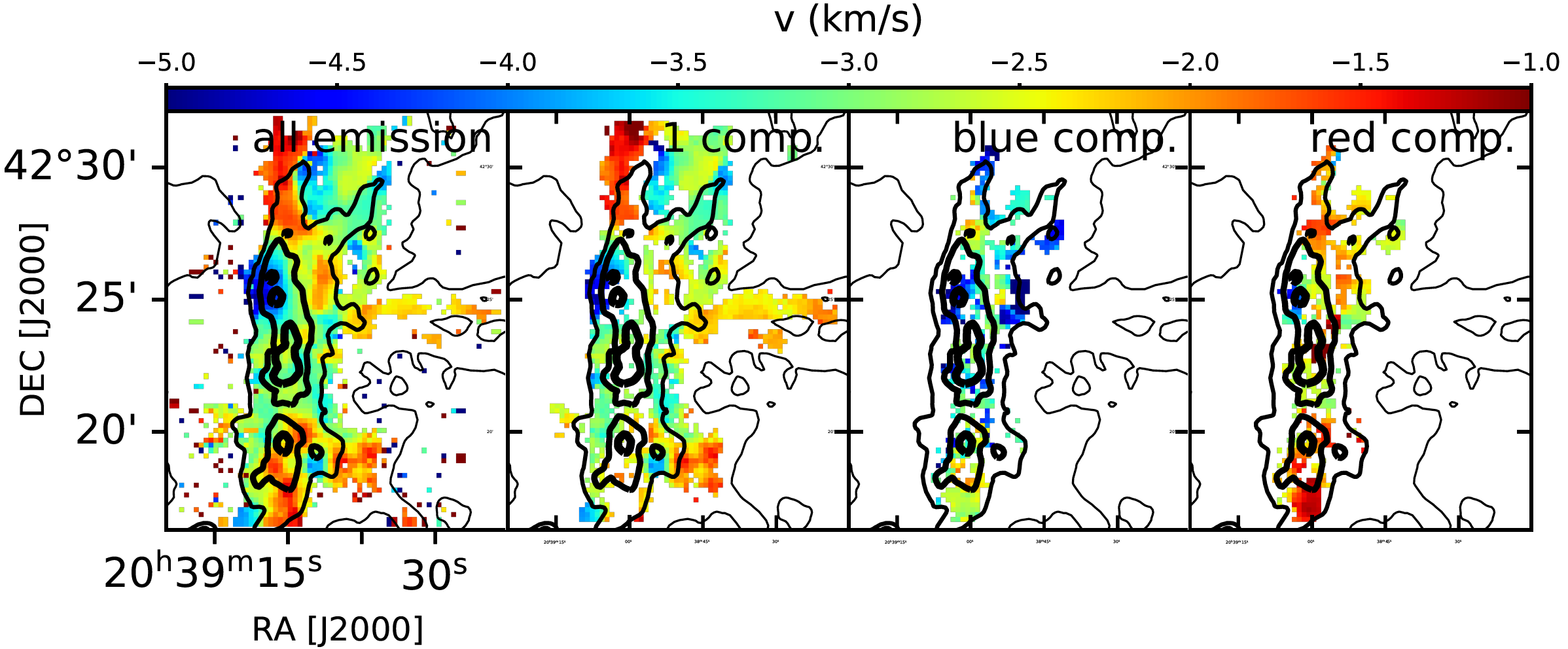}
    \includegraphics[width=\hsize]{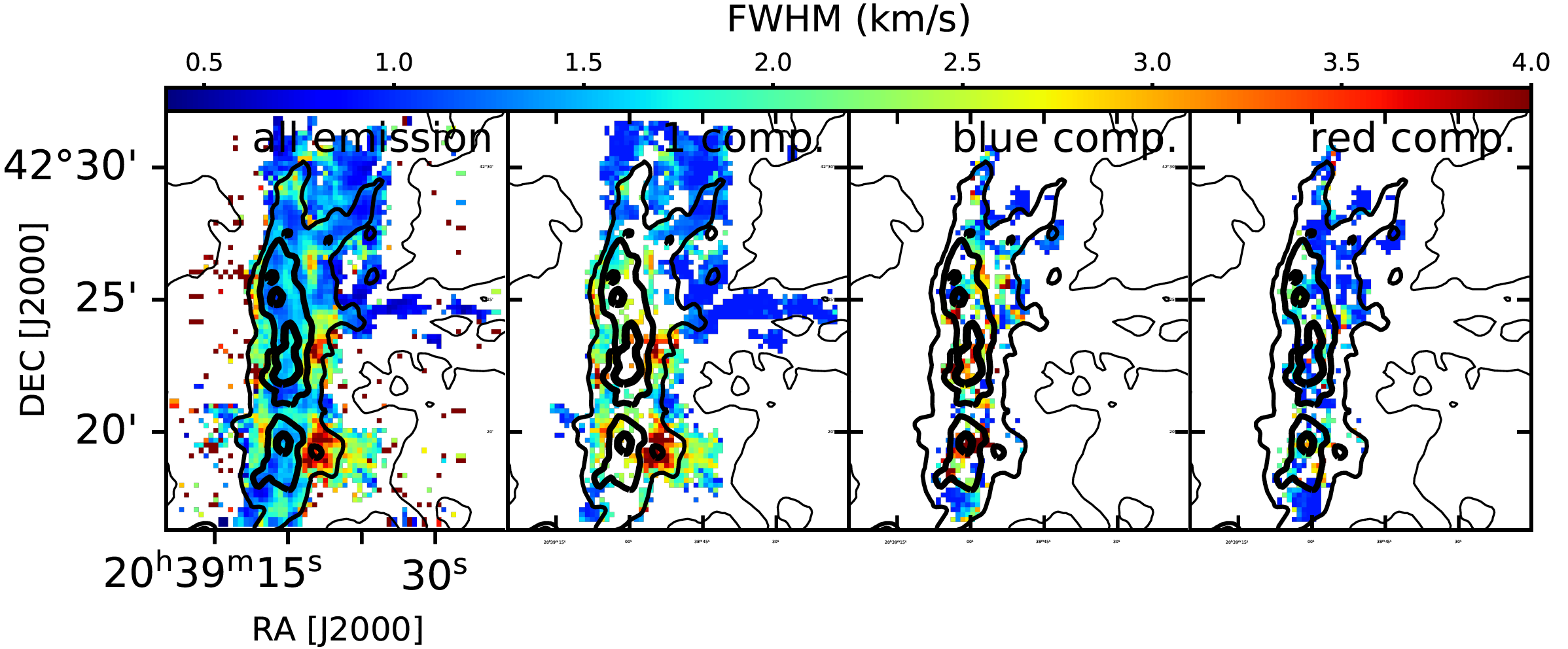}
    \caption{The same as Fig. \ref{fig:cii4figs} for C$^{18}$O(1-0).}
    \label{fig:c18o4figs}
\end{figure}

\begin{figure}
    \centering
    \includegraphics[width=0.325\hsize]{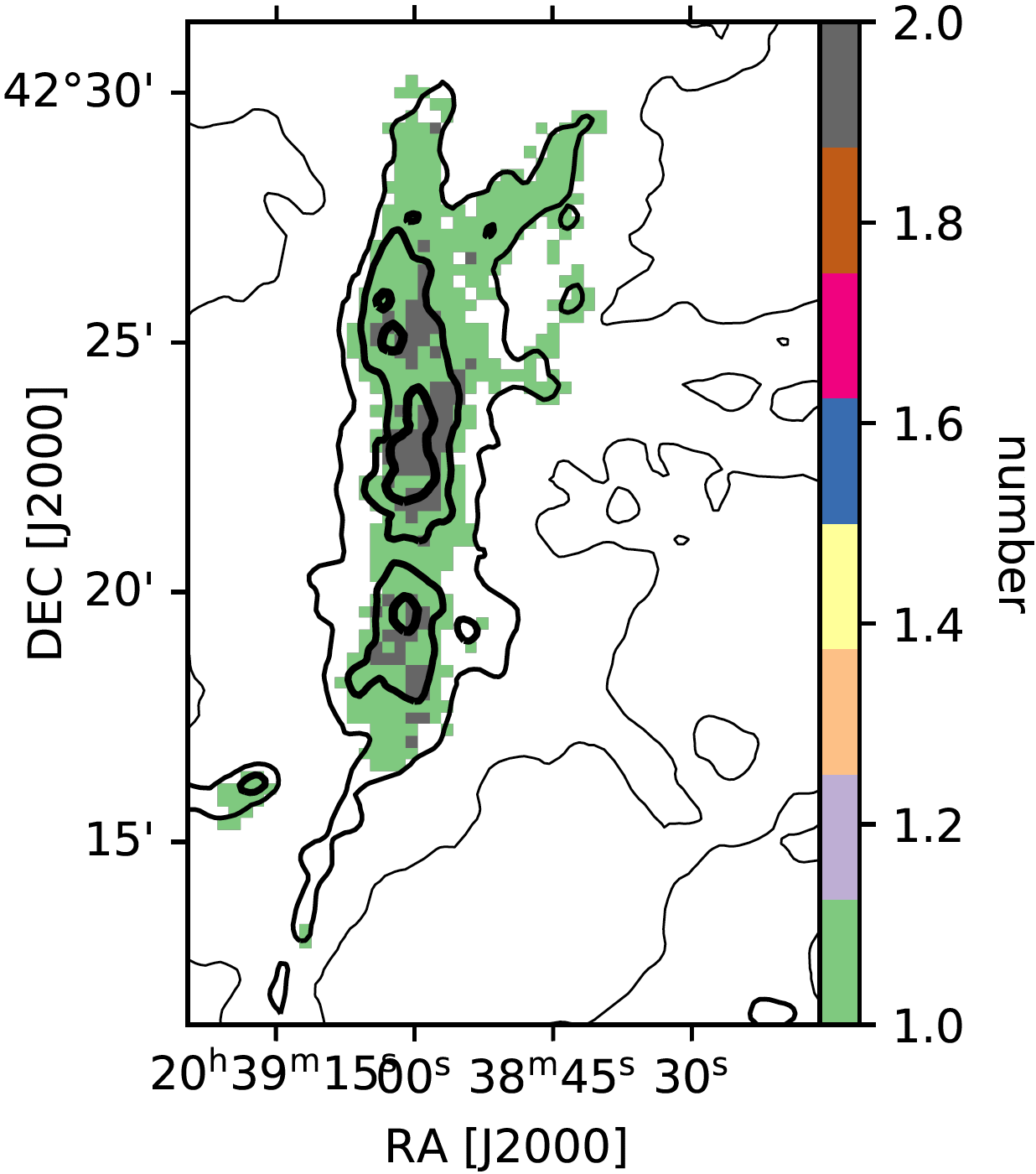}
    \includegraphics[width=0.32\hsize]{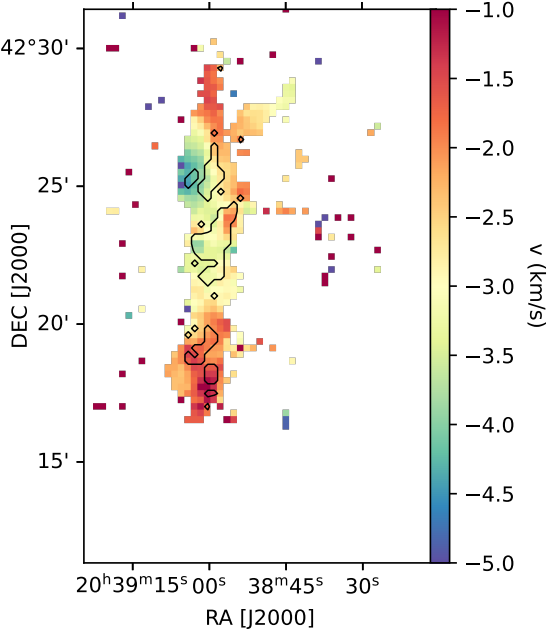}
    \includegraphics[width=0.31\hsize]{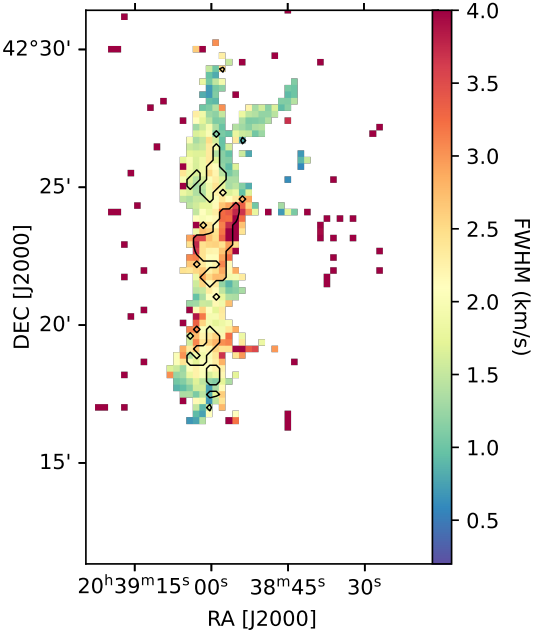}
    \caption{The same as Fig. \ref{fig:locationsC18Ofit} for H$^{13}$CO$^{+}$(1-0)}
    \label{fig:locationsH13COpfit}
\end{figure}

\begin{figure}
    \centering
    \includegraphics[width=0.5\hsize]{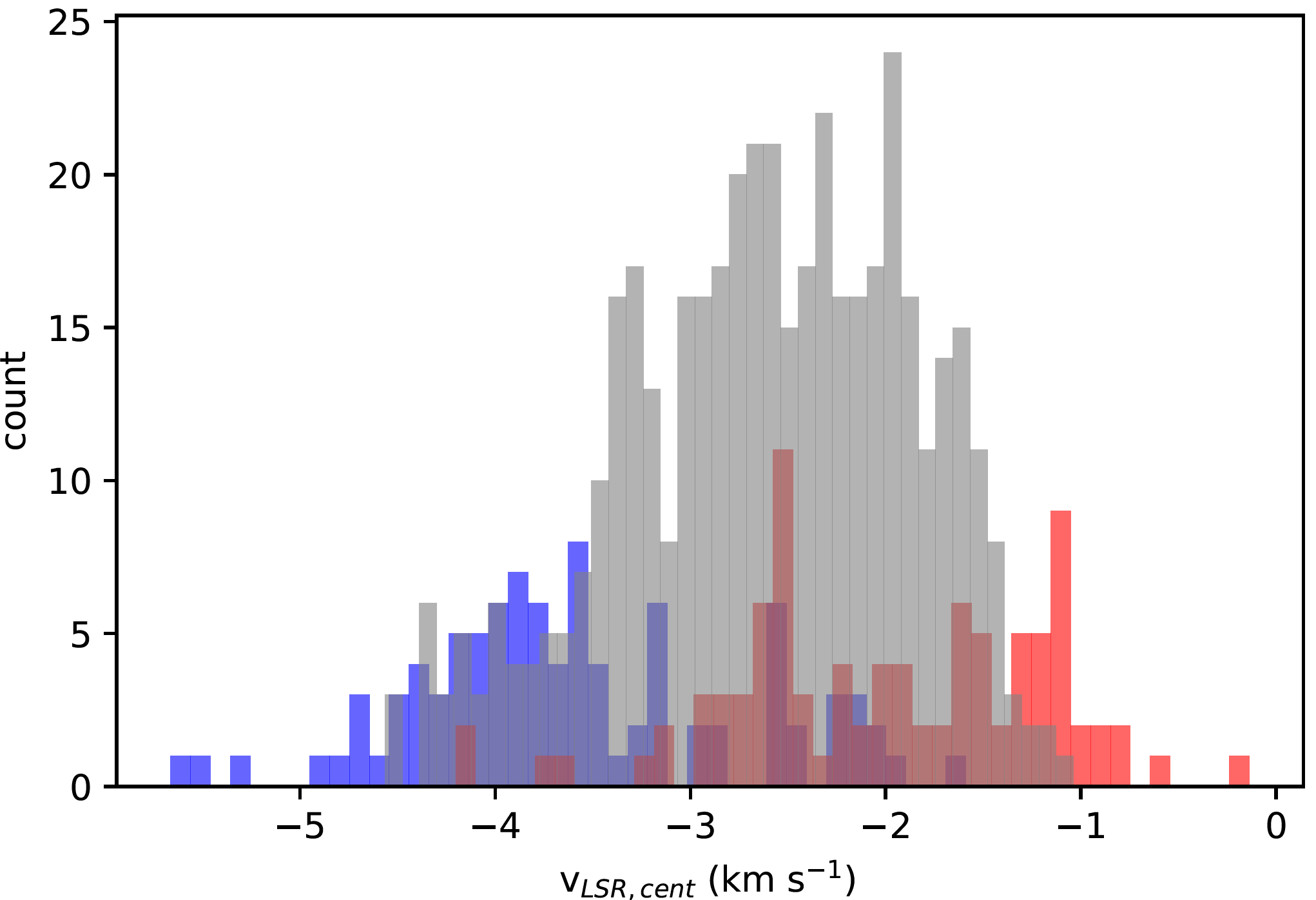}
    \caption{The same as Fig. \ref{fig:velHistC18O} for H$^{13}$CO$^{+}$(1-0)}
    \label{fig:velHistH13COp}
\end{figure}

\begin{figure}
    \centering
    \includegraphics[width=\hsize]{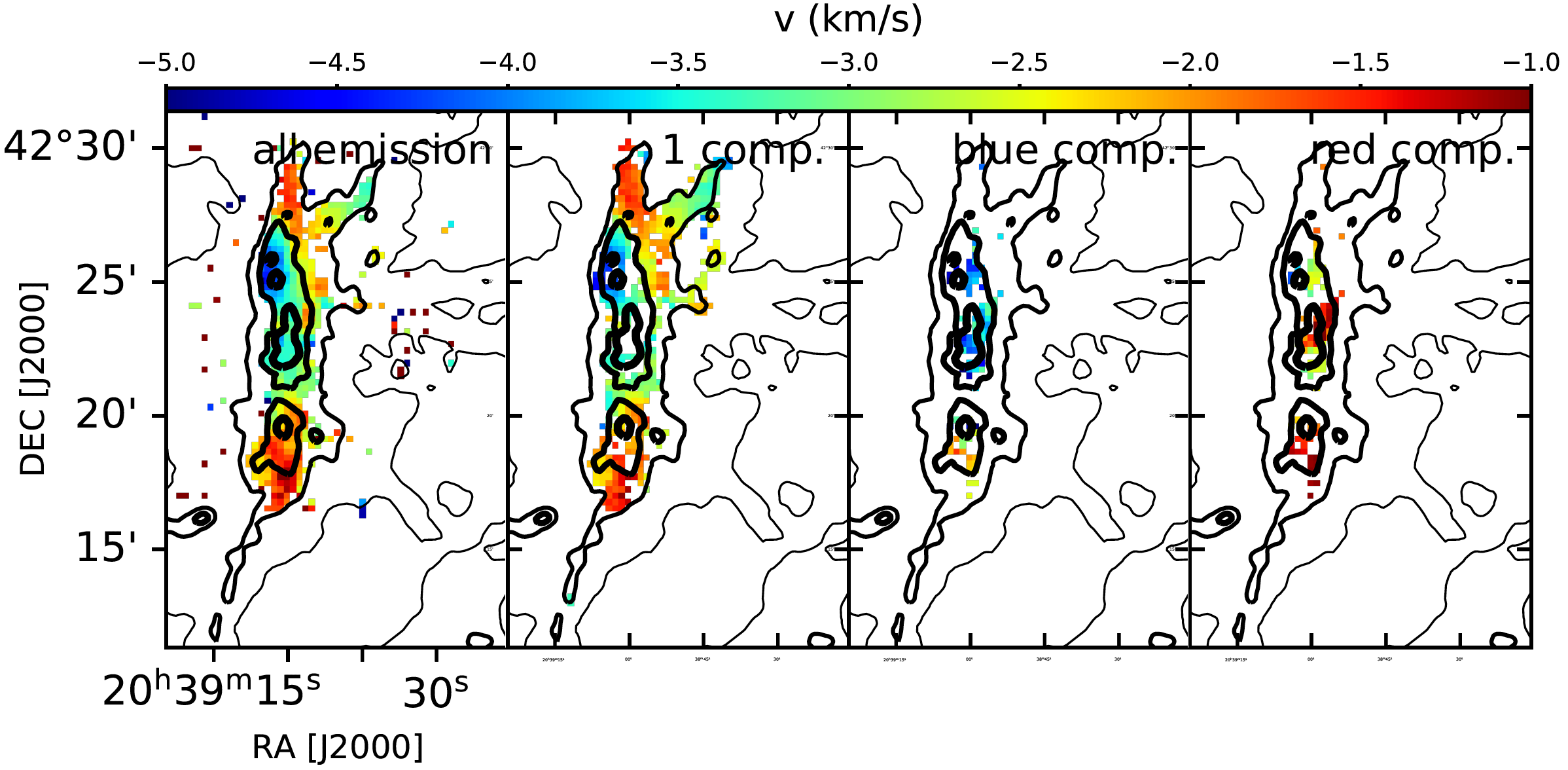}
    \includegraphics[width=\hsize]{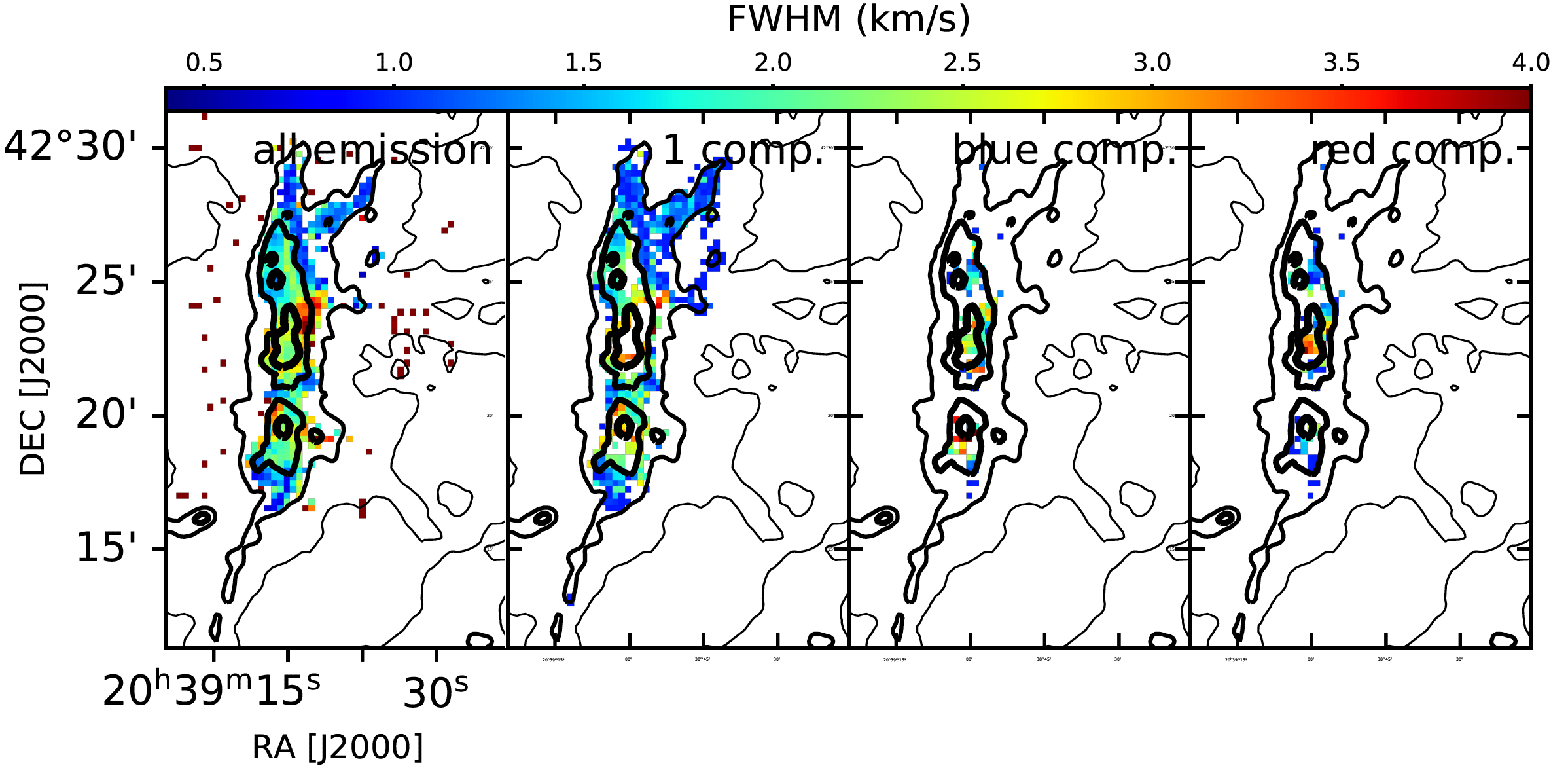}
    \caption{The same as Fig. \ref{fig:cii4figs} for H$^{13}$CO$^{+}$(1-0).}
    \label{fig:h13cop4figs}
\end{figure}

We also performed a multi-component fit with the BTS fitting tool to the C$^{18}$O(1-0) and H$^{13}$CO$^{+}$(1-0) molecular lines between -8 and 1 km s$^{-1}$ as they are not affected by strong opacity or self-absorption effects. This shows that there are significant regions that host two resolved velocity components in both lines, see Figs. \ref{fig:locationsC18Ofit} \& \ref{fig:locationsH13COpfit}. These regions are particularly in and at the edge of the DR21 ridge. It is noteworthy when inspecting these figures that the regions where the subfilaments connect to the ridge, which is associated with a high linewidth (FWHM $>$3 km s$^{-1}$), are fitted with a single velocity component. Either this suggests that several velocity components blend closely together there or that there might be a rapid reorientation in the line-of-sight velocity of the inflow or accretion shocks. From the velocity field maps, it can be observed that these regions are associated with a significant changes in velocity, see Figs. \ref{fig:c18o4figs} \& \ref{fig:h13cop4figs}. Inspecting the velocity distribution for the regions with two fitted components, see Figs. \ref{fig:velHistC18O} \& \ref{fig:velHistH13COp}, it is found that the interval for the redshifted velocity component is very similar for \CII\ and the molecular lines. This tends to confirm that the redshifted filaments are embedded in the redshifted inflowing mass reservoir. However, for the blueshifted velocity components, it appears that \CII\ emission is shifted towards more blueshifted velocities with respect to the emission in the molecular lines. This shift for the blueshifted \CII\ velocity component fits with the fact that there are no blueshifted subfilaments and thus that the blueshifted envelope of the DR21 ridge has a different morphology than the redshifted envelope.

\bibliography{sample631}{}
\bibliographystyle{aasjournal}

\end{document}